\documentclass[fleqn,usenatbib]{mnras}

\usepackage{newtxtext,newtxmath}

\usepackage[T1]{fontenc}
\usepackage{ae,aecompl}

\usepackage{graphicx}	
\usepackage{amsmath}	
\usepackage{color} 
\usepackage{rotating}
\usepackage{lscape}
\usepackage[normalem]{ulem} 

\def\ms{\mbox{$M_{\ast}$}}
\def\msun{\mbox{M$_{\odot}$}} 
\def\type{\mbox{\textsc{T-type}}}

\def\age{\mbox{$Age_{\rm lw}$}} 

\defcitealias{Fischer2019}{F19}
\defcitealias{Nair2010}{N10}
\defcitealias{Rodriguez-Puebla+2020}{RP20}
\defcitealias{DominguezSanchez2022}{DS22}

\definecolor{applegreen}{rgb}{0.55, 0.71, 0.0}
\definecolor{orange}{rgb}{1.0, 0.5, 0.0}
\definecolor{teal}{rgb}{0.0, 0.5, 0.5}

\title[MaNGA visual morphology]{Visual morphological classification of the full MaNGA DR17 sample: a general characterization.}

\author[J. A. V\'azquez-Mata et al.]{
J. A. V{\'a}zquez-Mata,$^{1}$\thanks{E-mail: jvazquez@astro.unam.mx}
H. M. Hern\'andez-Toledo,$^{1}$
V. Avila-Reese,$^{1}$ 
\newauthor{
A. Rodr\'iguez-Puebla,$^{1}$ 
L. A. Mart\'inez-V\'azquez,$^{1}$
M. Herrera-Endoqui,$^{1}$ 
I. Lacerna,$^{2}$ 
}
\newauthor{
L. C. Mascherpa,$^{3}$ 
and
D. F. Morell$^{1}$
}
\\
$^{1}$Universidad Nacional Aut\'onoma de M\'exico. Instituto de Astronom\'ia. A.P. 70–264, 04510. Ciudad de M\'exico, M\'exico\\
$^{2}$Instituto de Astronom\'ia y Ciencias Planetarias, Universidad de Atacama, Copayapu 485, Copiap\'o, Chile\\
$^{3}$Departamento de F\'isica, Facultad de Ciencias, Universidad Nacional Aut\'onoma de M\'exico, Ciudad Universitaria, CDMX, 04510, M\'exico\\
}

\date{Accepted XXX. Received YYY; in original form ZZZ}

\pubyear{2025}

\begin{document}
\label{firstpage}
\pagerange{\pageref{firstpage}--\pageref{lastpage}}
\maketitle


\begin{abstract}

We present the MaNGA Visual Morphology (MVM) catalogue, featuring a visual morphological classification of 10,059 galaxies in the final MaNGA sample. By combining SDSS and DESI Legacy Survey (DLS) images, we classified galaxies into 13 Hubble types, detected tidal features, categorized bars into different families, and estimated concentration, asymmetry, and clumpiness. 
The depth of the DLS images allowed us to identify structural details that were not evident in the SDSS images, resulting in a more reliable classification. 
After correcting for volume completeness, we find a bimodal distribution in galaxy morphology, with peaks in S0 and Scd types, and a transition zone around S0a–Sa types. Bars are present in 54\% of disc galaxies with inclinations $< 70^\circ$, following a bimodal trend with peaks in Sab-Sb and Scd-Sd types. Tidal structures are identified in $\sim$13\% of galaxies, particularly in massive E-Sa and low-mass Sdm-Irr galaxies. 
We derive the galaxy stellar mass function (GSMF) and decompose it into each morphological type. Schechter functions accurately describe the latter, while a triple Schechter function describes the total GSMF, associating three characteristic masses with different galaxy types. The abundance of early-type galaxies remains constant at low masses; they are predominantly satellites. 
We confirm that later-type galaxies are generally younger, bluer, more star-forming, and less metal-rich compared to early-type galaxies. Additionally, we find evidence connecting morphology and stellar mass to the star formation history of galaxies.
The MVM catalogue provides a robust dataset for investigating galaxy evolution, secular processes, and machine learning-based morphological classifications.

\end{abstract}




\section{Introduction}

The interplay among dynamic processes, star formation history, and the effects of the cosmic environment all shape the current morphology of galaxies. A robust morphological classification is therefore essential for comprehensively analyzing and interpreting the physical properties of galaxies in an evolutionary context. However, disentangling that evolutionary history, based on galaxy morphology, is a complex task that requires knowledge of other physical properties, estimated internally or globally, which are pieced together to devise evolutionary paths for different classes of galaxies.  

In recent years a series of Integral Field Spectroscopic (IFS) surveys, including CALIFA (Calar Alto Legacy Integral Field Area, \citealt{Sanchez2012,Sanchez2023}, SAMI \citep[Sydney-AAO Multi-object Integral field spectrograph,][]{Allen2015}, and MaNGA (Mapping Nearby Galaxies at Apache Point Observatory, \citealt{Bundy2015}),  
have conducted observations of galaxies within the local galaxy population. These observations enable the estimation of essential galaxy parameters on both global and spatially resolved scales \citep{GonzalezDelgado2015, Westfall2019, Law2021, Sanchez2022, Lacerda2022}. These data enables us to examine links between inner physical properties at various scales and the morphological features of galaxies \citep[e.g.,][]{CanoDiaz2019, Barrera-Ballesteros2022,Camps-Farina+2024}, as well as the evolutionary processes that transformed high redshifted galaxies into their current diverse morphological forms. Additionally, this new generation of IFS data offers new approaches to understanding the physical mechanisms behind the dynamical and chemical evolution across the Hubble sequence.

In this context, the MaNGA project is particularly relevant as it conducts IFS observations for the largest sample of local galaxies to achieve statistical significance. Throughout the development of this survey, and given the importance of having detailed morphological and structural information, different working groups have attempted to infer these properties, resulting in various catalogues delivered as Value Added Catalogues (VACs) and reported in the last data release, DR17, for the MaNGA survey \citep[][]{Abdurro-uf2022}. Among these, we mention:

\begin{itemize}

\item Galaxy Zoo 2 \citep[GZ2;][]{Willett2013}. GZ2 is a visually-based method that takes advantage of human pattern recognition capabilities using a decision tree scheme over the SDSS data release 7 (DR7) images. In this catalogue, the GalaxyZoo team (e.g. \citealt{Lintott2008}, \citealt{Willett2013}) cross-matched the GZ2 catalogue with MaNGA galaxies, covering approximately 80\% of the sample. This collective classification effort provides only a few broad morphological classes, not the detailed T-types of the Hubble sequence. 

\item Galaxy Zoo DECaLS \citep[][]{Walmsley2022} re-designed the original Galaxy Zoo decision tree and considered the deeper Dark Energy Camera Legacy Survey \citep[DECaLS][]{Dey2019} images ($r$ = 23.6 versus $r$ = 22.2 from SDSS) to improve the citizen classifications, helping to identify weak bars and minor mergers better. Using these data, they trained a Bayesian convolutional neural network (CNN) to predict classifications for an additional 174,000 galaxies, achieving 99\% accuracy for each question in the decision tree. Of this sample, 3620 galaxies are also in common with MaNGA.

\item The MaNGA Morphology Deep Learning DR17 Catalog (MDLM-VAC) provides morphological classifications for the final MaNGA DR17 galaxy sample using CNNs \citep[][]{DominguezSanchez2022}, based on the work by \citet{Nair2010} 
and the GZ2 catalogues. Among the morphological results, they report T-Type values (trained in regression mode, with values and errors coming from a k-folding method after training 10 models) with several additional advantages, such as improved recovery of the low-end of T-Types, better separation between early and late types, ellipticals and lenticulars, and the identification of edge-on and barred galaxies. 

\item Galaxy Zoo: 3D \citep[GZ:3D;][]{Masters2021} provides the identification of galaxy centres, foreground stars, galactic bars, and spiral arms for 29,831 galaxies using spatial pixel maps. Nearly all MaNGA galaxies are included in this sample. These masks can be used to select spectra or map quantities related to different structures and could serve as a training set for automated identification of spiral arms features.

\item The MaNGA PyMorph DR17 Photometric Catalog \citep[MPP- VAC;][]{Fischer2019, DominguezSanchez2022}, which provides the photometric parameters for MaNGA galaxies obtained from S\'ersic and S\'ersic+Exponential fits to the 2D surface brightness distributions in the $g, r$ and $i$ bands, using PyMorph \citep[][]{Vikram2010}, Source Extractor \citep[SExtractor;][]{Bertin1996} and GALFIT \citep[][]{Peng2002}. Among these parameters, we mention total fluxes, bulge-disc fractions, and position angles, among others.  

\item Visual Morphologies from SDSS + DESI Images (DR16+) \citep[][hereafter Paper I]{VazquezMata2022}. Unlike previous studies, the authors chose to perform a direct visual morphological classification for the MaNGA sample, presenting their results for the first half, approximately $\sim$4600 galaxies, from the MaNGA DR15 sample. They identified 12 morphological subtypes, ranging from ellipticals to irregulars in the standard Hubble scheme.
The classification was conducted after examining a series of image mosaics assembled by combining different post-processed images from SDSS and the DESI Legacy Imaging Surveys \citep[][]{Dey2019}. They listed the presence of bars, bright tidal features, inclined galaxies, systems with moderate and strong interactions, and also provided tentative classifications for cases with low signal-to-noise images. 

\end{itemize}

The results in Paper I demonstrate the advantages of using direct visual morphological classification of galaxies compared to previous methods. Among these, they mention: (i) the recognition of a wealth of morphological added value information identified after the post-processing of the DESI Legacy Survey images, (ii) the adoption of a purely light-based classification scheme for early type galaxies that, when combined with our post-processing image results, enables a more efficient separation among early E-S0-Sa types. Such a classification scheme was established on a more quantitative basis, retrieving information on the semi-minor to semi-major axes ($b/a$) and bulge to total ($B/T$) light ratios, resulting in a $b/a-B/T$ diagram, suited for large samples in upcoming surveys, and (iii) a detailed census of bars, bar families, and tidal features across the entire MaNGA sample, just to mention a few.  

In this paper, we build on the methods used in Paper I and present the most detailed and consistent visual morphological classification for the entire sample of 10,059 galaxies from the MaNGA survey. For the first time, we combine detailed morphological data with stellar population properties of MaNGA galaxies, derived through spectral inversion fitting using stellar population synthesis models \citep[pyPipe3D;][]{Sanchez2022}. After applying appropriate volume corrections, we conduct a statistical analysis of the MaNGA sample focusing on the Galaxy Stellar Mass Function, broken down by morphological types. We also consider available information on whether galaxies are central or satellite. 

The content of this paper is as follows. Section~\ref{sec:manga} outlines the procedures for morphological classification, the identification of various structural components, and the volume-completeness corrections. Section~\ref{sec:results} presents the results of our visual morphological classification for the 10,059 MaNGA galaxies. We also include a quantitative morphological analysis using our estimates of the non-parametric ($CAS$) structural parameters, ending with a description of the MaNGA Visual Morphology Value Added Catalogue (MVM-VAC). Section~\ref{sec:SMF} provides a statistical characterization of the MaNGA sample based on the Galaxy Stellar Mass Function (GSMF) and Morphology. Section~\ref{sec:stellar_prop} explores the relationships between stellar population properties and galaxy morphology, stellar mass, and classification as centrals and satellites. Section \ref{sec:discussion} compares our morphological findings with other related works, examining the connection of bars and tidal features with stellar mass and morphology in a two-dimensional context. Finally, Section~\ref{sec:conclusions} presents our conclusions. 
A series of Appendices complement our morphological results.
Appendix~\ref{App:morph-mass} summarizes in a table the relationship between stellar mass and morphology. Appendix~\ref{App:colour} demonstrates the method used to colour-code the various joint (bivariate) distributions with physical and morphological properties. Appendix~\ref{sec:TableBars} offers a more detailed summary of our visual identification of bars, including bar families and morphological types. Finally, Appendix~\ref{Sec:diffuse} provides information on the MaNGA galaxies that are difficult to classify.

For this work, we assume the cosmological parameters $\Omega_M$ = 0.3, $\Omega_{\Lambda}$ = 0.7, and $h$ = 0.71.


\section{Data Sample and Procedures}
\label{sec:manga}

MaNGA is one of the main projects of the Sloan IV phase, designed to conduct IFU observations using a set from 19 to 127 fibres covering from 12 to 32 arcseconds in projected diameter, respectively \citep{Drory+2015}. The MaNGA survey utilizes the two BOSS spectrographs at the 2.5-meter Sloan Foundation Telescope at Apache Point Observatory in New Mexico. The spectra cover a wavelength range of 3600-10300\AA, reaching a minimum signal-to-noise ratio of 5\AA$^{-1}$ at 1.5$R_e$ \citep{Law2015}, where $R_e$ is the $r-$band effective radius of each observed galaxy.

The observations cover up to $1.5R_e$ for $\sim 45\%$ of the sample (Primary) and up to $2.5R_e$ for $\sim 35.6\%$ of the sample (Secondary). They are supplemented by a colour-enhanced sample (up to $1.5R_e$; $\sim 15\%$ of targets) that over-represents unusual regions of the $(NUV - i)-M_i$ diagram, such as high-mass blue galaxies, low-mass red galaxies, and the “green valley” \citep{Law2015}, along with a small percentage ($\sim 5.3\%$; 537 galaxies) of the galaxies from ancillary programs.

The MaNGA survey \citep[][]{Bundy2015} has successfully achieved its goal of obtaining integral field spectroscopy for over 10,000 nearby galaxies. Out of 10,296 galaxy cubes, 10,145 have the highest data quality and from these, 10,126 correspond to unique targets identified by their MANGAID, forming our final data sample in the present paper.

\subsection{The Visual Morphological Classification}
\label{sec:methods}

This section briefly summarizes the procedures presented in \citet[][here after Paper I]{VazquezMata2022} for the visual morphological classification of 10,126 galaxies in the final MaNGA DR17 sample. It covers the classification of bars into bar families, the identification of tidal features, and our image processing methods for estimating non-parametric $CAS$ morphological parameters, based on the DESI images.

The morphological classification relies on visual inspection of the $r-$band SDSS DR7 images and the ninth public data release (DR9) of the DESI Legacy Imaging Surveys \citep[][]{Dey2019}, referred to as DESI images. These images are (i) post-processed to emphasize internal structures such as bars, spiral arms, rings, etc., as well as external features like outer spiral arms, rings, and low surface brightness tidal debris, and (ii) combined into mosaics that include colour composite images from SDSS and DESI, along with residual images from the post-processing catalogue for the DESI images \citep[see Paper I and][for more details]{Dey2019}. Note that the visual inspection was solely performed on these mosaics, which are static images. These working mosaics are accessible from the MVM-VAC website (mentioned above) and can be used for several purposes, including training algorithms for image classification.

We used images from three wide-area optical imaging surveys designed to provide targets for the Dark Energy Spectroscopic Instrument project \citep[DESI;][]{DESI2016}. In principle, DESI quality requirements lead to deeper images than those from SDSS, reaching a median surface brightness limit of $\sim$ 27.2 mag arcsec$^{-2}$ for a 5$\sigma$ detection \citep{Cano-Diaz2022}, which is more suitable for detecting morphological details. All post-processing was done while preserving the native pixel size of the DESI images and using a sufficiently large image area to ensure accurate background estimation and improved performance of contrast enhancement procedures.

\subsubsection {Morphological Considerations}
\label{sec:steps}

\begin{figure*}
\centering
 \includegraphics[width=0.7\textwidth]{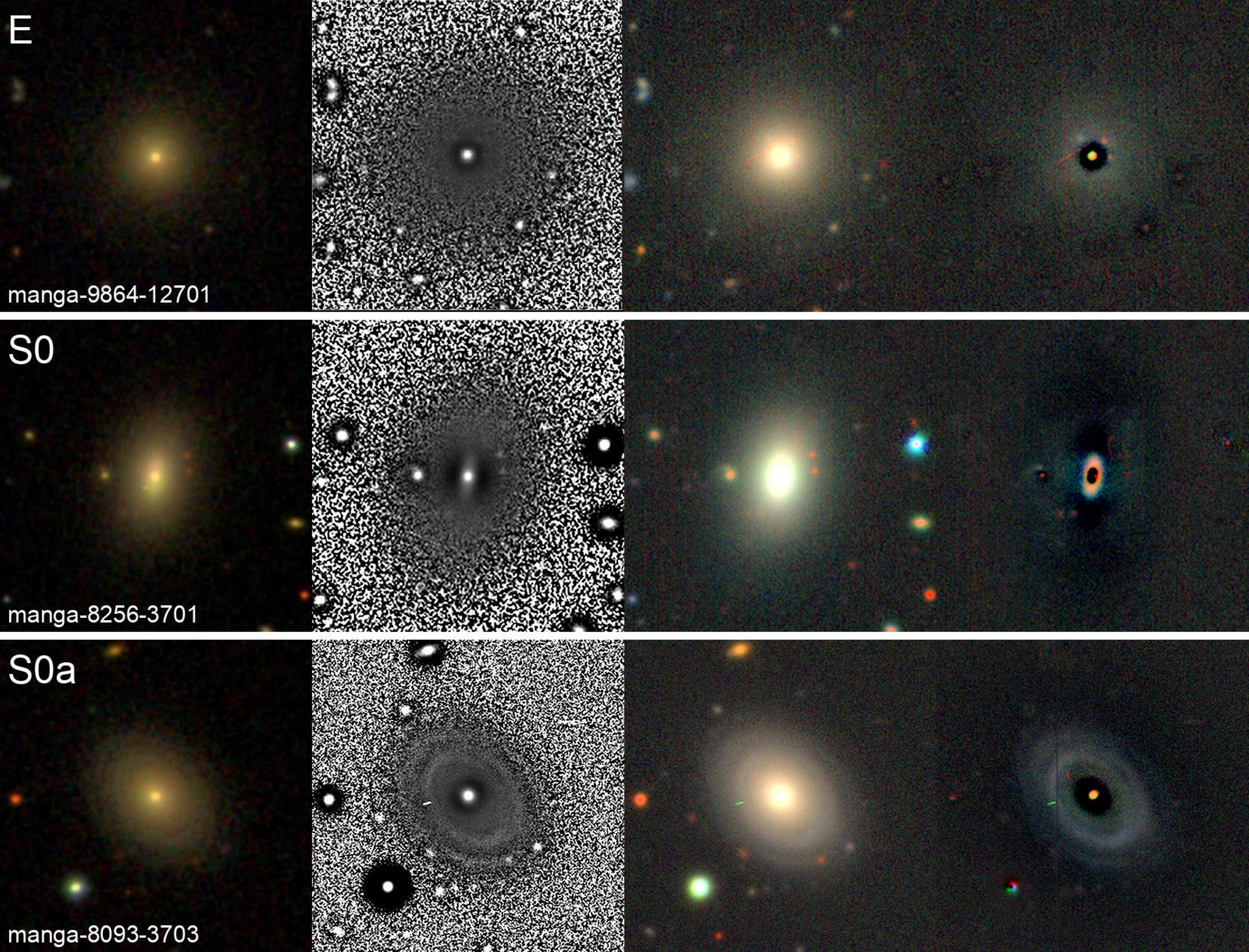}
 \caption{Poster-child of our classification scheme for early E, S0 and S0a types. For every mosaic, columns from left to right show the $gri$ SDSS image, the filtered-enhanced $r$-band DESI image, the $grz$ DESI image and the residual DESI image after a best 2D surface brightness model subtraction.} 
 \label{fig:morpho_early}
\end{figure*}

\begin{figure*}
\centering
 \includegraphics[width=0.7\textwidth]{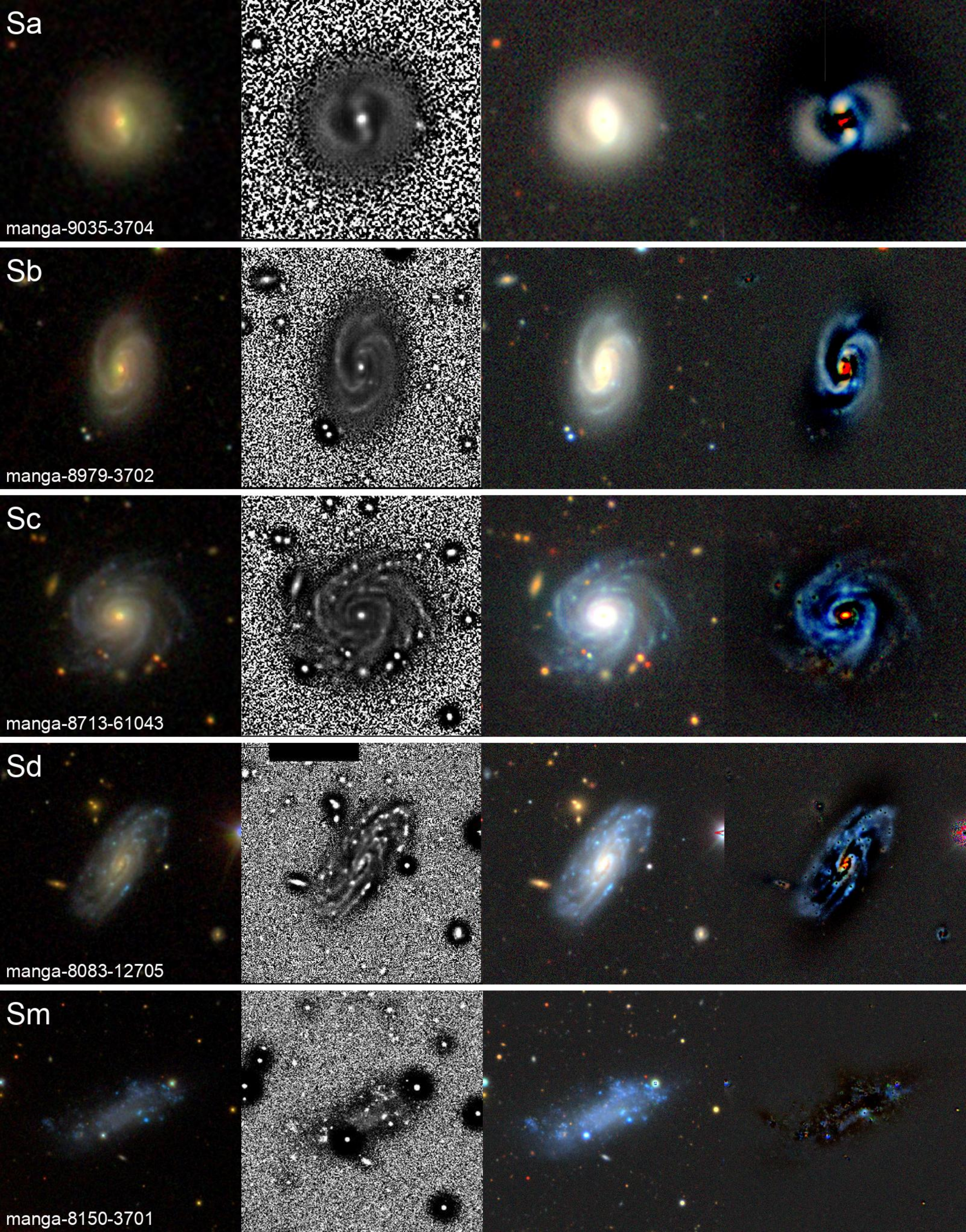}
 \caption{Poster-child of our late type classification scheme from Sa to Sm. Mosaics are similar to those in Figure~\ref{fig:morpho_early}.} 
 \label{fig:morpho_mosaic}
\end{figure*}

Our morphological classification of the DR17 MaNGA sample followed the basic Hubble scheme (E, S0, Sa, Sb, Sc, Sd, Sm, Irr, and their intermediate types), resulting in 12 identified types. 
Figures~\ref{fig:morpho_early} and \ref{fig:morpho_mosaic} display two mosaic images summarizing our morphological classification for early- and late-type galaxies, respectively. In these figures, each row contains four images: a colour-composite $gri$ SDSS image (left), a filter-enhanced $r-$band DESI image (middle left), a colour-composite $grz$ DESI image (middle right), and the corresponding residual DESI image (right) after subtracting a best-fitting model of its light distribution.  
As shown in Paper I, the filter-enhanced images combined with DESI residual images are valuable for identifying inner and outer morphological features. 

For early-type galaxies, we combined the procedures described in \citet[][]{HernandezToledo2010} with a purely light-based approach from \citet{Cheng2011} to classify early types into E, S0 and S0a Hubble types, respectively (see Sec 3.1.4 in Paper I for more details). 
This strategy, along with post-processing of the images, allowed us to identify early-type E galaxy candidates with potential inner embedded discs. These discs are mostly small, but in a few cases relatively larger, which helps distinguish between E, Edc, and S0 types \citep[see also][]{Graham2019}. 
This separation was further implemented in the $B/T-b/a$ diagram\footnote{B/T was obtained from the MPP-VAC, \citep{DominguezSanchez2022}.} obtaining a purity of 81\% and a completeness of 72\%, comparable to those values reported by \citet{Cheng2011} for bulge-dominated galaxies. Alternatively, we used the $C-b/a$ diagram\footnote{C: Petrosian 90\% light radius/Petrosian 50\% light radius. b/a: semi-minor to semi-major axis ratio. Both were obtained from the NSA catalogue.}, obtaining a purity of 78\% and completeness of 78\%, which we suggest as an effective alternative diagram for early-type separation suitable for large surveys. 

For disc galaxies, we evaluated the prominence of the central light concentration and the presence of weak to strong disc-like features. We also examined the resolution of the spiral arms and how tightly or loosely they are wound. At the end of the late-type spiral sequence, we noted whether faint central bulges are present or absent, as well as the flocculence of the arms made up of individual stellar clusters and clumps, typical of Sd (SBd) types, to irregular-shaped bulgeless galaxies of the Sm (SBm) type, and highly irregular shapes of Im type.

The MaNGA sample also includes a significant portion of moderate and highly inclined galaxies ($\sim$20\%). In these cases, the depth and post-processing of the DESI images provide abundant morphological information in the outskirts of discs, revealing faint outer spiral arms and other low-surface brightness structures.

We emphasize that the AGN flux contribution to the central regions of galaxies is generally low and highly concentrated, with minimal impact on our visual assessment of the central light prominence in galaxy images. \citet{HernandezToledo2023} analysed the relative flux contribution of type I and II AGNs to the optical light in the central regions of MaNGA galaxies through detailed 2D B/D/Bar/PSF image decomposition, finding that the point-source AGN flux contribution never exceeded 15-20\%,  except in two cases where it approached nearly 50\%. It is also important to note that the presence of AGN (type I and II) in the overall MaNGA sample is around 4-7\%\citep[e.g.,][]{Sanchez2018,Comerford+2020,HernandezToledo2023, Comerford2024}. We conclude that neither the frequency of AGNs, nor their flux contribution to the light in the central regions of optical galaxy images, significantly affects our criteria for visually classifying galaxies. 

\subsubsection{Bar Classification}
\label{Sec:BarClasificaion}

The identification of bars presented here is based on visual assessment of the most obvious and easily recognisable cases. In practice, an initial classification considered no bars (SA), suspected bars (SAB) and confirmed bars (SB) across different Hubble types (for S0 and spiral galaxies). Using filter-enhanced and residual images helped improve bar detection. This approach includes identifying small bars hidden by clumpy or dusty structures or by dominant light concentrations in the central regions of galaxies. Bars aligned with spiral arms and or along the major axis of discs, in moderately to highly inclined galaxies, were also considered as much as possible. 

Once bar candidates were identified, a new visual inspection was performed to classify the bar families following the Comprehensive de Vaucouleurs revised Hubble-Sandage (CVRHS) system \citep{Buta2015}. Galaxies were reclassified as unbarred (A), weakly barred (AB), strongly barred (B), or intermediate cases ($\underline{A}$B and A$\underline{B}$). The classifications are as follows:

\begin{itemize}

\item A: no visible trace of a bar.

\item $\underline{A}$B: usually a roundish feature, such as an oval, that is not significantly brighter than the surrounding disc.

\item AB: a clearly visible bar in the inner regions of the galaxy, often formed by the inner parts of spiral arms.

\item A$\underline{B}$: a prominent, noticeable bar, which may be surrounded by structures such as lenses.

\item B: a conspicuous, straight bar with high contrast compared to other structures like the disc.

\end{itemize}

\begin{figure}
\centering
 \includegraphics[width=0.9\columnwidth]{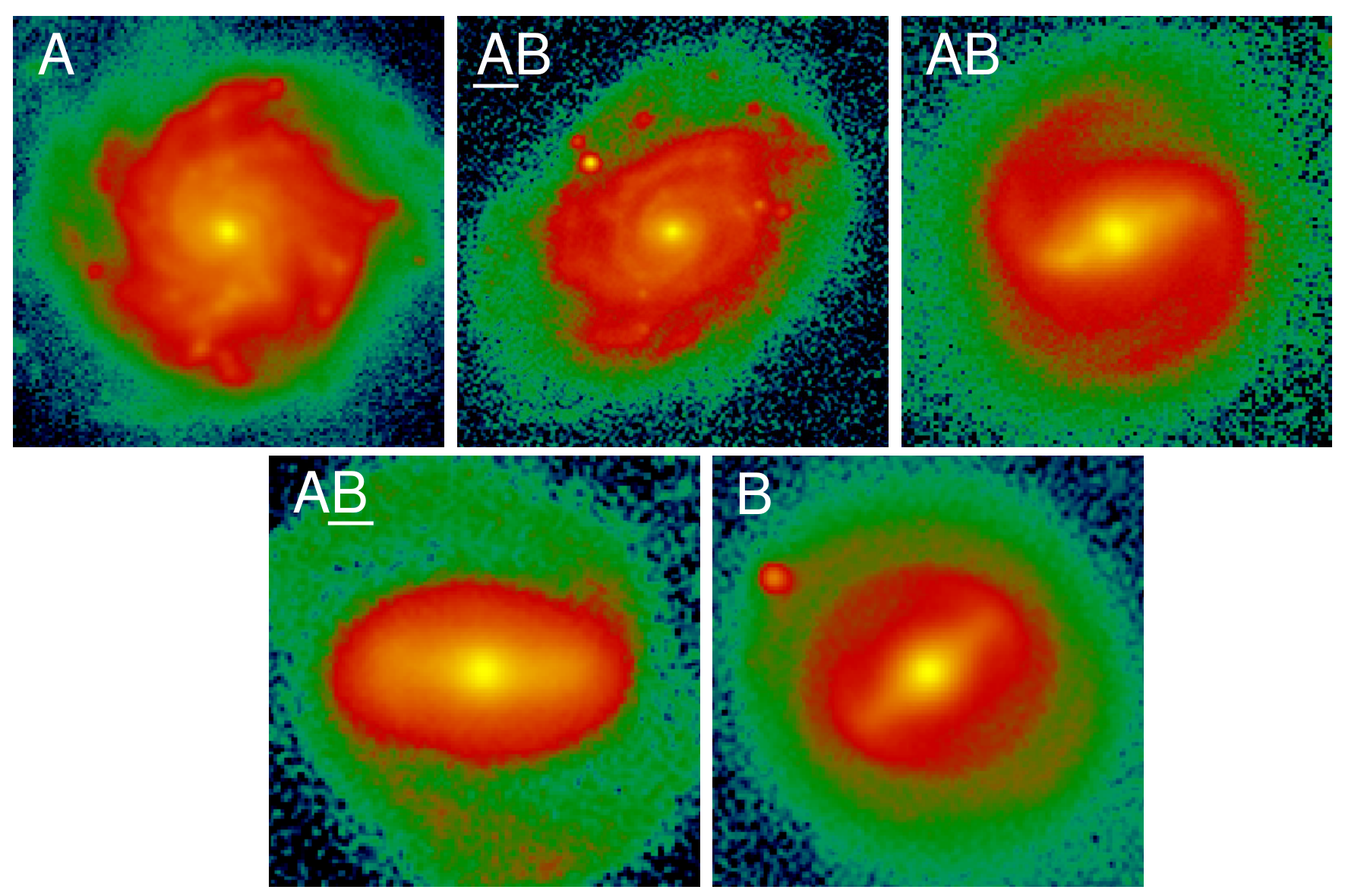}
 \caption{Bar family classification examples following the CVRHS system \citep{Buta2015}. From left to right: (A) no-barred galaxy, ($\underline{A}$B) weak bar showing a roundish structure, (AB) bar in the inner region of the galaxy, (A$\underline{B}$) clear conspicuous bar, and (B) bar with high contrast with respect to other structures. Note that AB, A$\underline{B}$ and B families are strong bars. See more details in Sec. \ref{Sec:BarClasificaion}.} 
 \label{fig:Bar_class}
\end{figure}

Figure \ref{fig:Bar_class} is a mosaic of images illustrating the bar family classification scheme. The top three panels show a non-barred galaxy (A), a weakly barred galaxy where the bar appears as an oval structure ($\underline{A}$B), and a barred galaxy where the bar is formed by the inner parts of the spiral arms, so that these bars may have a curved appearance (AB). The bottom two panels depict a bar 
surrounded by a lens-like structure of similar size and orientation (A$\underline{B}$), and finally, a strongly barred galaxy (B) where a straight bar has swept all the material inside its radius. These strong bars typically feature barlenses (roundish central components embedded in bars) in their inner regions \citep[e.g.,][]{HerreraEndoqui2017}.

It is important to consider the weakly barred family, as weak bars (ovals) also create a non-axisymmetric potential that influences the dynamical and morphological properties of galaxy discs. For instance, observations show that the fraction of inner structures such as rings and lenses is higher in intermediate and weak families than in very strong bars \citep{Laurikainen2013, HerreraEndoqui2015}. 
The filtered images, combined with the residual DESI images, are valuable tools for visually identifying disc structures closely associated with bars, such as X/Boxy/Peanut-bulges, ansae, barlenses, inner lenses, and rings, which can be studied further in MaNGA galaxy research.

\subsubsection{Tidal Features identification}
\label{Sec:Tides}

As part of our release, we report the visual identification of the most prominent and brighter tidal features, including streams, filaments, shells, fans, and structures likely caused by strong mergers and/or close or nearby interactions, either from the primary galaxy or from the companion(s) being stripped. For this, only the original colour composite $gri$ SDSS and $grz$ DESI images were used, without additional post-processing. In this case, we assign a binary flag (\emph{Tides}), with \emph{Tides} $=$ 1 indicating the presence of any of the mentioned tidal features, while \emph{Tides} $=$ 0 means no visual evidence of them. A more detailed analysis, using enhanced image post-processing techniques to better detect tidal features, relevant to understanding their origin or nature, will be reported elsewhere.

\subsection{Non-parametric morphology: CAS parameters}\label{subsec:non-param morph}

Concentration (C), asymmetry (A), and clumpiness (S), hereafter called the \emph{CAS system} \citep[e.g.][]{Conselice2003}, form a set of parameters designed to capture major features of the structures in galaxies without assuming any specific form, unlike parametric fitting. To estimate the \emph{CAS} parameters for the  full MaNGA sample, we also used the $r$ band images from the DESI images (DR9), which include an improved version of the reduction procedures for the Legacy Surveys.

\begin{figure}
\centering
 \includegraphics[width=0.9\columnwidth]{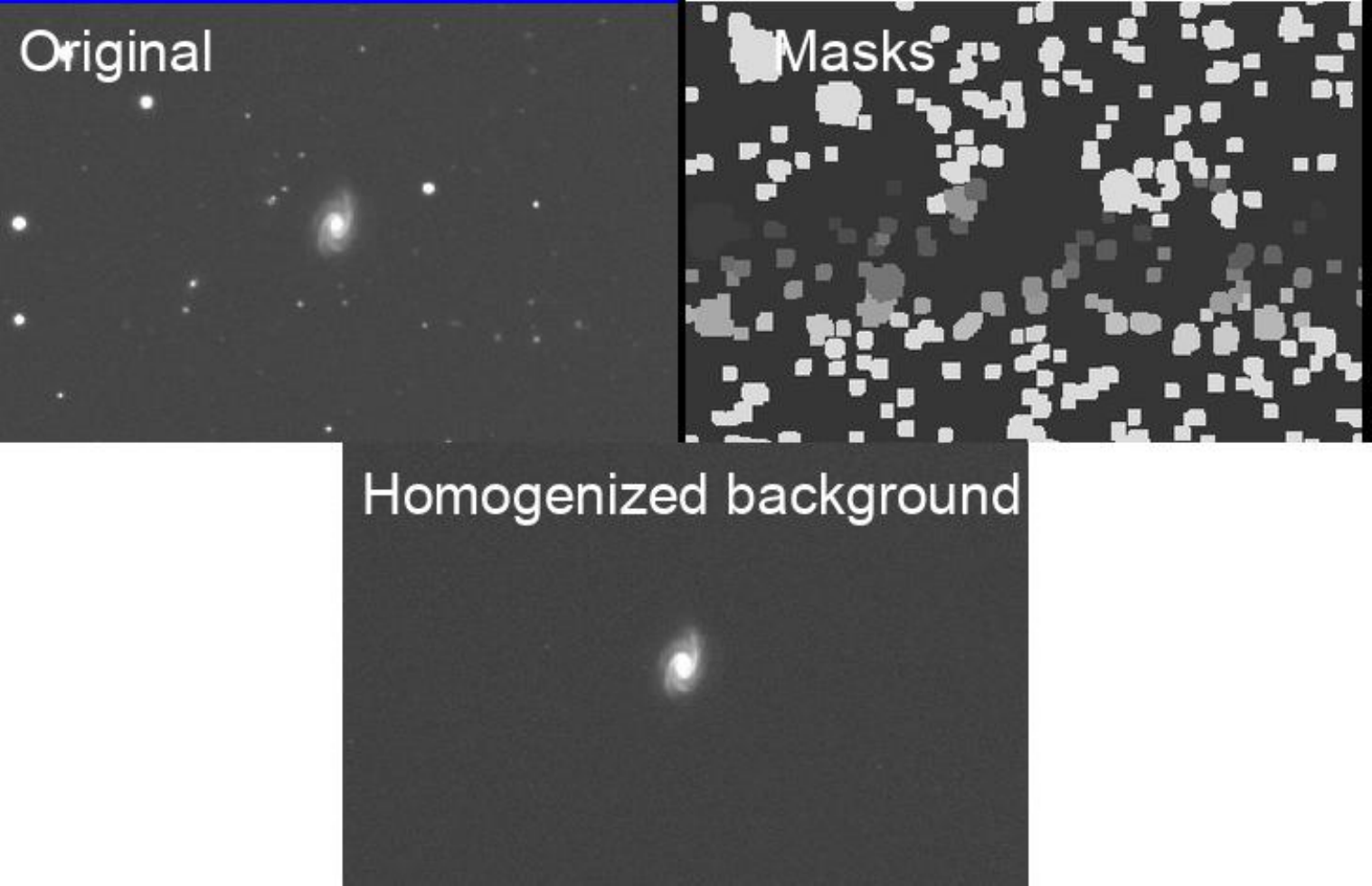}
 \caption{ 
 Background replacement and homogeneization of images (see subsection \ref{subsec:non-param morph}) previous to $CAS$ estimates. Panels: original image, computed masks, background replacement and homogeneization.}
 \label{fig:cas_img}
\end{figure}

The images for each MaNGA target were retrieved centred on a 800 x 800 pixel frame using the native scale (0.262 arcsec/pix). A mask was created by defining an annulus enclosing each target at two radii, rmin and rmax, where pixel values are used to replace those inside the mask. 
Once the mask and the target regions rmin and rmax are set, the pixel values belonging to the background were saved in a list. The next step is to replace all the pixel values within the mask with a random selection of pixels from the background list. For this random pixel selection, the \texttt{ran3} function described in the Numerical Recipes library \citep{Press2007} was used to avoid using a linear congruential method. This process results in images free of stars, nearby galaxies, and other contaminant objects, with a homogeneous background around each MaNGA galaxy from which the \emph{CAS} parameters were measured. Figure~\ref{fig:cas_img} shows an example of this procedure. 

\emph{CAS} parameters were estimated within 1.5 times the Petrosian inverted radius at r($\eta = 0.2$), following \citet[][]{Conselice2003}. The values of the \emph{CAS} parameters and their corresponding errors are reported in the MVM-VAC DR17 presented in this work.

\subsection{Volume completeness corrections}
\label{sec:Vcorr}

The MaNGA sample was selected to have a roughly flat distribution in log\ms \citep[][]{Wake2017}, making this sample incomplete in properties such as stellar mass, colour and morphology. Another key goal of this paper, beyond presenting our detailed morphological analysis, is to conduct a statistical characterization of the MaNGA sample through the Stellar Mass Function divided by morphological types. Achieving this requires addressing these incompleteness issues by introducing appropriate volume corrections. 
Volume corrections are defined as the expected fraction of galaxies as a function of stellar mass, colour and redshift, needed to recover the local Galaxy Stellar Mass Function (GSMF), as explained in e.g., \citealt[][and Calete et al. (in prep.)]{Sanchez+2019, Rodriguez-Puebla+2020, VazquezMata2022}. For detailed information on how the volume corrections were calculated, we refer the reader to those papers, especially Section 3.2 of \citet{VazquezMata2022}. It is important to note that these corrections are reliable for galaxies with $\log(M_\ast/M_\odot)>$ 9.2. Therefore, in the following sections, when we mention volume-corrected fractions, the galaxy count involved totals 8,763, excluding galaxies below this stellar mass.

\section{Morphological and structural properties of MaNGA galaxies}  
\label{sec:results}
 
\subsection{MaNGA DR17 Morphological Content}
\label{sec:morpho_results}

The morphological evaluation of the entire DR17 MaNGA sample included as many galaxies as possible. As explained in Sect.~\ref{sec:manga}, we assembled our final sample for morphological classification from 10,126 datacubes identified with a unique ID. Different subgroups were identified based on specific properties. For example, about 1,602 highly inclined galaxies, $> 70^{\circ}$, were identified using their aspect ratio ($b/a$) from the NSA Atlas \citep{Blanton2011}. If the high inclination is relaxed to $i> 60^{\circ}$, this number increases to $\sim$ 3,200 galaxies. These numbers are important for our analysis, as described below.

Galaxies displaying light contamination from scattered light haloes, which originate from large nearby galaxies or cases of weakly interacting galaxies, were tentatively classified when possible. In instances where faint or diffuse objects had ambiguous morphological features, a tentative type was assigned and flagged as “unsure”. The "unsure" galaxies constitute $\approx10$\% of the full MaNGA DR17 sample (1,114 galaxies). The reader is referred to Appendix~\ref{Sec:diffuse} for more details on galaxies that are difficult to classify. 
Additionally, 67 galaxies show clear signs of interaction and strong perturbations, making them unclassifiable and thus excluded from our morphological analysis. However, they are listed in Table~\ref{tab:morpho} for reference and included in the final catalogue presented in Section~\ref{sec:VAC}.

\begin{figure*}
 \centering
 \includegraphics[width=0.7\textwidth]{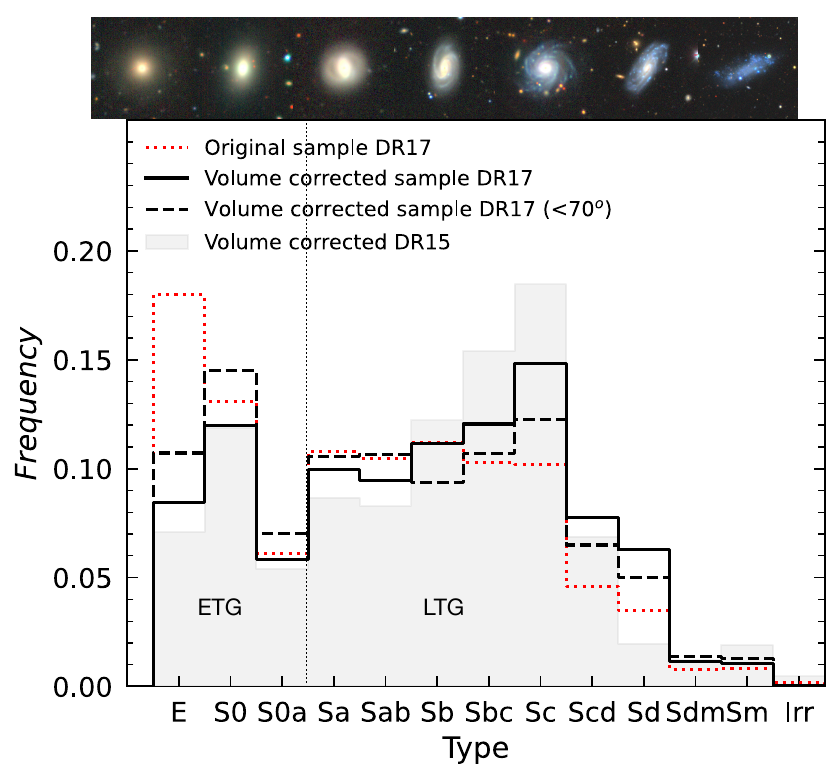}
 \caption{Morphology histogram showing the results of the classification carried out in this work. Every distribution represents the complete MaNGA sample (dotted-red line), the volume-corrected sample (solid-black line), the corresponding volume-corrected sample excluding galaxies with inclinations $> 70^\circ$ (dashed-black line), and the previous distribution presented in Paper I for the DR15 MaNGA sample (shadow gray area). Note that i) these distributions are normalized to the total number of galaxies in each sample, and ii) the volume-corrected distributions, only galaxies with $\log (\ms/M_{\sun})\geq$9.2 are considered. The vertical dotted line illustrates the separation between early and late type galaxies adopted in this work.
 }
 \label{fig:hist_morpho}
\end{figure*}

Therefore, this work provides a morphological classification for 10,059 galaxies. Figure~\ref{fig:hist_morpho} shows the normalized morphological distribution of the original (red-dotted line) and volume corrected (black-solid line) samples. The main effects of the volume corrections, which aim to match the local GSMF, are to decrease the frequencies of early-type galaxies and increase those of types later than Sb.
For comparison, the gray-shaded histogram corresponds to the volume-corrected morphological distribution presented in Paper I for the MaNGA DR15 sample (4611 galaxies). We observe some differences between these morphological distributions. The DR17 distribution is now less steep from Sa to Sc galaxies and less peaked at the Sbc-Sc types than in the DR15, also showing a higher frequency of Sd galaxies compared to DR15. As stated in Sec~\ref{sec:steps}, the morphological assignment at the end of the late-type spiral sequence was based on the presence or absence of faint central bulges and the flocculence of the arms. Based on that, we found that a subsample of Sbc and Sc types in DR15 actually has fainter bulges than previously assigned, now corresponding to Scd and Sd types (a correction of the order $\Delta \type \pm 1$). This correction has been made and reported in the latest version of the MVM-VAC (see subsect. \ref{sec:VAC}). For reference, we also present the volume-corrected sample excluding galaxies with inclinations $> 70^{\circ}$.

Figure~\ref{fig:hist_morpho} shows evidence of a bimodal morphological distribution in the volume-corrected DR17 sample, with E (8.4\%) and S0 (11.9\%) types making up 20.3\% on one side, and a broader distribution spanning from Sa to Irr types (73.9\%) on the other, with a peak in Sc type. The ``valley'' in the distribution corresponds to S0a type (5.8\%), similar to what was observed in the DR15 sample. Note that if we do not distinguish S0a galaxies from S0 galaxies, the first peak would be in the S0 type with 17.7\%, with a valley at the Sa-Sab types and the second peak in the Sc types (15\%). From this point forward, we will refer to galaxies of types E--S0a as early-type galaxies (ETG), and those of types Sa--Irr as late-type galaxies (LTG).

Table~\ref{tab:morpho} presents the morphological evaluation for 10059 MaNGA galaxies, including 67 'unclassified' cases labelled as S*. \textit{Col. 1} lists the discrete morphological types. \textit{Col. 2} shows the number of galaxies in each type. \textit{Col. 3} displays the fraction relative to the total sample for each type. \textit{Col. 4} provides the volume-corrected fractions for the MaNGA DR17 galaxies.
The lower panel of Table~\ref{tab:morpho} reports the 
fractions of edge-on galaxies ($> 70^{\circ}$), galaxies with visible tidal features, barred galaxies, and galaxies with a Faint-Diffuse-Compact appearance that were flagged as difficult to classify, and that are discussed in Appendix~\ref{Sec:diffuse} in more detail.

\begin{table}
  \begin{center}
    \caption{The morphological evaluation of the MaNGA sample. Number and Fraction are referred to the total number in the sample, as well as the inferred fraction when volume-corrections are applied. The bottom panel show the results for edge-on galaxies, galaxies with tides or bars, and galaxies showing a faint-diffused-compact aspect on the images and that were difficult to classify (see Appendix A). Notice that Volume-corrected fractions are reliable only for galaxies with $M_{\ast}>10^{9.2}M_{\odot}$ and thus only galaxies above this limit are accounted (number in parenthesis).}  
    \label{tab:morpho}
    \begin{tabular}{c c c c}
    \hline
      \textbf{Type} & \textbf{N} & \textbf{Fraction} & \textbf{Volume-corr} \\
              &   &  & \textbf{fraction} \\
      \hline
E & 1809 & 0.179 & 0.084 ( 1597 ) \\
S0 & 1317 & 0.13 & 0.119 ( 1175 ) \\
S0a & 615 & 0.061 & 0.058 ( 540 ) \\
Sa & 1085 & 0.107 & 0.099 ( 948 ) \\
Sab & 1051 & 0.104 & 0.094 ( 948 ) \\
Sb & 1126 & 0.111 & 0.111 ( 990 ) \\
Sbc & 1034 & 0.102 & 0.12 ( 910 ) \\
Sc & 1026 & 0.101 & 0.148 ( 881 ) \\
Scd & 462 & 0.046 & 0.077 ( 372 ) \\
Sd & 352 & 0.035 & 0.063 ( 265 ) \\
Sdm & 79 & 0.008 & 0.011 ( 48 ) \\
Sm & 84 & 0.008 & 0.011 ( 40 ) \\
Irr & 19 & 0.002 & 0.0 ( 3 ) \\
S$^*$ & 67 & 0.007 & 0.005 ( 46 ) \\
    \hline
    Edge-on ($> 70^{\circ}$) & 1602 & 0.19 & 0.23\\
    Tides & 1807 & 0.18 & 0.13\\
    Barred & 3779 & 0.45 &  0.46\\
    Faint-Diffuse-Compact & 1114 & 0.11 & 0.17\\
    \hline
    \multicolumn{4}{l|}{$^*$S refers to galaxies that were not classified but showing an} \\
    \multicolumn{4}{l|}{apparent disc structure, or are strong mergers.} \\
    \end{tabular}
  \end{center}
\end{table}

\begin{figure*}
 \includegraphics[width=0.4\textwidth]{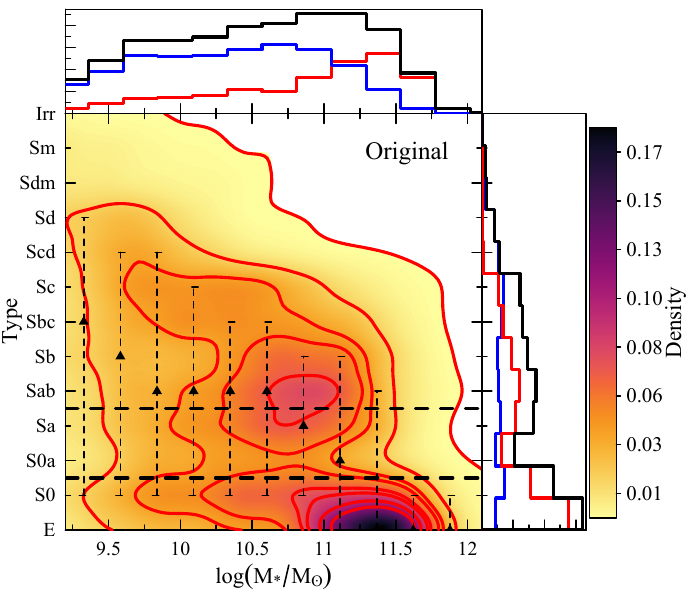}
 \qquad
 \includegraphics[width=0.4\textwidth]{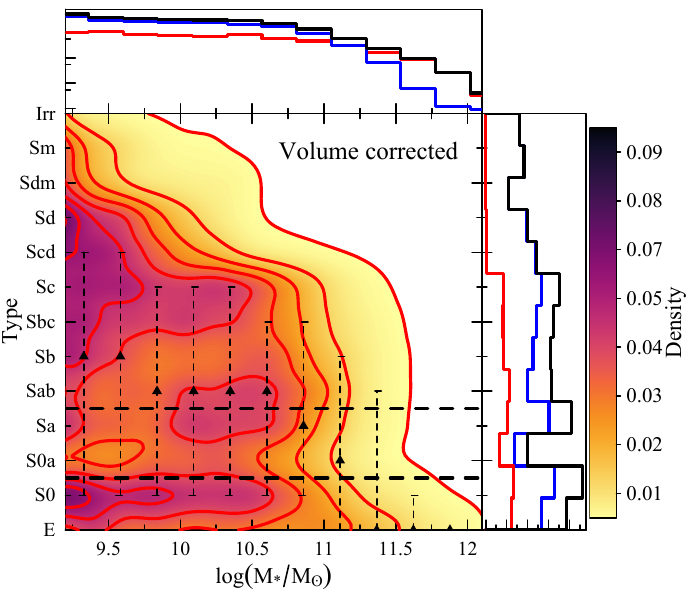}
  
 \caption{Bivariate distribution of the MaNGA galaxies with $i\le70^\circ$ in the morphological type versus \ms\ diagram, not taking (\textit{left panels}) and taking (\textit{right panels}) into account volume completeness corrections. The bivariate densities are shown with the isocontours and colours in linear scale representing a fraction from the total sample, with the last isocontour equivalent to 0.0005 of the normalized number density. Filled triangles and bars show the median and quartiles in morphology given a stellar mass bin. Horizontal dashed lines show the transition zone of the bimodal distribution.
 The top and right histograms in each panel show the 1D distributions of stellar mass and morphology. These are divided by ETG (red) and LTG (blue) and by massive (log(\ms/\msun)$\ge 10.5$; red line), and low massive ($9.2\le$log(\ms/\msun)$<10.5$; blue line) galaxies.
 The raster figures are constructed from a regular grid of 10 x 10 bin in all the diagram intervals. At the centre of each pixel, we obtained the weighted median of the number density galaxy population, considering the entire sample, by associating weights to each galaxy according to a bivariate normal distribution. The kernel widths of the bivariate distribution in the \textit{x} and \textit{y} directions are obtained automatically following \citet[][]{Silverman1986} rule. Note the volume corrections are additionally considered. For more details see Appendix~\ref{App:colour}.}

 \label{fig:mass_morpho}
\end{figure*}

It is important to note that the morphological type frequencies shown in Fig. \ref{fig:hist_morpho} are volume-complete only for masses log(\ms/\msun)>$ 9.2$. As is well known, lower mass galaxies tend to be later types on average (see also Fig. \ref{fig:mass_morpho}) \citep[e.g.][]{Nair2010}. Consequently, very late-type galaxies are strongly underrepresented in our mass-limited sample. The volume-corrected frequency of galaxies by type heavily depends on the mass limit. That is why it is important to study the bivariate distribution of mass and type (see below).

Figure~\ref{fig:mass_morpho} displays the joint (bivariate) distribution of \ms\ and \ \type\ for the full MaNGA DR17 sample as isocontours of relative densities {\it linearly spaced}, with labels shown in the colour palette (see App~\ref{App:colour} for more details). 
The stellar masses were adopted from the NASA-Sloan Atlas Catalogue \citep[NSA,][]{Blanton2011}, derived by a S\'ersic fit and adjusted to match the cosmology used in this paper. Two distributions are presented: one for the original sample (left panel) and another for the volume-corrected sample (right panel) for $\log(\ms/\msun)\geq 9.2$. Galaxies with inclinations $i > 70^\circ$ were omitted from both distributions.

Table~\ref{tab:mass} in Appendix \ref{App:morph-mass} presents the numbers and fractions of galaxies based on the \type\ across several stellar mass bins, summarizing the relationship between mass and morphology for both the original and volume-corrected data. The volume-corrected data in the right panel of Fig.~\ref{fig:mass_morpho} reveal an approximately bimodal grouping in the \ms\--\type\ diagram,  with a transitional area (fewer objects) around S0a-Sa galaxies. The starry symbols and their error bars emphasize the well-known trend that lower mass correlates with later morphological types, although with significant scatter. At the high-mass end, where $\log(\ms/\msun)>11$, most galaxies are early types, while at the low-mass end, where $\log(\ms/\msun)<10$, galaxies of nearly all morphologies are present, exhibiting an approximate bimodal distribution with the highest abundances in the Sc-Scd and in the S0 types.

The last contour in the \ms--\type\ diagram roughly corresponds to $5\times10^{-4}$ of the normalized number density, and indicates a limit beyond which local galaxies are very unlikely to exist (see Section \ref{sec:implications} for a discussion). 
This limit shows that at low masses, with log(\ms/\msun)$\lesssim 9.7$, all morphological types are possible. However, the higher the mass, the less likely late types are to be present. For log(\ms/\msun)$>10.5$, types later than Sc are extremely rare, and for log(\ms/\msun)$>11.5$, the limit applies for types later than S0, so that, very massive galaxies can only be of early types.

The 1D mass and type distributions are shown on the top and right-hand sides of each panel in Fig.~\ref{fig:mass_morpho}. Note that, for the volume-corrected data, the 1D mass distribution actually represents the galaxy stellar mass function (GSMF; omitting high inclined galaxies), shown by the black line. This is broken down into LTGs (Sa--Irr) and ETGs (E--S0a), depicted by the blue and red lines, respectively. In Section \ref{sec:SMF}, we discuss the GSMF and its breakdown into morphological types in more detail. 
The 1D distributions of \type's are displayed in Fig. \ref{fig:hist_morpho}, but here they are split into high-mass galaxies with log(\ms/\msun)$\ge 10.5$ (red line) and low mass galaxies with $9.2\le$log(\ms/\msun)$<10.5$ (blue line). Note that only galaxies with $i<70^\circ$ are included. For volume-corrected data, the frequency of massive galaxies is roughly similar across all types up to type Sc; later types are very rare among massive galaxies. For less massive galaxies, the morphological distribution peaks at Scd--Sd and also at S0 types. Additionally, S0 galaxies show a relatively constant density at intermediate masses ($<3\times 10^{10}$\msun), which declines sharply at higher masses ($>5\times 10^{10}$\msun).

\subsection{Bars}
\label{sec:bars}

\begin{figure*}
 \includegraphics[width=0.8\textwidth]{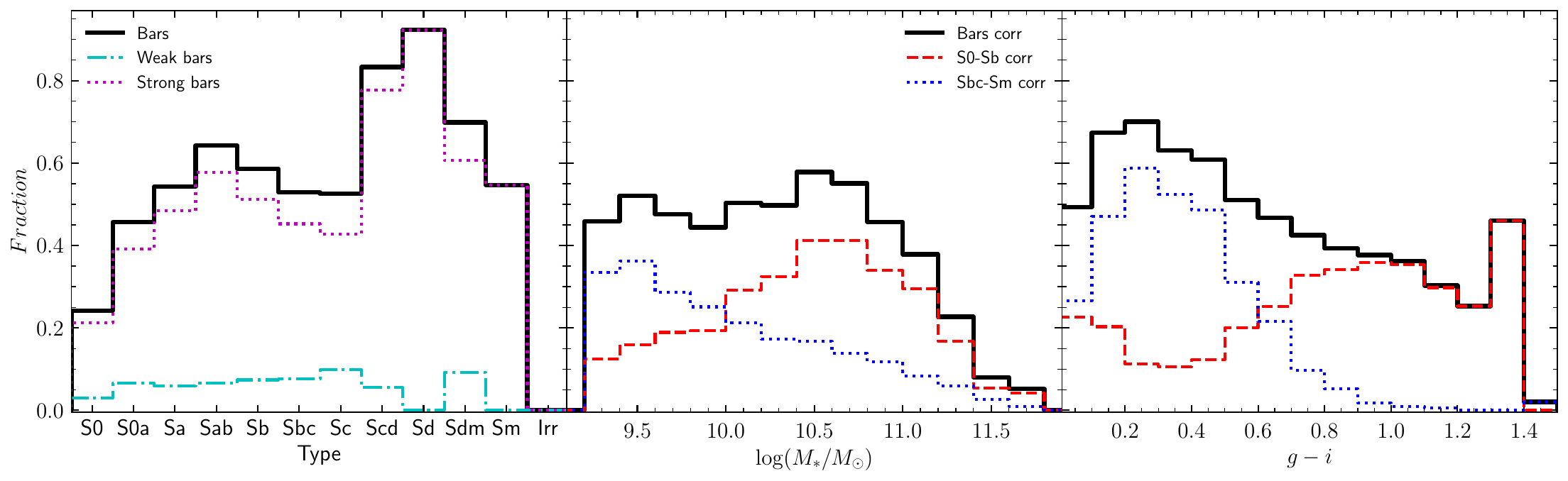}
 \caption{
 Fractions per bin (volume-corrected) of barred galaxies (solid-black lines) as functions of morphology, stellar mass and \emph{g-i} colour. 
 The first panel show the fraction distribution separated in Bar families: strong bars (AB, A$\underline{B}$, B) in dotted-magenta and weak bars ($\underline{A}$B) in cyan-dashed line. 
 The last panels present the fraction distributions divided into early (S0-Sb; red-dashed line) and late (Sbc-Sm; blue-dotted line) type discs. Note: fraction distribution are defined as the number of barred galaxies in each bin divided by the total number of galaxies in that bin. Only galaxies with inclinations $<$70$^\circ$ are considered. Galaxy colours are corrected as described in Sec~\ref{sec:masstypes}.}
 \label{fig:Bar_fam}
\end{figure*}

Out of 8,317 \textit{disc} galaxies (S0-Irr) in the MaNGA DR17 sample (regardless of inclination), 3,779 galaxies show evidence of a bar, resulting in a visually identified bar fraction of 45.4\%. This figure slightly increases to 46.2\% when considering volume-corrected fractions. When an inclination restriction ($i < 70^\circ$) is applied, these fractions become 52\% and 54\%, respectively.

Dividing the sample into two morphological groups -earlier type discs (S0-Sb) and later type discs (Sbc-Irr)- at all inclinations, there are 2291 (60.6\%) and 1488 (39.4\%) galaxies, with corresponding volume-corrected fractions of 49.4\% and 50.6\%. When additionally imposing an inclination restriction ($i < 70^\circ$), these fractions adjust to 61.7\% and 38.3\% for earlier and later-type discs, with volume-corrected fractions of 50.8\% and 49.2\%, respectively.

The resulting global bar fraction in the MaNGA sample is not significantly different from the 48-52\% reported by \citet{Barazza2008} and 45\% by \citet{Aguerri2009}, who analysed optical SDSS images in local samples, despite using different methods such as ellipse fitting and Fourier analysis. More recently, \citet{Quilley2022} report that 52\% of the EFIGI \citep[Extraction de Formes Id\'ealis\'ees de Galaxies en Imagerie,][]{Baillard2011} galaxies from types S0 to Im have a bar feature, which aligns with our findings, although this last is a magnitude-limited sample.  

Our morphological evaluation also categorizes bars into bar families, as described in subsect.~\ref{Sec:BarClasificaion}. Figure~\ref{fig:Bar_fam} shows the volume-corrected fractions per bin of barred galaxies (the number of barred galaxies divided by the total galaxies in each bin), based on their host morphological type (left panel), stellar mass (middle panel), and $g-i$ colour (right panel). The first panel also displays the contribution from weak ($\underline{A}$B) and strong families (AB, A$\underline{B}$, B). As observed, the contribution of the weak family to our bar distribution is underrepresented. This matches the findings from other bar family identifications \citep[e.g.,][]{Buta2015}, which suggest that weak bar families are underrepresented due to the limitations of visual classification. For completeness, we include our visual classification of bar families in Appendix~\ref{sec:TableBars}. The following panels show the contribution of two groups to these distributions: (S0-Sb) and (Sbc-Sm) morphological types groups. 

There is a clear bimodality in the bar fraction based on the morphological type (left panel), with a first peak ($\sim$ 65\%) in early Sab type discs, a second peak ($\sim$ 85\%) in late Sd type discs, and a transition region ($\sim$ 50\%) at Sbc/Sc types.
The positions of the two morphological peaks, as well as the size of the associated fractions, match the results from the Spitzer Survey of Stellar Structure in Galaxies (S$^4$G) barred galaxies, as described in \citet{Buta2015} and \citet{DiazGarcia2016}, based on a visual classification. An important point from this last paper is that the amplitude of the bar fraction distribution, after visual identification, is consistently higher than that obtained through other quantitative methods. However, the overall shape of the distribution (bimodality) remains stable, regardless of the bar identification method used.

The middle panel shows a more consistent pattern of the bar fraction with stellar mass (around 50\%), exhibiting a decreasing trend for masses above $\log (M_{\ast}/M_{\sun}) \approx$ 10.6 up to our highest mass limit. The reader should notice the sharp cut-off at the lower mass end, $\log (M_{\ast}/M_{\sun}) \sim 9.2$ (our completeness limit), which may hinder the detection of trends at lower stellar masses. When looking at two broad morphological groups, there appears to be an initial bimodality with possible mass peaks at $\log (M_{\ast}/M_{\sun}) \approx$ 9.5 and 10.5, corresponding to the contributions of late and early type discs, respectively.  

Finally, the right panel of Figure~\ref{fig:Bar_fam} shows a monotonic decreasing trend of the bar fraction with ($g-i$) colour, with a slight increase only in the last colour bin, ($g-i$)$> 1.3$.

\citet{Erwin2018} used distance- and mass-limited subsamples of the S$^4$G sample to investigate the properties of bars and their dependences on mass and colour, providing us with a framework to compare our results. They find that the bar frequency peaks at about 70\% at $\log (M_{\ast}/M_{\sun}) \sim 9.7$, decreasing at both lower and higher masses, while regarding colour dependence, they find that it remains roughly constant over a wide range of colours ($g-r \sim$ 0.1–0.8), although a slight decrease trend with colour is also observed. 
Note, however, that these findings are based on samples where S0 galaxies were excluded due to incompleteness. When they include the S0s, the distribution trend becomes flatter in the mass range  9.2 < $\log (M_{\ast}/M_{\sun}) < 11.5$, which is somewhat similar to our results. Our findings on the colour dependence of the bar fraction are also roughly consistent with  \citet{Erwin2018}, although we observe a steeper decrease in our data. This difference may be due to the absence of S0 galaxies in their samples, which makes their slope less steep at redder colours than ours. 

On the other hand \citet{Geron2021}, using a volume-limited sample from  Galaxy-Zoo DECaLS \citep[][]{Walmsley2022}, found a bar fraction dependency with stellar masses similar to that in our work, identifying two peaks at stellar mass values similar to our peaks observed from the contribution of late and early type discs. However, note that \citet{Geron2021} evidences a bimodality in the bar fraction with $g-r$ colour, in contrast to \citet{Erwin2018} and our work.

Differences in bar detection methods and bar definitions may account for the variations in the observed distributions, particularly since certain methods may be more sensitive to weaker or smaller bars than others. Our image processing, in contrast to the decision three and the preparation of the images in GZ-DECaLS, may contribute to this, for example, in cases where the light of a prominent bulge screens or hides a bar, particularly for bars smaller in angular size. The effect of this on the observed distributions will depend on whether the sizes vary significantly with galaxy properties, like colour, which is an analysis at the moment outside of the scope of the present paper.

The presence of significant features in the bar distribution, such as bimodalities based on properties like stellar mass and colour, may be relevant. Other studies, such as \citet{Baldry2004}, have reported a characteristic mass $\log (M_{\ast}/M_{\sun}) = 10.2$, where a bimodality in galaxy properties has been observed, suggesting a link between the red and blue sequences and the origin and evolution of bars.

\subsection{Tidal features}

\begin{figure}
\centering
 \includegraphics[width=0.8\columnwidth]{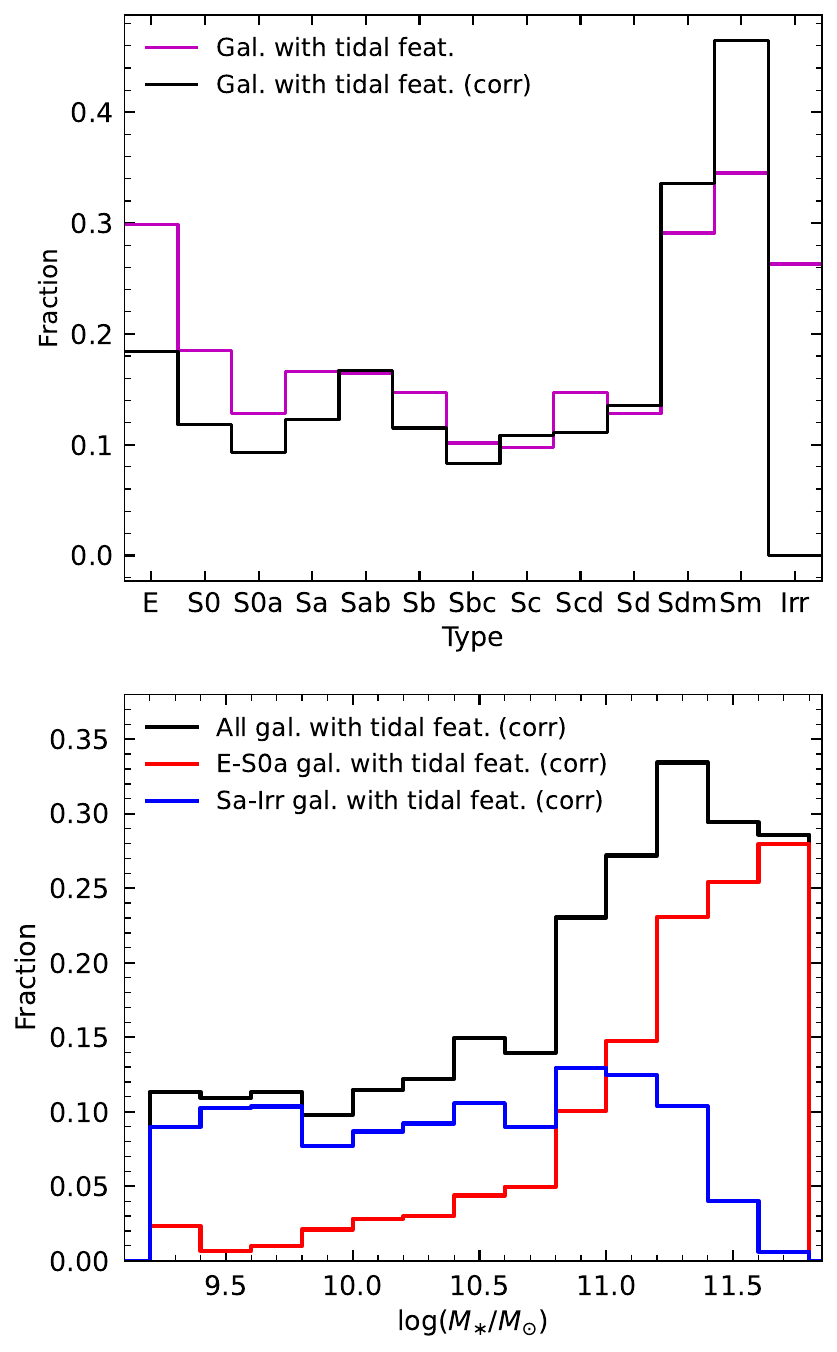}
 \caption{Distribution of the fraction of galaxies with tidal features per bin of morphology (upper panel) and stellar mass (lower panel), for the original (solid-magenta-line) and volume-corrected (solid-black-line) samples.  
 Tidal debris are divided into early (red) and late (blue) types in the lower panel. Note that these fractions mean the number of galaxies with tidal features in each bin divided by the total number of galaxies in the same bin. }

 \label{fig:tides_hist}
\end{figure}

Our results on the frequency of tidal features in the MaNGA sample are based on visual inspection of the most noticeable (bright) tidal features near each galaxy, using the available DESI composite $grz$ colour images. Note that no additional post-processing was applied to the images to highlight tidal features, so the reported fraction should be viewed as a lower limit on their actual frequency. Out of 1,0059 galaxies of all morphological types in the MaNGA sample, 1,807 (17.9\% or 13\% in the volume-corrected case) show bright tidal features of any of the types previously mentioned (see Section~\ref{Sec:Tides}).

The upper panel of Figure~\ref{fig:tides_hist} displays the original (solid magenta line) and volume-corrected (solid black line) fractions of tidal features around galaxies for each morphological type, while the lower panel shows the volume-corrected fractions in stellar mass bins for all galaxies (black line), separated into LTGs and ETGs, represented by blue and red lines, respectively. 
Our findings reveal that tidal features are present throughout the entire morphological domain in the MaNGA sample. The fractions range from about 20\% for E galaxies to roughly 40\% for Sdm--Sm types. However, the reader should note that the sample size in the extreme late type bins is quite small, so these fractions might be overestimated. Although the volume-corrected fraction for irregular galaxies approaches zero, 
the lower panel indicates that the fraction of tidal features steadily increases with stellar masses (black line). Additionally, when dividing this distribution into two  morphological groups, we observe a relatively constant contribution from LTGs (blue line) across most mass ranges, whereas for ETGs (red line), there is a consistent increase in the fraction as stellar mass increases. 

Although comparing our results on tidal features with other studies is challenging due to differences in sample selection, environment, observational definitions, and surface brightness and depth limits, we performed a first-order comparison with various theoretical and observational works.
The fraction of tidal features in the ETG group is 23\% (13.4\% volume-corrected), which exceeds the 15\% (12.9\% volume-corrected) in the LTG group. As a rough comparison, \citet{Hood2018} reported a fraction of 13 $\pm$ 3\% in their gas-poor sample and 19 $\pm$ 2\% in their gas-rich sample. \citet{Bilek+2020} found that the incidence of shells, streams, and tails is approximately 10-15 \% for each category, in their volume-limited sample of 177 nearby massive ETG galaxies from the MATLAS survey \citep{Duc+2015}. The overall fraction rises to about 30\% for galaxies displaying tails, streams, or similar features.

Recently, \citet{Rutherford+2024} found that 31 $\pm$ 2\% of ETGs with $\log (M_{\ast}/M_{\sun}) >$ 10 exhibit tidal features (shells and/or streams) in the SAMI Galaxy Survey \citep{Croom+2021}, using deep imaging from the Subaru-Hyper Suprime-Cam. This fraction is higher compared to our 18.4\% when applying the same mass limit, most likely due to differences in the images depth used in each sample.
In numerical simulations, \citet{Valenzuela_Remus_2022arXiv220808443} reported that 18 $\pm$ 3 of ETG galaxies with stellar mass $M_{\ast} \geq 2.4 \times 10^{10} M_{\sun}$ show at least one type of tidal feature (shells, streams, or tidal arms) in the hydrodynamical cosmological simulation Magneticum Pathfinder\footnote{http://www.magneticum.org}. This fraction is similar to ours 22.7\% in the same mass limit. The variation in these fractions is caused by differences in morphological definitions and the thresholds set for surface brightness, stellar mass, and galaxy distance.  
More recently, \cite{YoonY+2024} reported that the fraction of tidal features in nearby massive ($M_{\ast} > 1.6 \times 10^{11} M_{\sun}$) early-type galaxies, from the NSA catalogue, is three times higher in lower-density environments than in higher-density ones, highlighting that the tidal fractions strongly depend on the environment of the sample.

\subsection{Non-Parametric Structural Parameters}
\label{sec:cas}

\begin{table}
  \begin{center}
    \caption{CAS parameters. Columns 2, 4 and 6 show the average CAS parameters by morphological type. Columns 3, 5, and 7 are the corresponding 1-$\sigma$ variation.}
    \label{tab:CAS}
    \begin{tabular}{c|c|c|c|c|c|c}
      \hline
Type & C & errC & A & errA & S & errS \\
\hline
E & 3.76 & 0.34 & 0.12 & 0.4 & 0.14 & 0.18 \\
S0 & 3.48 & 0.37 & 0.1 & 0.22 & 0.12 & 0.14 \\
S0a & 3.41 & 0.37 & 0.09 & 0.2 & 0.12 & 0.14 \\
Sa & 3.39 & 0.35 & 0.11 & 0.22 & 0.13 & 0.15 \\
Sab & 3.35 & 0.29 & 0.13 & 0.33 & 0.16 & 0.18 \\
Sb & 3.18 & 0.26 & 0.17 & 0.55 & 0.19 & 0.23 \\
Sbc & 2.96 & 0.22 & 0.18 & 0.33 & 0.21 & 0.29 \\
Sc & 2.9 & 0.18 & 0.21 & 0.26 & 0.25 & 0.34 \\
Scd & 2.83 & 0.21 & 0.24 & 1.47 & 0.24 & 0.42 \\
Sd & 2.78 & 0.19 & 0.26 & 0.72 & 0.31 & 0.55 \\
Sm & 2.82 & 0.26 & 0.25 & 0.03 & 0.22 & 0.33 \\
Irr & 2.41 & 0.14 & 0.29 & 0.03 & 0.12 & 0.31 \\
\hline
    \end{tabular}
  \end{center}
\end{table}

 \begin{figure}
 \includegraphics[width=0.48\textwidth]{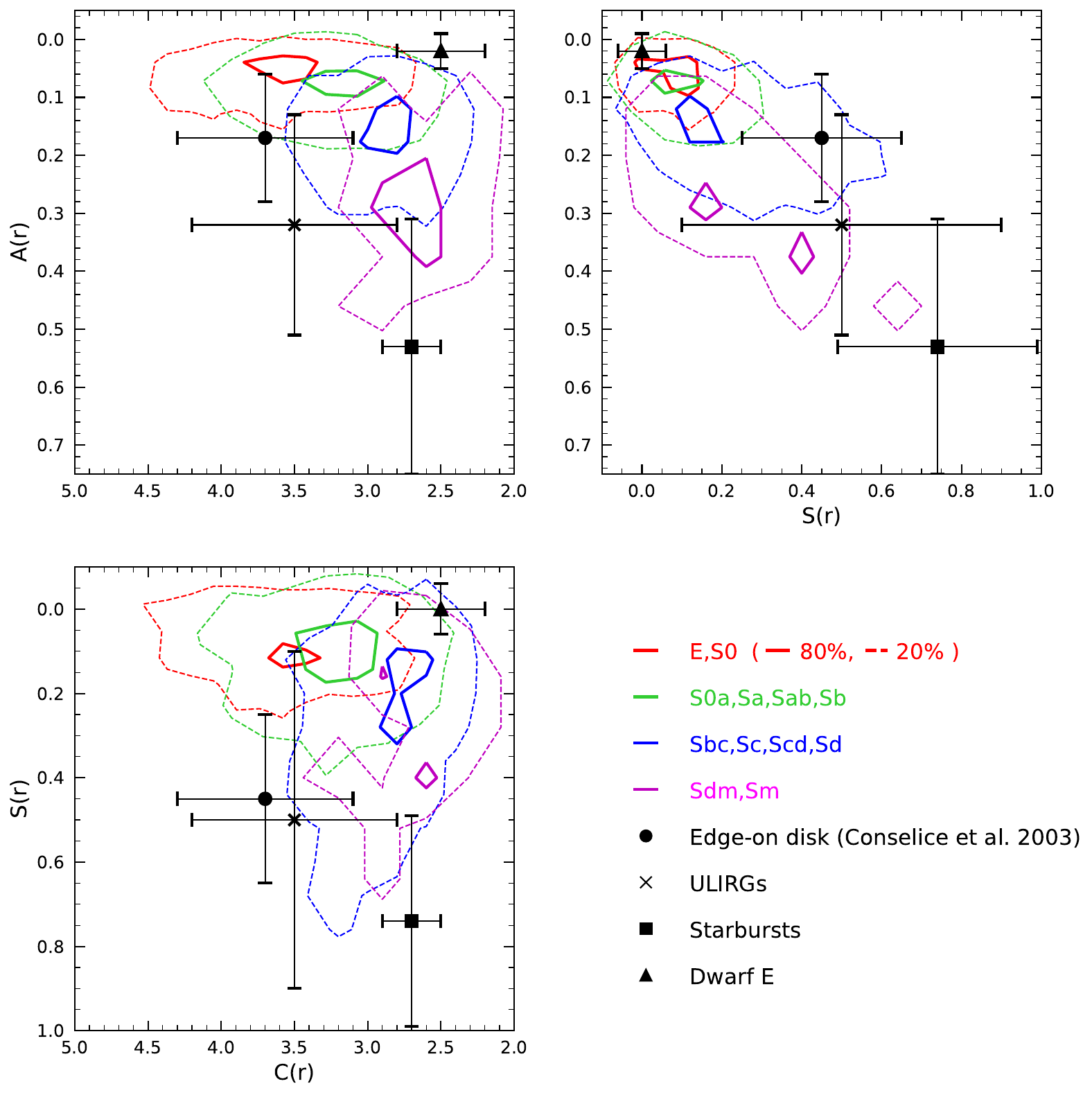}
 \caption{\emph{CAS} parameters in two levels contour plots (solid-lines = 0.8 and dotted-lines = 0.2), for early (red), intermediate (green), late (blue) and very late (magenta) types. As comparison (in black symbols) we mark the position of Edge-on, ULIRGs, Starbursts and Dwarfs Ellipticals according to \citet{Conselice2003}. Note that these plots only consider galaxies with low inclinations ($<$70$^o$) and within the following interval values: 0$<$A$<$0.5, 2$<$C$<$5, -0.2$<$S$<$0.8, according to different authors \citep[e.g.][]{Pawlik2016}.}
 \label{fig:CAS}
\end{figure}

Table~\ref{tab:CAS} shows the average $CAS$ parameters by morphological type along with their 1-$\sigma$ variations. The results indicate that our $CAS$ estimates generally follow the expected trends with morphology \citep[e.g.][]{Conselice2003}. ETGs have the highest light concentration and the lowest asymmetry and clumpiness values. In contrast, spirals from early to late types (LTGs) gradually exhibit less light concentration but higher asymmetries and clumpiness values.  

Figure~\ref{fig:CAS} illustrates different projections of the 3D \emph{CAS} space and their dependences on morphology. We present four morphological families shown with isodensity contours at 20\% (dashed-lines) and 80\% (solid-lines) levels. We only examine galaxies with low inclinations ($<$70$^o$) and within the following parameter ranges: 0$<$A$<$0.5, 2$<$C$<$5, and -0.2$<$S$<$0.8, which, according to various authors \citep[e.g.,][]{Pawlik2016}, helps prevent including spurious values caused by measurement errors. 
For reference, we include the average and corresponding 1 sigma values (solid black error bars) for edge-on galaxies, ULIRGs, Starbursts, and Dwarf Ellipticals from \citet{Conselice2003}. 
Although there is a significant overlap among these morphological families in the three panels, a clear transition from early to late-type galaxies is observed, emphasizing the usefulness of the \emph{CAS} parameters as indicators of morphology (see, for example, the A--C upper left diagram and their corresponding mean and 1-$\sigma$ values in Table~\ref{tab:CAS}).

It has been shown (e.g., \citealt{Conselice2003}, \citealt{HernandezToledo2005}; \citealt{Lotz2004}, \citealt{Lotz2008}, \citealt{Lotz2011}, see also \citealt{Nevin2023} and references therein) that strongly disturbed systems occupy a specific region of the \emph{CAS} morphological space parameters. In this context, the aperture size may also be significant. For example, \citet{Pawlik2016} use a maximum radius to include most of the pixels associated with faint structures, which are likely connected to different merger stages; this radius is larger than the typical 1.5 $R_e$ used by \citet{Conselice2003} and other authors. The \emph{CAS} data provided for the MaNGA DR17 galaxies can also help compare structural results from galaxy evolution models and simulations \citep{RodriguezGomez2019}.

The reader interested in using the catalogue values of the \emph{CAS} parameters may find it helpful to review the flagging codes described in Table~\ref{tab:VACdesc}. These include galaxies with low inclination (Edge-On = 0), galaxies showing no evidence of tidal features (Tides = 0), and faint, diffuse (low S/N) galaxies (Diffuse = 0).

\begin{table}
  \begin{center}
    \caption{MVM-VAC columns description.}
    \label{tab:VACdesc}
    \begin{tabular}{l|l}
      \hline
      Column name & Description \\
\hline
NAME      & MaNGA name of the mosaics \\
PLATEIFU  & MaNGA plate-ifu \\
MaNGAID   & MaNGA identification \\
OBJRA	  &	Object right ascension (degrees) \\
OBJDEC	  &	Object declination (degrees)  \\
TYPE      & Standard Hubble classification. AB=possible bar. \\
          & B=Barred galaxy.\\
TTYPE     & Morphological code number after the following \\
          & convention: E = -5, S0 = -2, S0a = 0, Sa = 1, \\
          & Sab = 2, Sb = 3, Sbc = 4, Sc = 5, Scd = 6, Sd = 7,\\
          & Sdm = 8, Sm = 9, Irr, BCD, dwarfs = 10, S = 11. \\
DIFFUSE   & 1=diffuse, faint or compact galaxy, 0=otherwise. \\
          & Caution must be taken when using the Hubble \\
          & classification for these galaxies. \\
BARS      & 1=conspicuous straight bar, 0.75=clear conspicuous \\
          & bar, 0.5=clear bar in the inner regions of the \\
          & galaxy, 0.25=typically a roundish structure, \\
          & 0=no trace of bar, -0.5=difficult to distinguish \\
EDGE\_ON   & indicates the appearance to be edge-on \\
          & (absence = 0, definite = 1) \\
TIDES	  &	1=tidal debris presented; 0=no tide \\
C	      &	Concentration. (-999=null value)  \\
E\_C	      &	Concentration error. (-999=null value)  \\
A	      &	Asymmetry. (-999=null value)  \\
E\_A	      &	Asymmetry error. (-999=null value)  \\
S	      &	Clumpiness. (-999=null value)  \\
E\_S	      &	Clumpiness error. (-999=null value) \\
cas\_flag & 1=reliable estimation of the CAS parameters; \\
         & 0=biased CAS values due to the presence of light \\
         & coming from neighbouring stars and galaxies that \\
         & was not well masked. \\
\hline
    \end{tabular}
  \end{center}
\end{table}

\begin{table*}
  \caption{The first 10 entries of the MaNGA Visual Morphology VAC. Columns are described in detail in Table~\ref{tab:VACdesc}.}
    \label{tab:VAC}
    \setlength\tabcolsep{2pt}
    \begin{tabular}{*{18}{c}}
      \hline
name & plateifu & mangaid & objra & objdec & Type & TType & diffuse & bars & edge & tides & C & E\_C & A & E\_A & S & E\_S & CAS\\
manga- &  &  &  &  &  &  &  &  & on &  &  & &  &  &  &  & flag \\
\hline
  10001-12701 & 10001-12701 & 1-48157 & 133.37109061 & 57.59842514 & Sbc & 4 & 0 & 0.0 & 1 & 0 & 2.419 & 0.26 & 0.191 & 0.012 & 0.26 & 0.2555 & 1\\
  10001-12702 & 10001-12702 & 1-48188 & 133.68566986 & 57.48025032 & SABbc & 4 & 0 & 0.5 & 0 & 0 & 2.882 & 0.195 & 0.094 & 0.046 & 0.01 & 0.1085 & 1\\
  10001-12703 & 10001-12703 & 1-55648 & 136.01715996 & 57.09232917 & SABbc & 4 & 0 & 0.5 & 1 & 0 & 3.249 & 0.261 & 0.135 & 0.013 & 0.43 & 0.334 & 1\\
  10001-12704 & 10001-12704 & 1-55616 & 133.98996686 & 57.67796766 & Sd & 7 & 0 & 0.0 & 1 & 0 & 3.38 & 0.11 & 0.212 & 0.005 & 0.85 & 0.1105 & 0\\
  10001-12705 & 10001-12705 & 1-55784 & 136.75137451 & 57.45143692 & Sbc & 4 & 0 & 0.0 & 0 & 0 & 2.883 & 0.246 & 0.152 & 0.009 & 0.15 & 0.1835 & 1\\
  10001-1901 & 10001-1901 & 1-55567 & 133.33002800 & 57.04115537 & Sbc & 4 & 0 & 0.0 & 0 & 0 & 2.709 & 0.47 & 0.14 & 0.007 & 0.18 & 0.1805 & 1\\
  10001-1902 & 10001-1902 & 1-48201 & 134.19392335 & 56.78674699 & SB0 & -2 & 0 & 1.0 & 0 & 0 & 3.964 & 0.344 & 0.1 & 0.004 & -0.01 & 0.0255 & 1\\
  10001-3701 & 10001-3701 & 1-48111 & 132.46564676 & 57.14372790 & SAB0 & -2 & 0 & 0.5 & 0 & 0 & 2.853 & 0.582 & 0.05 & 0.003 & 0.27 & 0.14 & 1\\
  10001-3702 & 10001-3702 & 1-48136 & 132.91276824 & 57.10742355 & E & -5 & 0 & 0.0 & 0 & 0 & 4.31 & 0.319 & 0.112 & 0.005 & 0.26 & 0.141 & 1\\
  10001-3703 & 10001-3703 & 1-55612 & 134.59149894 & 57.68496532 & SABb & 3 & 0 & 0.25 & 0 & 0 & 2.953 & 0.466 & 0.066 & 0.006 & 0.17 & 0.1585 & 1\\

\hline
    \end{tabular}
\end{table*}

\subsection{MaNGA Visual Morphology Value Added Catalogue (MVM-VAC)}
\label{sec:VAC}

The morphological classification detailed in this paper has been integrated into a MaNGA Visual Morphology Value Added Catalogue (MVM-VAC), one of the publicly accessible catalogues from the SDSS DR17 release. This catalogue includes 10,126 MaNGA galaxies (entries) with unique ID, after removing duplicates, blank entries, or star targets from the original list of 11,273 targets. Table~\ref{tab:VACdesc} describes the content of each column in the VAC: the detailed morphological classification (\emph{TYPE}), which includes the AB and B types to indicate galaxies with potential bar structures and confirmed bars; the corresponding code number (\emph{TTYPE}) is also provided. The VAC also provides a flag on galaxies that, under our scheme, are challenging to classify (\emph{DIFFUSE}), such as diffuse, faint, compact, or S types (disc-like structure suspected). 

Bar families (\emph{BARS}) are reported following the procedures described in Section~\ref{sec:bars}. These are indicated by a code number for easier handling: B=1, A$\underline{B}$=0.75, AB=0.5 and $\underline{A}$B=0.25. An additional code (0) indicates no trace of a bar. Edge-on galaxies (\emph{EDGE\_ON}) are those with inclination $>$60$^{\circ}$ based on semi-axis $b/a$ ratio, as described in Section~\ref{sec:steps}. After a visual inspection of the DESI images, tidal structures (\emph{TIDES}) were identified in a binary mode (0; no tides) or (1; tides) without specifying the type of tides. Values for the structural parameters Concentration (\emph{C}), Asymmetry (\emph{A}) and Clumpiness (\emph{S}), along with their errors, are also available (see Section~\ref{sec:cas} for details). As an example, the first 10 rows of the MVM-VAC are presented in Table~\ref{tab:VAC}, the full table is available through the SDSS-IV and SDSS-V websites \footnote{\url{https://www.sdss.org/dr18/data_access/value-added-catalogs/?vac_id=80}}.

\begin{figure}
 \includegraphics[width=\columnwidth]{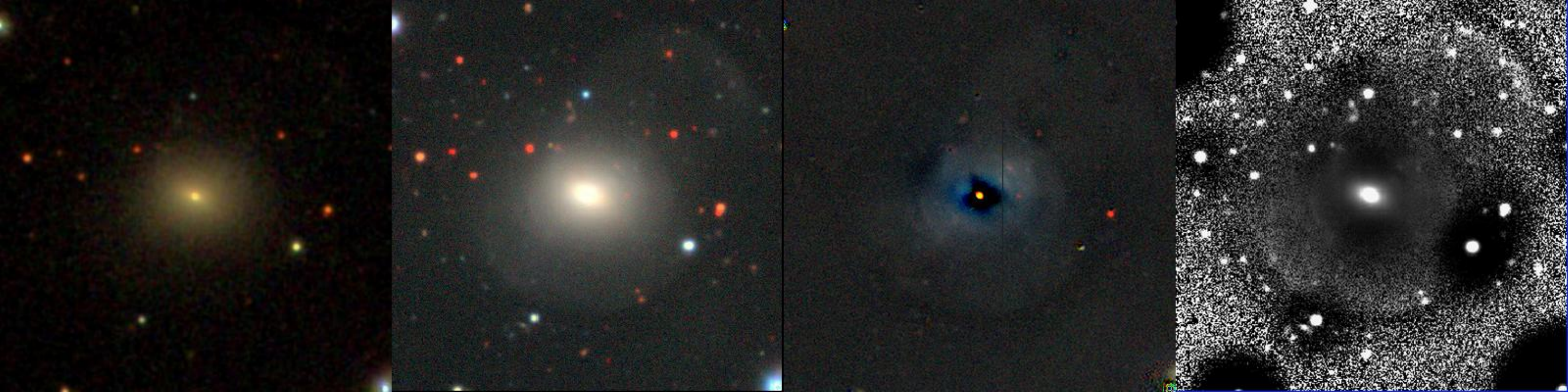}
 \caption{Manga-7968-9102 mosaics. From left to right hand panels: $gri$ colour composite SDSS image, $grz$ colour composite DESI image, the residual DESI image after subtracting the best model fit and the filter-enhanced $r-$band image.}
 \label{fig:VACmosaic}
\end{figure}

As described in Section~\ref{sec:steps}, preparing images and conducting visual analysis are essential for morphological classification. Figure~\ref{fig:VACmosaic} shows a representative mosaic of each MaNGA galaxy. The left panels display the SDSS $gri$ and DESI $grz$ colour composite images, while the right panels show the residual image (after subtracting the best-fitting model), taken from the available post-processing catalogue for the DESI images, followed by the filter-enhanced DESI $r$-band image. Our mosaics include images from the three DESI Legacy Imaging Surveys: DECaLS, BASS and MzLS. These mosaics can provide valuable morphological and structural information for various purposes; for instance, they can serve as input images for training of Convolutional Neural Network algorithms. Mosaics for 10126 galaxies are available in the SDSS website mentioned earlier.

\begin{figure} 
\centering
 \includegraphics[width=\columnwidth]{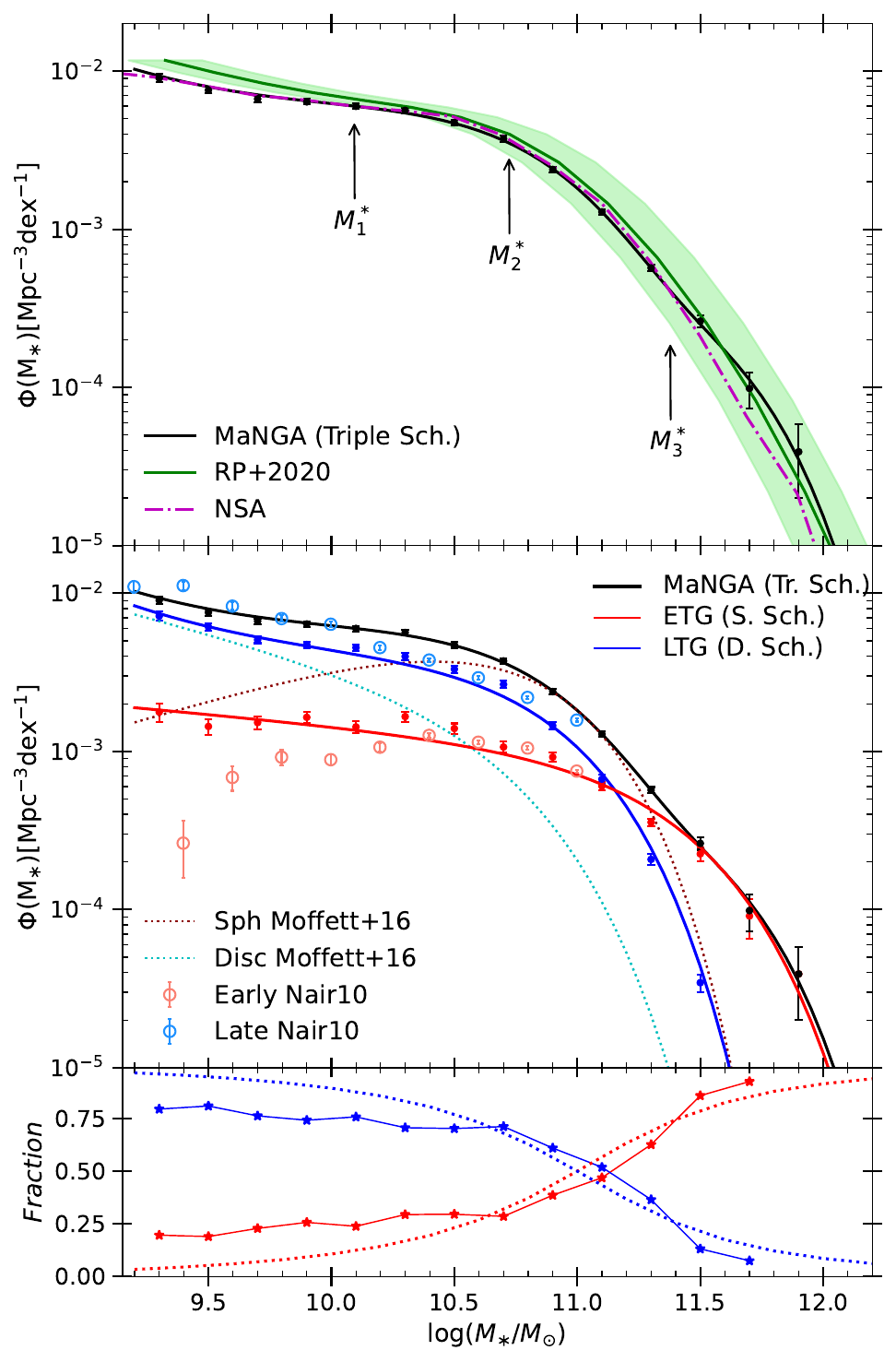}
 \caption{Upper panel: GSMF for the MaNGA sample (black dots) with its best-fitting (black line) considering a triple Schechter function. Errors are estimated through a bootstrap approach in a set of 8,000 randomly selected samples, with the size of the original sample. The arrows indicate the three corresponding characteristic masses. As comparison, the the NSA catalogue GSMF (magenta dot-dashed line) and SDSS DR7 + low-z SDSS GSMF fit according to \citetalias{Rodriguez-Puebla+2020} (green line) are plotted. The shadow region represents the uncertainty due to the systematic uncertainties in the \ms\ estimate. Middle panel: The MaNGA GSMF divided into ETGs (red circles), and LTGs (blue circles), fitted by a single and double Schechter function, respectively (blue and solid lines). As comparison, the GAMA GSMFs for spheroids and discs \citep[dotted lines;][]{Moffett2016} and the \citet{Nair2010} GSMF for early- and late-type galaxies (open circles) are reproduced. Lower panel: The early and late GSMF divided by the total MaNGA GSMF (red and blue asterisks and solid lines respectively), and as comparison the fractions reported by \citetalias{Rodriguez-Puebla+2020} (dotted lines). Note that estimations and comparisons are corrected to the cosmology used in this paper.}
 \label{fig:SMF}
\end{figure}

\section{The MaNGA Stellar Mass Function} 
\label{sec:SMF}

In this section, using the volume corrections described in section \ref{sec:Vcorr}, we estimate the GSMF of all galaxies in the MaNGA sample, breaking it down by morphological \type.

The GSMF is estimated using the standard 1/$V_{\rm max}$ method \citep[][]{Schmidt1968}: 
\begin{equation}
     \phi(\ms) =\frac{1}{\Delta \log \ms}\sum_{i=1}^{N}\frac{w_i}{V_{\rm max,i}},
\end{equation}
where $w_i$ is the volume-corrected weight of the $i$-th galaxy given its \ms\ and $g-i$ colour, and $V_{\rm max,i}$ is the maximum volume at which this galaxy can be observed in the MaNGA sample within the redshift interval 0.005 $< z <$ 0.15 (see section \ref{sec:Vcorr}). The $\Delta \log \ms$ is the mass bin width, which is 0.2 dex in our case. 
Notice that the weights are reliable only for log(\ms/\msun)$>$9.2.   
To estimate the GSMF errors, we use a bootstrap approach in a set of 8000 randomly selected samples, each of the same size as the original sample. The errors are determined as the standard deviation across all the bootstrap subsamples produced by this technique.

\begin{table*}
  \begin{center}
    \caption{GSMF fits considering a composited triple Schechter function (for all galaxies), a double Schechter function (LTG) and a single Schechter function (ETG). Subindexes indicate the parameters related to first, second or triple Schechter function. Notice that $\alpha_1 = \alpha$, $\alpha_2 = \alpha + 1 $ and $\alpha_3 = \alpha + 1 $.}
    \label{tab:fits-triple}
    \begin{tabular}{c|c|c|c|c|c|c|c}
      \hline
Type & $\log(\phi_1 \, /\, \text{Mpc}^{-3} \, \text{dex}^{-1})$ & $\alpha$ & $\log (M^*_1 / \msun)$ & $\log (\phi_2 \, /\, \text{Mpc}^{-3} \, \text{dex}^{-1})$ & $\log (M^*_2 / \msun)$ & $\log (\phi_3 \, /\, \text{Mpc}^{-3} \, \text{dex}^{-1})$ & $\log (M^*_3 / \msun)$ \\ 
\hline
All & -3.331 $\pm$ 0.831 & -1.843 $\pm$ 0.205 & 10.093 $\pm$ 0.506 & -2.452 $\pm$ 0.107 & 10.722 $\pm$ 0.052 & -3.456 $\pm$ 0.274 & 11.378 $\pm$ 0.088 \\
LTG & -3.740 $\pm$ 0.389 & -2.000 $\pm$ 0.062 & 10.201 $\pm$ 0.024 & -2.696 $\pm$ 0.056 & 10.831 $\pm$ 0.024 & - & - \\
ETG& -3.383 $\pm$ 0.107 & -1.137 $\pm$ 0.077 & 11.382 $\pm$ 0.051 & - & - & - & - \\
\hline
    \end{tabular}
  \end{center}
\end{table*}

Figure~\ref{fig:SMF} displays the total GSMF for the volume-corrected MaNGA sample (black dots with error bars). The black solid curve is a fit to these data and will be explained in more detail below. For comparison, we show the total GSMF derived from the SDSS-based DR7 NSA catalogue (magenta dot-dashed line) and the GSMF reported in \citet[][here after RP20]{Rodriguez-Puebla+2020}, represented by the green solid line, including its uncertainty due to the systematic uncertainty in the stellar mass estimate (green-shaded area). The middle panel breaks down the GSMF into ETGs (E--S0a) and LTGs (Sa--Irr), shown as red and blue filled dots, respectively. The red and blue solid lines are the best-fittings to these data, which will be discussed below. 

The description of the GSMF using an analytic function has been a long-standing challenge in extragalactic astronomy. Since the influential proposal of the Schechter function in \citet{Schechter1976}, various composite functions have been suggested to describe the GSMF across dwarf to giant galaxy scales. For instance, \citet{Rodriguez-Puebla+2020}, using SDSS DR7 data based on the photometric catalogue from \citet{Meert2015,Meert+2016} and the SDSS DR4 NYU-VAGC \citep{Blanton+2005} low-z sample for low masses, and after applying K+E corrections to $z=0$ and adjustments for low-surface brightness incompleteness, found that the combination of a sub-exponential Schechter function and a double power law function fits well their obtained GSMF. In this study, we fitted the MaNGA GSMF with this composite function but also found that a triple Schechter function performs equally well and, additionally, better captures some GSMF features observed at three characteristic masses. The fits were carried out using the Bayesian approach described in \citet{Rodriguez-Puebla+2013}, with Table~\ref{tab:fits-triple} summarizing the best-fitting parameters for the three Schechter functions. For the LTG GSMF, we fitted a double Schechter function, while a single Schechter function was used for the ETG GSMF. The corresponding best-fitting parameters are also listed in Table~\ref{tab:fits-triple}. 

The solid black, red, and blue lines in the upper and middle panels of Fig.~\ref{fig:SMF} represent the best-fittings, respectively. The three Schechter characteristic masses from the fit to the MaNGA GSMF are indicated with arrows in the top panel. In subsection \ref{sec:implications}, we will discuss the implications of these characteristic masses.
The MaNGA GSMF is expected to agree with the NSA GSMF (black solid and magenta dot-dashed curves in Fig.~\ref{fig:SMF}), since the weights used for volume correction in the MaNGA galaxies were estimated to reconstruct this function. As seen in Fig.~\ref{fig:SMF}, there is also good agreement with the \citetalias{Rodriguez-Puebla+2020} fit to the local GSMF (green line). 
The MaNGA GSMF closely follows the \citetalias{Rodriguez-Puebla+2020} GSMF despite differences in how the K+E corrections and stellar masses are calculated.  
The difference at the low-mass end is most likely due to the surface brightness corrections in \citetalias{Rodriguez-Puebla+2020}. 

\begin{figure*}
 \includegraphics[width=0.8\textwidth]{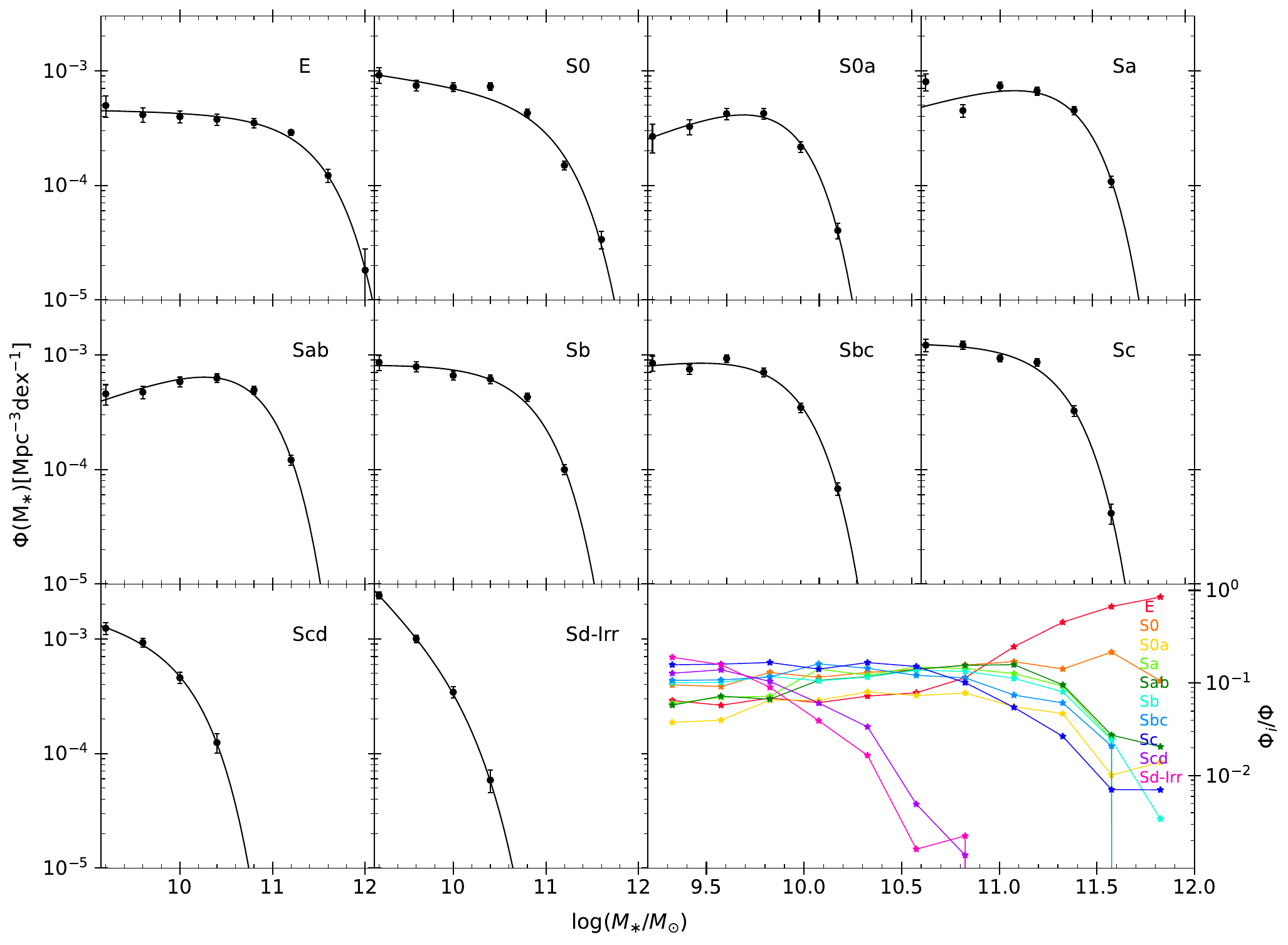}
 \caption{ 
 The GSMF of each morphological type and their fits using a simple Schechter function (solid black line); for the best-fitting parameters, see Table \ref{tab:SMF_all}. 
 \textit{Bottom-right panel} shows the fraction of each type as a function of \ms$, \phi_{i}/\phi$, where $i=\type$.}
 \label{fig:SMF_all}
\end{figure*}

The complexity of the GSMF suggests that it is composed of contributions from various galaxy populations, such as different morphological types, as will be shown below. The best-fittings to the LTG/ETG GSMFs displayed in the middle panel of Figure~\ref{fig:SMF}, using the double and single Schechter functions, are presented by solid blue and red lines, respectively. The GSMF of LTGs increasingly dominates at lower masses. Interestingly, the abundance of ETGs does not decline at low masses but remains roughly constant. At the high-mass end, ETGs dominate the GSMF starting from $\ms\sim 10^{11.1}$ \msun. We compare our GSMF and its decomposition into LTGs and ETGs with data from \citealt{Nair2010}, as processed by \citetalias{Rodriguez-Puebla+2020}, and with the GSMFs of spheroid- and disc-dominated galaxies from \citet{Moffett2016}. Our results roughly align with those derived from the \citet{Nair2010} data and morphologies, except for ETGs, which do not decrease at $\ms\lesssim 10^{10.2}$ \msun\ in our case. However, our results strongly conflict with those from \citet{Moffett2016}. This discrepancy may result from using a selection based on structural parameters rather than on a morphological classification.

The lower panel of Fig.~\ref{fig:SMF} shows our estimated fractions of ETGs and LTGs, which are $\phi_{\rm ETG}/\phi$ and $\phi_{\rm LTG}/\phi$, respectively (stars connected with solid lines). The dotted lines indicate similar fractions reported in \citetalias{Rodriguez-Puebla+2020}, based on the \citet{Huertas-Company+2011} morphological classification using support vector machine algorithms. The agreement is quite good, although at low masses we observe a larger fraction of ETGs (and a correspondingly lower fraction of LTGs) compared to the \citet{Huertas-Company+2011} classification. This again shows that the fraction of ETGs does not decrease significantly at low masses according to our morphological classification. As we will discuss in subsection \ref{sec:environment}, these low-mass ETGs are mostly satellite galaxies.

\begin{table*}
  \begin{center}
    \caption{GSMF fits considering a single Schechter function for all morphological types (columns 2-4) and their corresponding mass densities (column 5). Columns 6 and 7 show the stellar mass fraction using the summation of the $V_{\rm max}$-weighted stellar mass and via the GSMF fits, using the mass densities ratios.}
    \label{tab:SMF_all}
    \begin{tabular}{c|c|c|c|c|c|c}
      \hline
Type & log ($\phi$ Mpc$^{-3}$ dex$^{-1}$) & $\alpha$ & log $M^*$ [\msun] & $\log \rho_\ast (>10^9 \msun)$ [\msun/Mpc$^3$] & Volume-corrected &  GSMF fit\\ 
 & & & & & Mass Fraction & Mass Fraction\\
\hline
 E & -3.747 $\pm$ 0.103 & -1.016 $\pm$ 0.093 & 11.511 $\pm$ 0.038 & 7.767 $\pm$ 0.111 & 0.226 $\pm$ 0.072 & 0.228 $\pm$ 0.061 \\
S0 & -3.622 $\pm$ 0.064 & -1.115 $\pm$ 0.047 & 11.153 $\pm$ 0.042 & 7.559 $\pm$ 0.078 & 0.187 $\pm$ 0.104 & 0.142 $\pm$ 0.028 \\
S0a & -3.446 $\pm$ 0.057 & -0.670 $\pm$ 0.066 & 10.662 $\pm$ 0.037 & 7.165 $\pm$ 0.07 & 0.043 $\pm$ 0.008 & 0.057 $\pm$ 0.01 \\
Sa & -3.287 $\pm$ 0.039 & -0.765 $\pm$ 0.045 & 10.787 $\pm$ 0.031 & 7.456 $\pm$ 0.050 & 0.082 $\pm$ 0.014 & 0.112 $\pm$ 0.015 \\
Sab & -3.274 $\pm$ 0.032 & -0.706 $\pm$ 0.037 & 10.793 $\pm$ 0.027 & 7.471 $\pm$ 0.042 & 0.088 $\pm$ 0.015 & 0.116 $\pm$ 0.014 \\
Sb & -3.426 $\pm$ 0.046 & -0.987 $\pm$ 0.042 & 10.871 $\pm$ 0.032 & 7.437 $\pm$ 0.056 & 0.094 $\pm$ 0.029 & 0.107 $\pm$ 0.016 \\
Sbc & -3.294 $\pm$ 0.047 & -0.901 $\pm$ 0.048 & 10.725 $\pm$ 0.031 & 7.405 $\pm$ 0.057 & 0.073 $\pm$ 0.012 & 0.099 $\pm$ 0.015 \\
Sc & -3.249 $\pm$ 0.084 & -0.992 $\pm$ 0.081 & 10.663 $\pm$ 0.045 & 7.403 $\pm$ 0.097 & 0.179 $\pm$ 0.131 & 0.099 $\pm$ 0.023 \\
Scd & -3.374 $\pm$ 0.132 & -1.180 $\pm$ 0.161 & 10.109 $\pm$ 0.093 & 6.734 $\pm$ 0.172 & 0.014 $\pm$ 0.003 & 0.021 $\pm$ 0.009 \\
Sd-Irr & -3.663 $\pm$ 0.263 & -1.756 $\pm$ 0.151 & 10.158 $\pm$ 0.141 & 6.702 $\pm$ 0.329  & 0.012 $\pm$ 0.002 & 0.02 $\pm$ 0.015 \\
\hline
    \end{tabular}
  \end{center}
\end{table*}

\subsection{Dissection by morphological types}

In Figure~\ref{fig:SMF_all} we display the contribution of 10 morphological types to the total GSMF (Sd-Irr galaxies were combined due to the small number of these galaxies). Note that such a detailed breakdown of the GSMF by morphological type is rarely shown in the literature. We fitted each GSMF with a single Schechter function using the same Bayesian method mentioned earlier, and plotted the fits with a solid black line in each case. The fits were plotted to the last mass bin where data are available, and the fitted parameters are listed in Table~\ref{tab:SMF_all}. See how, in most of the cases, a single Schechter function adequately describes the GSMF of galaxies of a given morphological type.

In the bottom-right panel of Figure~\ref{fig:SMF_all}, the fraction of each type, $\phi_i/\phi$ with $i=\type$, as a function of \ms\ is shown. In the low mass regime, $9.2<$log(\ms/\msun)$\lesssim 9.7$, late-type galaxies dominate, from Sb to Sd-Irr, although the abundance of S0 galaxies is close to that of Sb or Sbc galaxies. The Sc type is the most abundant up to log(\ms/\msun)$\approx 10.5$; beyond this mass, the abundance of these late-type galaxies strongly drops out. At masses larger than $10^{11}$ \msun\, E and S0 galaxies begin to dominate, with the fraction of the former increasing rapidly with \ms. In this mass range, there remains a small (and decreasing with mass) fraction of Sa--Sb galaxies.   
In Section \ref{sec:implications}, we will discuss the variation of the Schechter parameters across  different morphological types and explore our results in terms of galaxy evolution at various mass ranges. Specifically, we will examine how the total GSMF can be described by a triple Schechter function, with three different characteristic masses associated with distinct morphological groups.

\subsection{The MaNGA Stellar Mass Budget by morphological types}
\label{sec:masstypes}

We can determine the fractional contribution of each galaxy morphological type to the cosmic stellar mass budget. This information is shown in Table~\ref{tab:SMF_all}.
The fraction of stellar mass is estimated by summing the $V_{\rm max}$-weighted stellar mass of each galaxy within each type. Relative errors are calculated considering the standard deviation of the $V_{\rm max}$-weighted stellar mass distribution for each type, and are propagated using standard relative error formulas. 
For masses above log(\ms/\msun)=9.2, we find that approximately 22.6\% of the stellar mass is in E galaxies. This value is higher than the $13\% \pm 4\%$ reported in \citet{Driver2007} for red elliptical galaxies, which used GIM2D bulge–disc decomposition based on the deeper Millennium Galaxy Catalogue $B$-band data \citep[][]{Liske2003,Driver2005}. 

In contrast, \citet{Kelvin2014} reported a significantly higher elliptical stellar mass fraction, $34^{+9}_{-4}$\%, estimated from a visual evaluation based on SDSS/UKIDSS data with a lower stellar mass limit of log(\ms/\msun)=9, while \citet{Gadotti2009} reported 32\% based on a Petrosian concentration index cut, using a sample of galaxies more massive than log(\ms/\msun)=10. 
If we separate the galaxies into spheroid-dominated (E--Sa) and disc-dominated (Sab--Irr), we find stellar mass fractions of 53.8\% and 46.2\%, respectively. These numbers compare to $39\%$ of stellar mass in red ellipticals and bulges, and $58\% \pm 6\%$ in galaxy discs, also reported in \citet{Driver2007}. Note that these values were obtained for galaxies brighter than $M_B$<-17. 

These variations among different studies may reflect the different stellar mass limits in the samples, as well as the difficulty in previous works of distinguishing morphological types in more detail. This often results in a mixing of types, for instance, when selecting Elliptical galaxies (i.e., not properly separating E and S0s) or when grouping spheroids with discs. 
Alternatively, we inferred the stellar mass fractions by considering the mass density by morphological type, reported in Table~\ref{tab:SMF_all}, column 5) according to the following equation:
\begin{equation}
   M_{\rm frac_i} = \rho_i / \rho_T, 
\end{equation}
where $\rho_i$ is the density for each morphological type and $\rho_T$ is the total density. As mentioned above, these densities were estimated using a single Schechter function fit to the GSMF for each morphological type (see Fig.~\ref{fig:SMF_all} and the last column in Table~\ref{tab:SMF_all}). Notice that this alternative method recovers nearly the same mass fraction values as those from the direct summation ($V_{\rm max}$-weighted stellar mass) method, for both Ellipticals (22.8\%) and the separation into spheroid-dominated and disc-dominated galaxies (53.9\% and 46.1\%, respectively), reflecting the consistency of the results from both methods.

\section{Stellar population properties and morphology of central/satellite galaxies}
\label{sec:stellar_prop}

As the largest IFS survey of local galaxies, MaNGA provides detailed and valuable spectroscopic data to characterize the global and local properties of galaxy stellar populations. Many of these properties have been derived from spectral inversion fitting using stellar population synthesis models and reported in publicly available VACs. This includes the Pipe3D \citep[][]{Sanchez2018} and pyPipe3D \citep[][]{Sanchez2022} VACs. Additionally, since MaNGA galaxies are selected from the SDSS, they come with photometric data in several filters, reliable estimates of internal dust extinction, and classifications into centrals and satellites. 

The detailed visual morphological classification we provide here for the MaNGA survey enables us to examine how the global and local stellar population properties depend on galaxy stellar mass and morphology, as well as distinguish between central and satellite galaxies. To demonstrate the potential of these data, we present a preliminary study focused on the \textit{global} stellar properties.    

\begin{figure*}
\centering

 \includegraphics[width=0.3\textwidth]{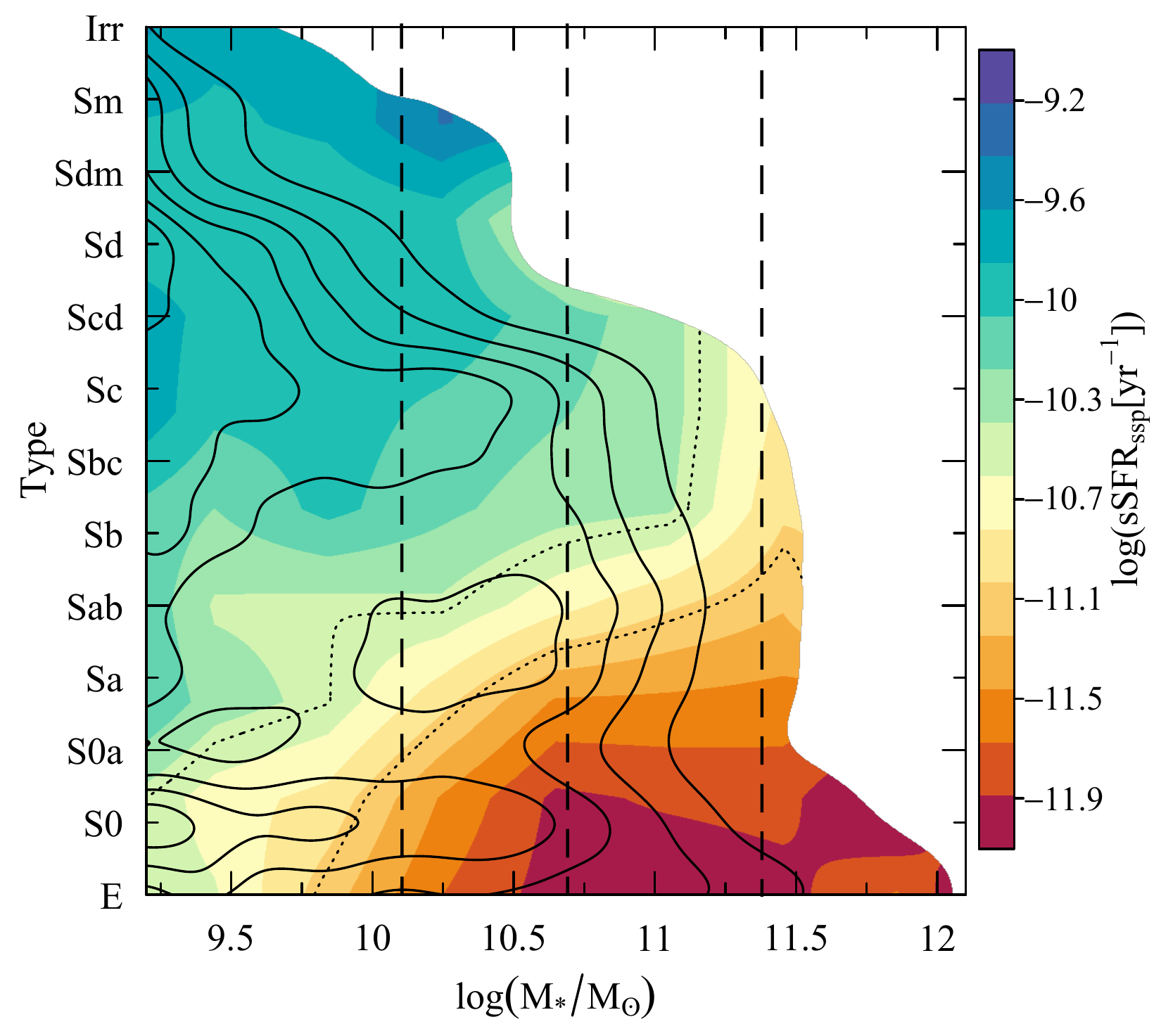}
 \includegraphics[width=0.3\textwidth]{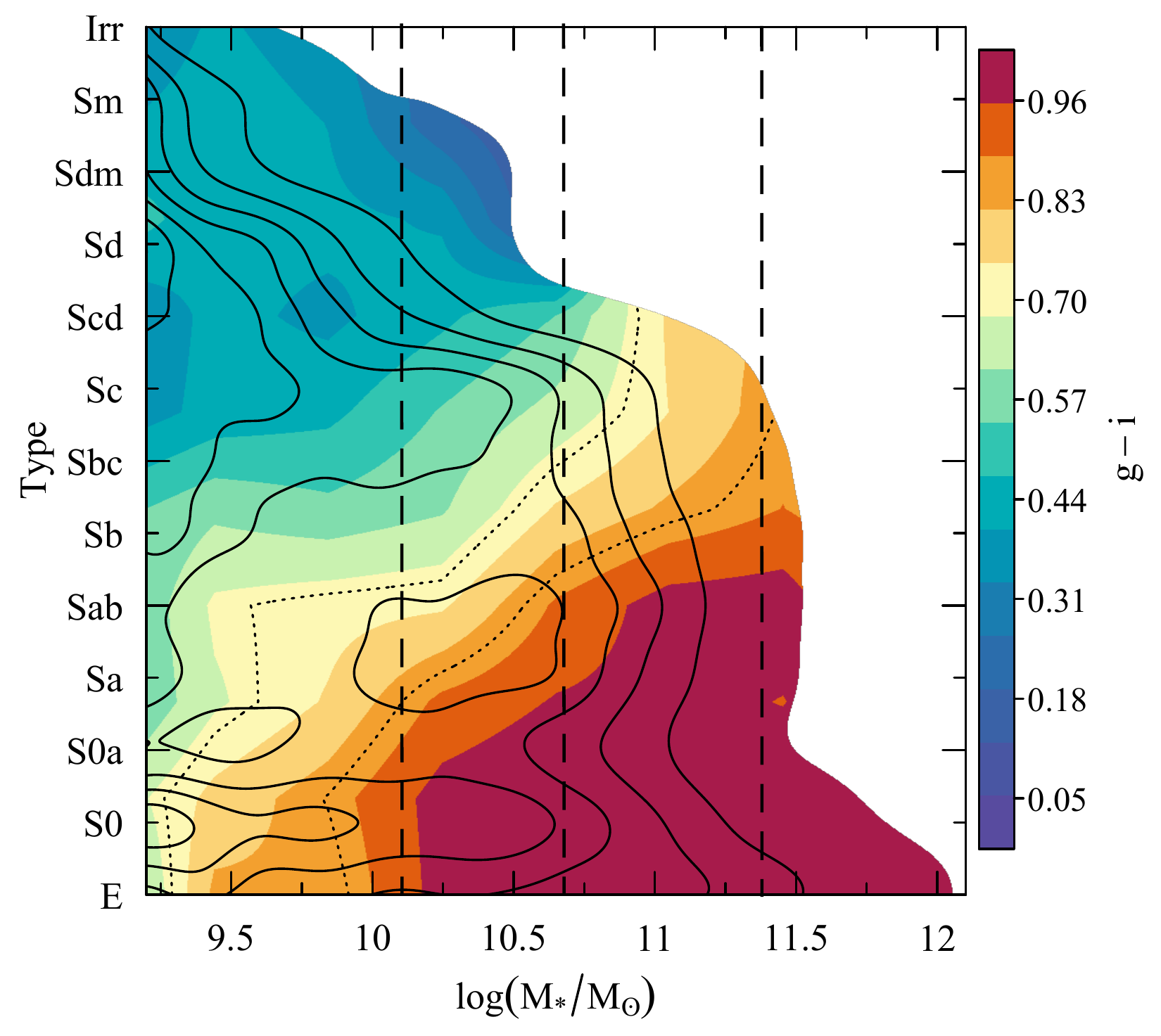}
 \includegraphics[width=0.3\textwidth]{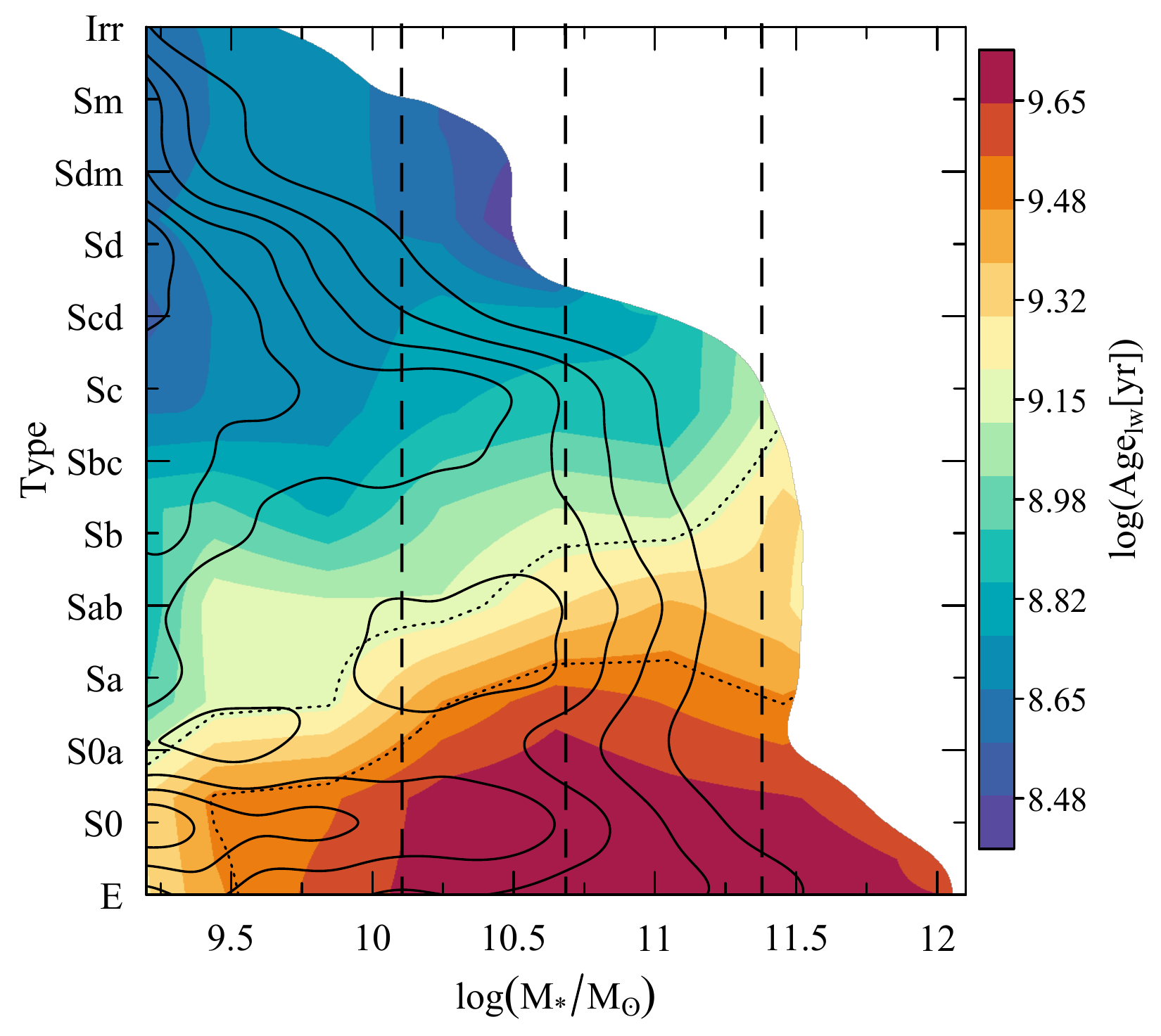}\\
 \includegraphics[width=0.3\textwidth]{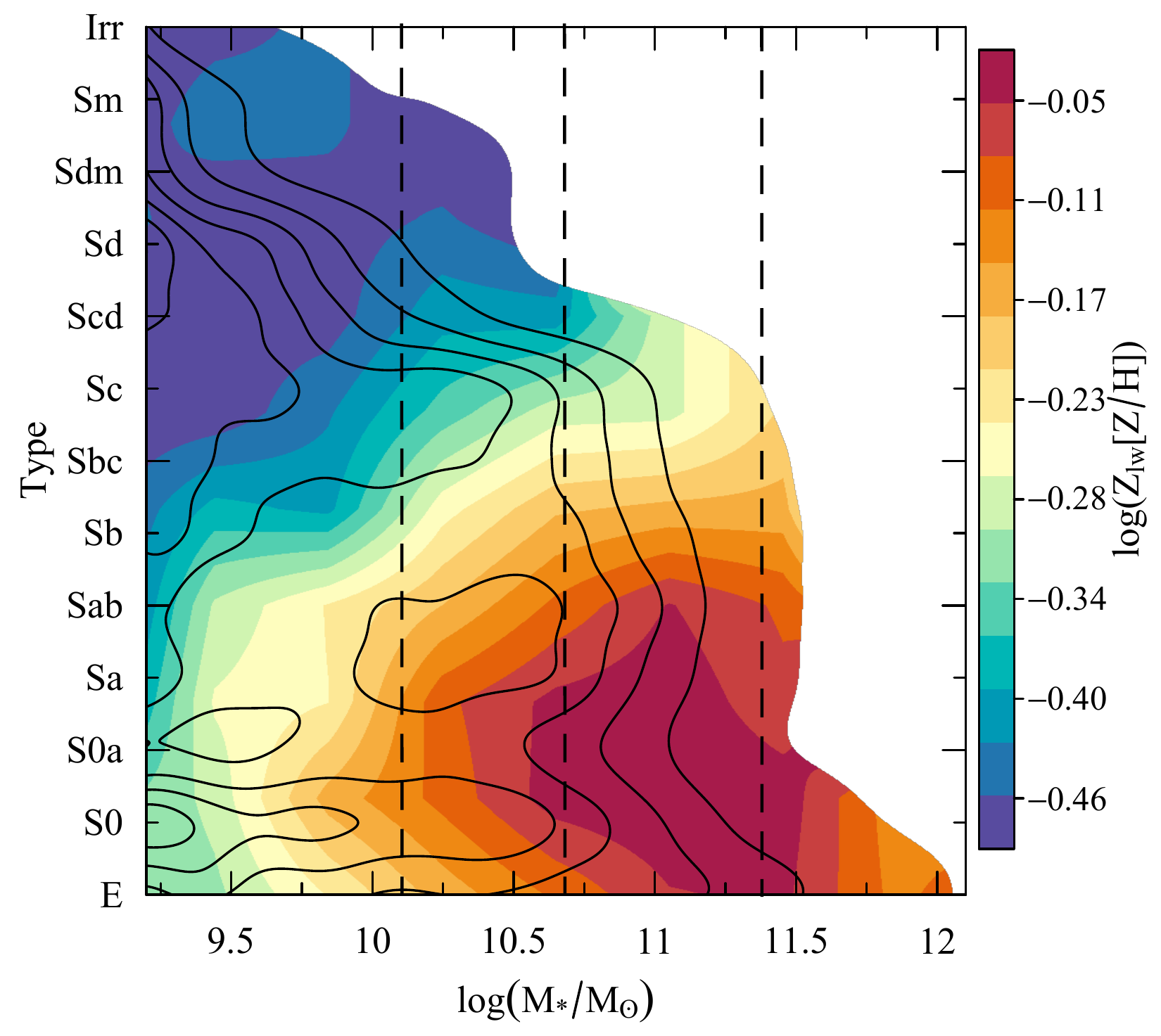}
 \includegraphics[width=0.3\textwidth]{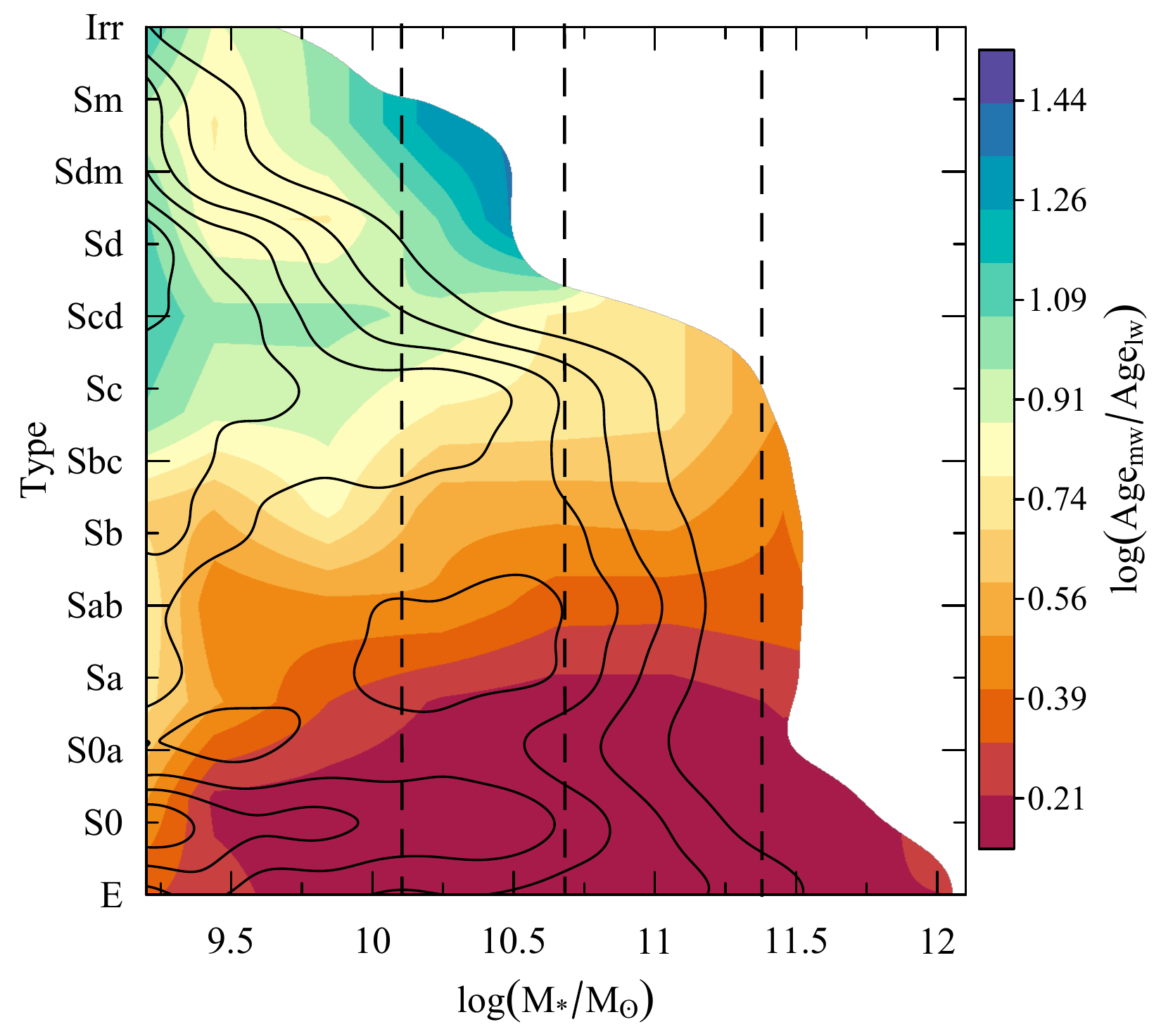}
 \includegraphics[width=0.3\textwidth]{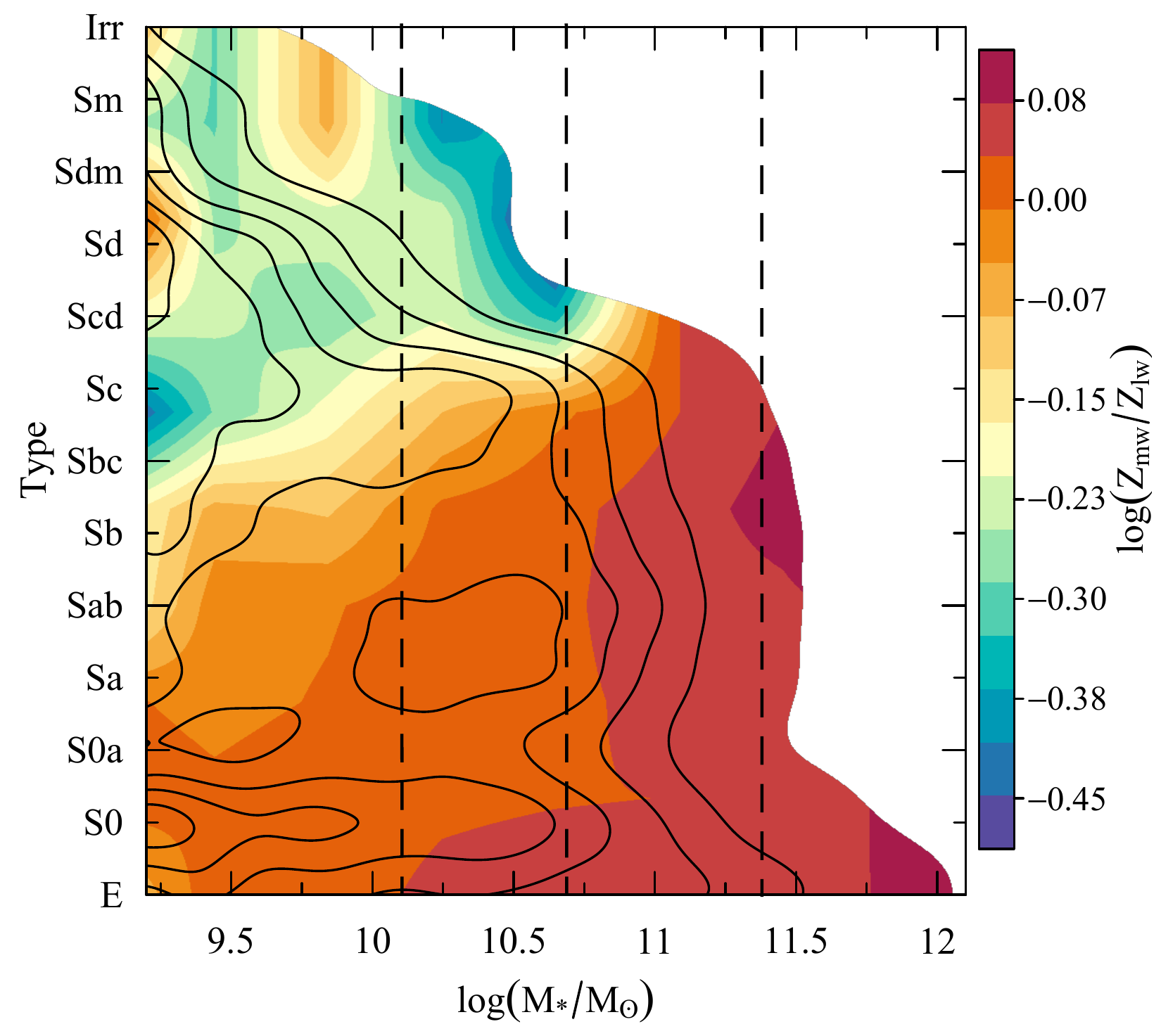}
 
\caption{Bivariate distributions in the \type--\ms\ diagram (isodensity contours and dashed lines are the same as right-hand panel in Figure \ref{fig:mass_morpho}) but coloured by the specific star formation rate (\textit{upper-left panel}), the dust-corrected $g-i$ colour (\textit{upper-central panel}), the luminosity-weighted age (\textit{upper-right panel}), the luminosity-weighted metallicity (\textit{lower-right panel}), the ratios of the mass- to luminosity-weighted mean ages (\textit{lower-central panel}) and metallicities (\textit{lower-right panel}) as indicated in the colour bars, respectively. The dotted lines indicate the transition zones in log(sSFR/yr) [-10.5,-11.1], $g-i$ colour [0.70-0.85] mag, and log($\age$/yr) [9.18-9.48].
}
\label{fig:AGE}
\end{figure*}

\subsection{Stellar population properties as a function of type and mass}
\label{sec:SPproperties}

In subsec. \ref{sec:morpho_results}, we showed that the distribution of local galaxies in the \ms--\type\ diagram is roughly bimodal, with a transition zone around S0a--Sa types that depends weakly on \ms. An important question is how the different stellar population properties of galaxies are distributed in this diagram. 
Figure \ref{fig:AGE} displays the smoothed and colour-coded median values (shown in the respective colour bars) of some of these properties in this diagram (galaxies with $i>70^\circ$ were excluded). From left to right, the top row shows the specific star formation rate (sSFR; derived from the SSP analysis), the $g-i$ colour, and the luminosity-weighted mean age (\age), while the bottom row displays the luminosity-weighted mean stellar metallicity ($Z_{\rm lw}$), and the ratios of the mass- to luminosity-weighted mean ages ($Age_{\rm mw}$/\age) and metallicities ($Z_{\rm mw}/Z_{\rm lw}$). The sSFRs (averaged in the last 32 Myr), as well as the mean ages and metallicities of the DR17 MaNGA galaxies, were obtained from the pyPipe3D VAC \citep{Sanchez2022}. These quantities were calculated from the data cubes of each galaxy through spectral inversion using stellar population synthesis models. Note that the ages are \textit{logarithmically} averaged. 
To determine the $g-i$ colour, the corresponding luminosities were taken from the NSA catalogue, with K and E corrections fixed to $z=0$ according to \citet{Dragomir2018} and \citetalias{Rodriguez-Puebla+2020}, and corrected by internal dust attenuation according to \citet{Salim2018}, using the extinction law of \citet[][]{Cardelli+1989}.

The global sSFR, $g-i$ colour, and \age\ are quantities that characterize the star formation (SF) histories of galaxies. The first refers to very recent stages of this history, the second mainly relates to the SF history over the last $\sim 1-2$ Gyr \citep[e.g.,][]{Kennicutt1998}, while the third provides an average of the entire SF history. However, since it is luminosity-weighted, it tends to be biased towards late SF and recent SF bursts if they occurred in the galaxy. Figure \ref{fig:AGE} shows that, in general, less massive and later-type galaxies have higher sSFRs, are bluer and younger than more massive and earlier-type galaxies. A closer look reveals that these quantities change with \ms\ for a given morphological type: the more massive the galaxy, on average, the lower the sSFR, the redder the colour, and the older the age. These trends are more pronounced for sSFR and colour than for \age, at least for types earlier than Sc. This could suggest that the processes of (late) cessation of SF are more dominated by the mass effect than by morphology, that is, mass quenching tends to dominate over morphological quenching. Also based on the IFS MaNGA survey, but with different analysis and arguments, some authors also have concluded that morphology or B/T ratio are less related to the (inside-out) quenching of galaxies than the mass \citep[e.g.,][]{Wang+2018,Bluck+2020}.

We can define \textit{nominal} approximate ranges of universal (mass-independent) values in sSFR, $g-i$ colour, and \age to establish transition zones. For sSFR, we follow the criterion by \citet{Pacifici2016} to classify galaxies as star-forming or quiescent (sSFR$_{th}=1/5t_H(z)$, where $t_H(z)$ is the Hubble time at redshift $z$); we consider a factor of two above and below this criterion to define a transition zone. Based on the cosmology used here and a mean $z$ of 0.045 for MaNGA galaxies, those with log(sSFR/yr)$>-10.5$ are considered star-forming, those with log(sSFR/yr)$<-11.1$ are quiescent, and those in between as transitioning. For the corrected $g-i$ colour, following the literature, we define the [0.70, 0.85] mag interval as the transition zone between the blue cloud and the red sequence, called the ``green valley''. For the luminosity-weighted age, we define the [1.5,3] Gyr interval as the transition zone between old and young galaxies. The dotted lines in the upper panels of Fig. \ref{fig:AGE} indicate these intervals in sSFR, colour, and age, respectively. 

According to Fig. \ref{fig:AGE}, \age\ depends more on \type\ than on mass, with the oldest galaxies being those with earlier-type morphology, and the nominal \age\ transition zone ranges from the S0a type at low masses to the Sb type at high masses, making the \age\ and \type\ transition zones roughly similar. 
Regarding the transition zones in colour (green valley) and in sSFR, they generally go from E--S0 types at low masses to Sb--Sbc types at high masses. This suggests that the late SF history of galaxies depends on both morphology and stellar mass. 
Although we did not present the $D_n$4000 index, it shows a similar trend in the \ms--\type\ diagram as \age, supporting the known correlation between $D_n$4000 and \age. 
Regarding the luminosity-weighted mean stellar metallicity, $Z_{\rm lw}$, less massive and later-type galaxies generally have lower metallicities than more massive and earlier-type galaxies. These trends are similar to those seen in sSFR and $g-i$ colour. For a given \type\ (for types earlier than Sd), the less massive the galaxy, the lower the $Z_{\rm lw}$ on average is. However, for ETGs, $Z_{\rm lw}$ reaches its maximum values in the mass range of 10.7--11.5 in $\log(\ms/\msun)$, with lower values at higher masses. This indicates that the younger stellar populations in the most massive galaxies could result from minor mergers with lower-metallicity galaxies or formed from low-metallicity accreted gas.  

\begin{figure*}
 \includegraphics[width=0.8\textwidth]{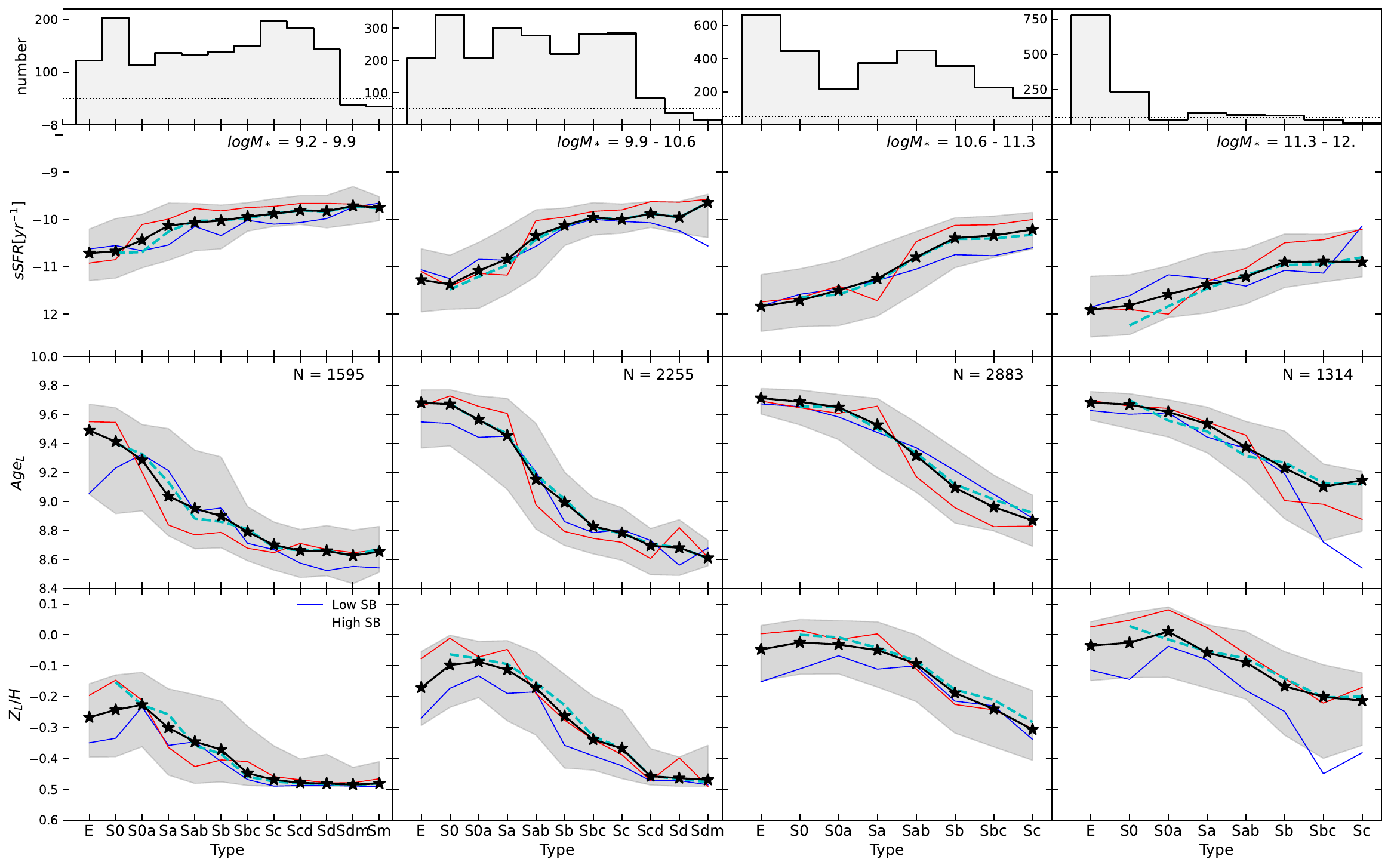}
 \caption{Upper panels: the morphological types histogram in four stellar mass bins. To avoid biased statistics, only bins with more than 10 galaxies are considered in this figure. As reference, dotted line shows the level at 50 galaxies. Middle-upper panels: specific star formation rate (sSFR) as a function of morphological type. Stars represent the median in each bin, and the grey shadow region is the error associated with the 0.16 and 0.84 percentiles. To see a correlation with the mean surface brightness (SB, $\mu = m + 2.5 \times\log(A)$), we consider the median of the sample below the 0.16 percentile of SB (blue lines) and the above the 0.84 percentile of SB (red lines). Dashed cyan lines show the median corresponding to the barred galaxies sample. Lower-middle and lower panels: stellar population light-weighted age and stellar population light-weighted metallicity in terms of the morphological types. Symbols and lines are similar to upper panels.}
 \label{fig:Age-ZH-sigma}
\end{figure*}

We have presented the luminosity-weighted mean ages and metallicities, which may differ from the mass-weighted ones, depending on the SF and chemical enrichment histories of a galaxy. In this sense, the $Age_{\rm mw}$/\age\ ratio measures how coeval or dispersed in time the SF history of a galaxy was, or it can also indicate the presence of very recent bursts of SF  
\citep[e.g.,][]{Plauchu-Frayn2012, Lacerna+2020}. The lower middle panel of Fig.~\ref{fig:AGE} shows the mean values of $\log$($Age_{\rm mw}$/\age) in the same \ms--\type\ diagram. By construction, $Age_{\rm mw}$>\age, but the larger the difference, the more dispersed the SF history and/or the more relevant the presence of recent SF bursts. 
We see that $\log$($Age_{\rm mw}$/\age) depends strongly on \type\ and much less on \ms. For E--Sa galaxies, the ratio is low, suggesting that the SF in most of these early-type galaxies was nearly coeval and poor in recent SF episodes, except for the least massive ones. Intermediate types, Sab-Sc, have values between 0.5 and 0.9 dex (differences of 3-8 Gyr), indicating more extended SF histories and possible contributions from recent SF bursts, especially for the less massive galaxies. Meanwhile, the very late type galaxies, Scd-Irr, which generally present young ages, also have the highest values of $\log$($Age_{\rm mw}$/\age), exceeding 1 dex ($\gtrsim$10 Gyr), suggesting very recent SF episodes for these galaxies. 

Regarding the $Z_{\rm mw}/Z_{\rm lw}$ ratio (last panel of Fig.~\ref{fig:AGE}), it is mostly related to the processes of metal outflow and inflow, as well as mergers during galaxy evolution. From Fig.~\ref{fig:AGE}, we see that for galaxies more massive than approximately $10^{11}$ \msun, $Z_{\rm mw}$ exceeds $Z_{\rm lw}$, regardless of their type. This indicates that in massive galaxies, recycling and chemical enrichment of gas have been very effective during the early active phases of SF, while younger populations formed from less metallic gas, some of which is likely pristine gas accreted by the system. Dry mergers with smaller galaxies, which may have younger and less metal-rich stellar populations than the primary galaxy, also work in this direction.  
For \ms$<10^{11}$ \msun, early-type galaxies still have $Z_{\rm mw}/Z_{\rm lw}$ ratios $\gtrsim1$, as do those from S0a at low masses to Sc at \ms$\sim10^{11}$. For other galaxy types, this ratio drops below 1 and decreases with later types and lower mass, meaning the youngest stellar populations in these galaxies are more metal-rich than the older ones. This suggests that, especially for types later than Sbc, young stellar populations formed from gas that experienced chemical enrichment. In contrast, older populations seem to have formed from more pristine gas. The above may also imply that these galaxies (later than Sbc) do not accrete pristine gas at late times.

\begin{figure*}
 \includegraphics[width=0.8\textwidth]{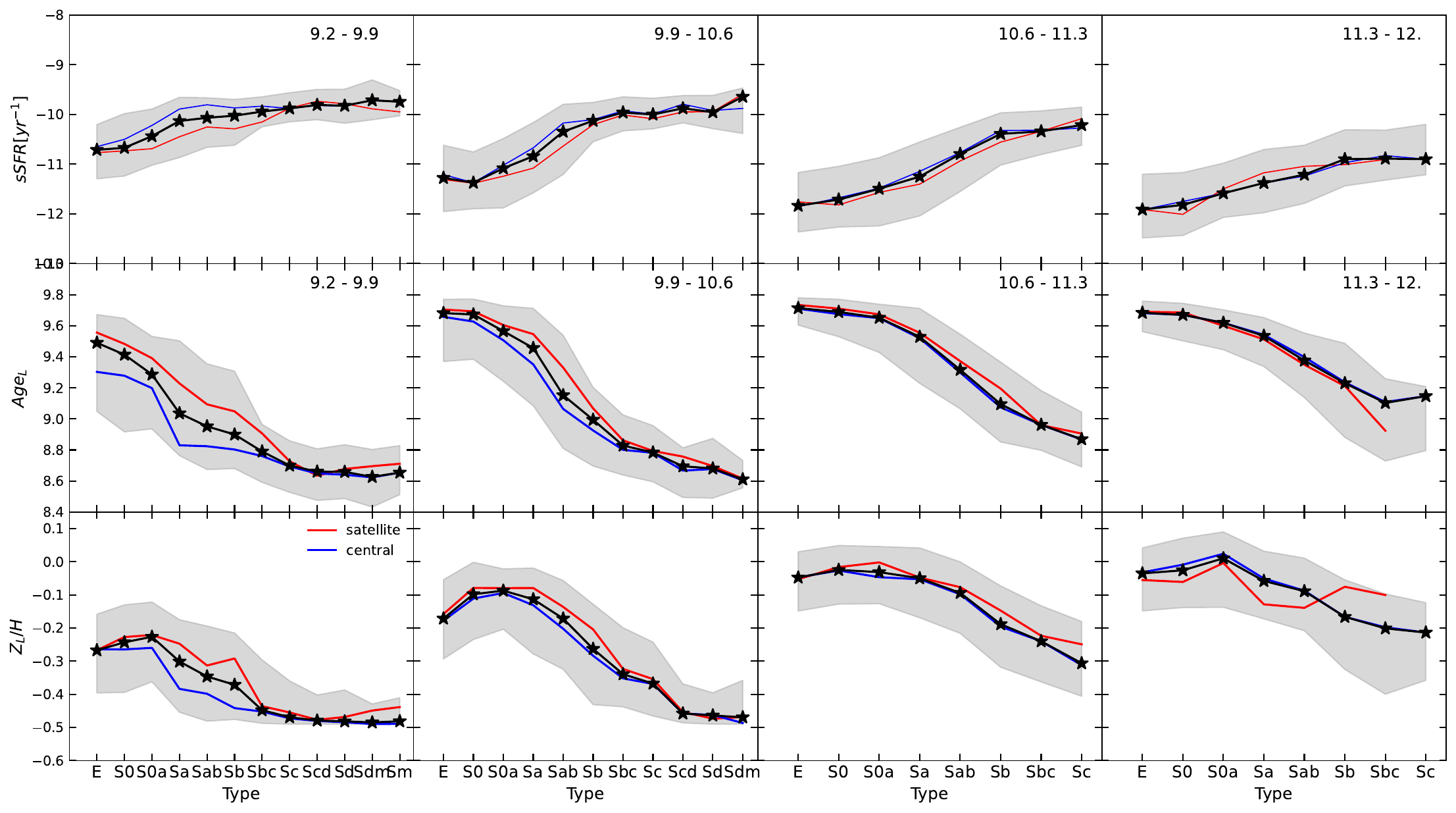}
 \caption{Stars and error region are the same as in Fig~\ref{fig:Age-ZH-sigma}. Red lines show the median of the satellite galaxies for every property as a function of morphology, and the blue line the median of the central galaxies sample. }
 \label{fig:Age-ZH-env}
\end{figure*}

To examine in more detail the dependence of sSFR, \age\, and $Z_{\rm lw}$ on morphology and \ms, we plot in Fig. \ref{fig:Age-ZH-sigma} the medians (stars) and 16--84th percentiles (shadowed grey region) of these quantities for each \type\ across four \ms\ bins. 
To minimize statistical bias, each bin contains at least 10 galaxies, with the number distribution shown on the upper panels.
The sSFR increases with \type\ in all mass bins (and decreases with \ms, on average). However, in the $9.9<\log$(\ms/\msun)$\le$10.6 bin, the correlation reveals two distinct regimes at the earliest and latest types, consistent with the well-known fact that galaxies in this mass range exhibit the strongest bimodality in SFR and colour distributions.  
Overall, \age\ and $Z_{\rm lw}$ decrease with \type\ in all mass bins (with both quantities increasing with \ms). The anticorrelation of \age\ shows two regimes within the $9.9<\log$(\ms/\msun)$\le$10.6 bin. For $Z_{\rm lw}$, the anticorrelations are pronounced from S0a to Sc types, with a flattening observed in later types. Interestingly, $Z_{\rm lw}$ peaks for S0a galaxies, while E and S0 galaxies display lower $Z_{\rm lw}$ than S0a galaxies, especially in the $9.9<\log$(\ms/\msun)$\le$10.6 bin. This may suggest some metallicity mixing in these spheroid-dominated galaxies, likely caused by minor mergers with smaller, less metal-rich, later-type galaxies.
The dashed-cyan lines in Fig. \ref{fig:Age-ZH-sigma} panels indicate the median values solely for barred galaxies. For all types and masses, barred galaxies do not show significant deviation in their sSFR, ages, or metallicities compared to the general galaxy population.  

As shown in Fig. \ref{fig:Age-ZH-sigma}, the scatter in the sSFR correlation and the anticorrelations of \age\ and $Z_{\rm lw}$ with \type\ are large. We have investigated whether these scatterings are segregated by the mean surface brightness (SB, $\mu = m + 2.5 \times\log(2\pi R_{\rm eff}^2)$) 
of galaxies. In Fig. \ref{fig:Age-ZH-sigma}, we plot the corresponding medians for the subgroups with the lowest SBs (LSBs; below the 16th percentile of the corresponding SB distribution) and highest SBs  (HSBs; above the 84th percentile of the corresponding SB distribution), blue and red lines, respectively. 
In the case of the \type--sSFR correlations in different mass bins, for types later than S0a-Sa (disc-dominated galaxies), the scatter appears to be related to the SB, with HSBs galaxies being above the main relation (higher sSFR), and LSB galaxies below it (lower sSFR), that is, lower SB galaxies are less active in forming stars than higher SB ones. In the case of the \type--\age\ correlations, the SB does not appear to be a driver of their scatter. If anything, for early-type galaxies less massive than $\sim 4\times 10^{10}$ \msun, there is some segregation with the SB, with HSB galaxies slightly above (older) the median relation and LSB galaxies below it (younger). In the case of the \type--$Z_{\rm lw}$ correlations, there is also some segregation with the SB for E--Sa galaxies at low masses, and for E to Sb at high masses: HSB galaxies have metallicities slightly above the median relation, while LSB galaxies have lower metallicities than the median relation. For later types, the scatter does not segregate with SB. Note that these findings do not depend on the very late type galaxies, where low numbers could bias the results.
In the next subsection, we explore whether the scatter of the correlations mentioned above could be related to the nature of galaxies as central or satellites.

\subsection{Central and satellite galaxies}
\label{sec:environment}

The environment plays an important role in the evolution of galaxies. One way the environment environment influences galaxies is when a galaxy becomes a satellite within a system dominated by a central galaxy. The gravitational field of the host system (primarily influenced by the dark matter halo of the central galaxy) impacts the evolution of both the dark matter and baryonic components of the accreted satellite galaxy. Using the satellite/central galaxy information provided in the MaNGA GEMA-VAC catalogue \citep[see][for more details]{ArgudoFernandez2015}\footnote{\url{https://www.sdss.org/dr18/data_access/value-added-catalogs/?vac_id=97}}, we examine how much the stellar population properties of central and satellite galaxies differ from each other as a function of \ms\ and morphological type, and how the fractions of central/satellite galaxies vary with morphology and \ms. The central/satellite galaxy distinction in the GEMA VAC is taken from the SDSS halo-based groups catalogue by \citet{Yang2007}. In any given halo-based group, the central galaxy is identified as the most massive galaxy, while the other group members are classified as satellites.

In Fig. \ref{fig:Age-ZH-env}, we analyse whether the sSFR, \age, and $Z_{\rm lw}$ of galaxies vary with their mass and morphological type between central and satellite galaxies. As in Fig. \ref{fig:Age-ZH-sigma}, we plot the median and 16-84th percentiles (black solid lines and shaded areas) of sSFR, \age, and $Z_{\rm lw}$ as a function of \type\ across four mass bins. The medians corresponding to central and satellite galaxies are shown with blue and red solid lines. It is observed that for galaxies more massive than approximately $4\times 10^{10}$ \msun, there are no significant differences in sSFR, \age, and $Z_{\rm lw}$ across any morphology type. If anything, for the most massive galaxies, there is a slight tendency for satellite galaxies with earlier types than Sab to be marginally less metallic than the central ones. However, as we will see below, the fraction of massive satellites is very low, which weakens the statistical significance of this observation. For galaxies less massive than approximately $4\times 10^{10}$ \msun\ and of earlier types than Sbc, satellites tend to have lower sSFRs, and are older and somehow more metallic than the median relations, while the opposite is true for centrals, indicating that environmental factors partially influence the scatter around these relations in these regimes. 
For Sbc and later types, differences between centrals and satellites are minimal. 
Our results suggest that the halo environment may enhance the cessation of SF and chemical enrichment in satellites less massive than approximately $4\times 10^{10}$ \msun\ and of earlier types than Sbc, which likely also experienced structural/dynamical transformations. As we will see below, the smaller the mass, the higher the fraction of satellites in earlier types. Satellite galaxies of types Sbc and later are probably recently accreted, thus still resembling their central counterparts. In contrast, massive satellites, which are relatively rare (see below), appear to be less influenced by the halo environment than their less massive counterparts. 

\begin{figure}
 \includegraphics[width=\columnwidth]{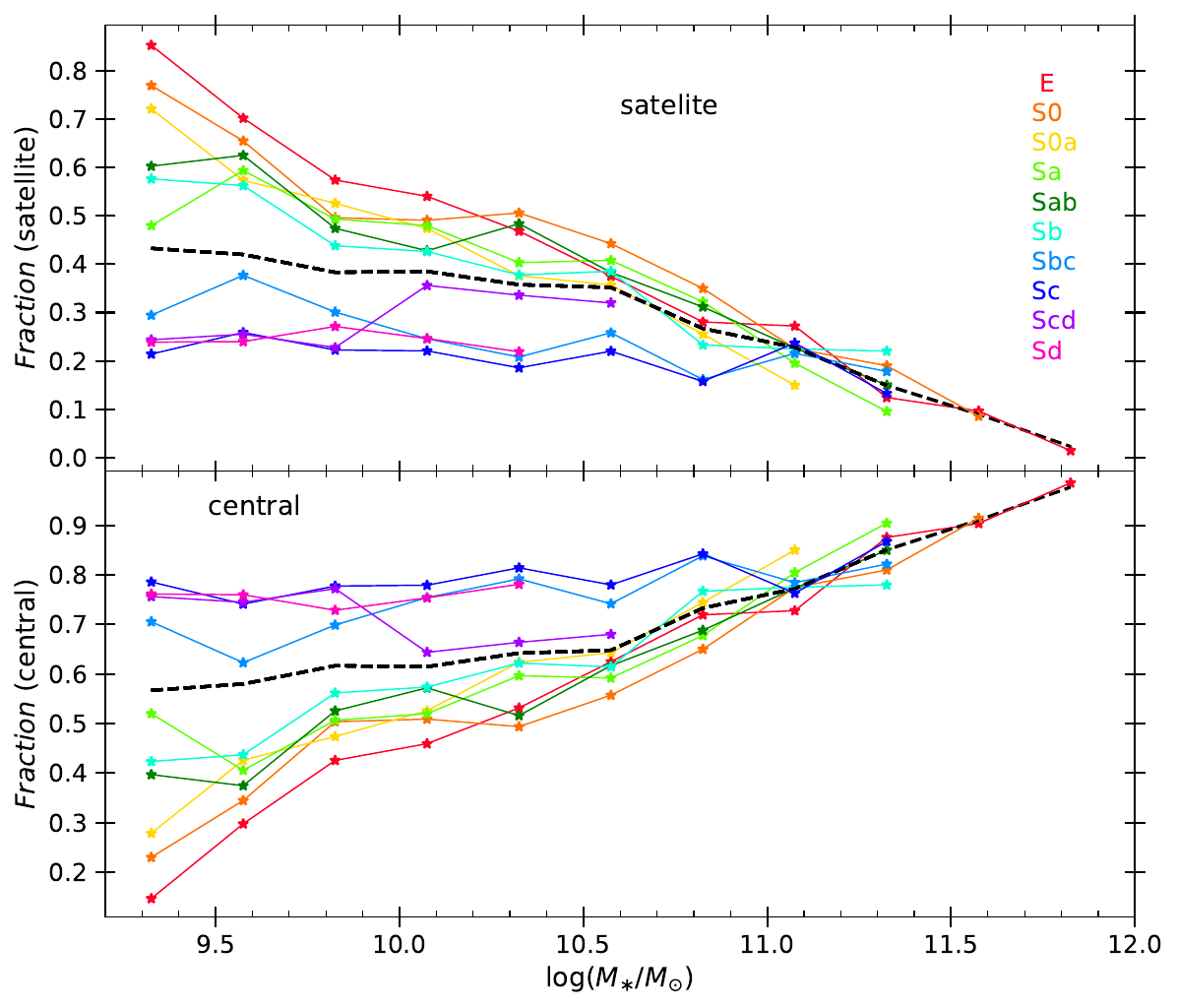}
 \caption{\textit{Top panel:} Fraction of satellite galaxies as a function of stellar mass (black dashed line) and for each morphological type (colour solid lines). \textit{Bottom panel:} Same as in the top panel but for the central galaxies.}
 \label{fig:frac-sat_all}
\end{figure}

In subsect.~\ref{sec:SMF}, we present the dissection of the GSMF by morphology types (see Fig. \ref{fig:SMF_all}). An important question is what fractions of central and satellite galaxies contribute to the GSMF of each morphology type. Figure~\ref{fig:frac-sat_all} shows these morphology volume-complete fractions in different mass bins. The top panel represents the fraction of satellites, while the bottom panel shows the fractional complement, that is, the fractions of central galaxies. The black dashed lines indicate the overall fractions of satellites and centrals as a function of mass, namely $\phi_{\rm sat}(\ms)/\phi(\ms)$ and $\phi_{\rm cen}(\ms)/\phi(\ms)$. For $\ms\lesssim 3\times 10^{10}$ \msun, the total satellite fraction is about $40\%$, but decreases to less than 5\% at higher masses. 

According to Fig. \ref{fig:frac-sat_all}, the satellite and central fractions of each morphological type have very different dependencies on mass. At high masses, satellite fractions are low across all galaxy types. However, as mass decreases, the satellite fraction increases for earlier types. In contrast, for later types, it stays roughly the same around $\sim 20$\% up to $\sim 2\times 10^{11}$ \msun. Conversely, for central galaxies, late types are dominant at all masses. The most massive galaxies, which are mainly early types, are almost all centrals. 
The rising fraction of satellites among early types at lower masses suggests that a primary mechanism by which galaxies become early types, in the low-mass range, is through environmental processes, especially when a galaxy becomes a satellite.

In general, the lower the mass, the more common late-type galaxies (see Figs. \ref{fig:mass_morpho} and \ref{fig:SMF_all}). However, as shown in Fig. \ref{fig:mass_morpho}, there is a significant population of low-mass early-type galaxies. Additionally, Fig. \ref{fig:SMF_all} indicates that the number density of E and S0 galaxies does not decrease at low masses. To add to these findings, Fig. \ref{fig:frac-sat_all} reveals that most of these low-mass early-type galaxies are satellites. Therefore, the presence of low-mass early-type galaxies, which are not too rare, can primarily be attributed to environmental effects when a low-mass galaxy becomes a satellite.

\section{Discussion}
\label{sec:discussion}

\subsection{Morphology comparisons}
\label{sec:MophComp}

\begin{figure}
 \includegraphics[width=0.95\columnwidth]{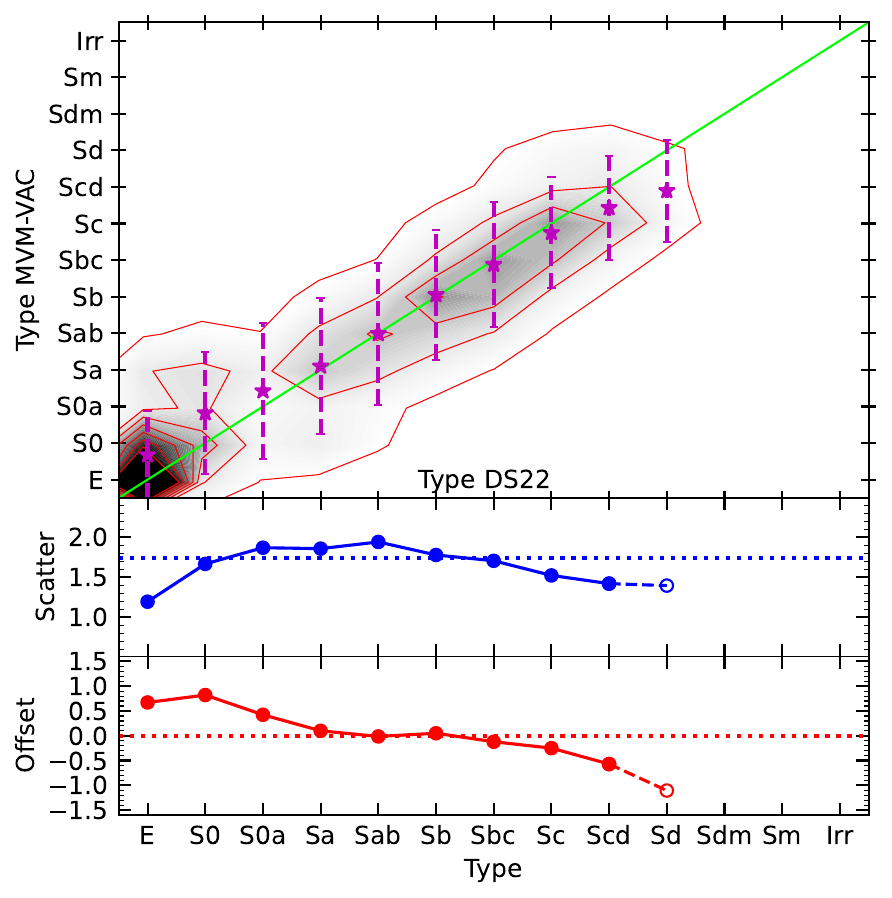}
 \caption{{\it Upper panel:} \citetalias{DominguezSanchez2022} classification vs. the visual classification in this work in 2D density histograms normalized to the total number of the sample. Isocontours are presented in lineal scale and the most external is formed with bins containing at least 0.5\% of the sample (50 galaxies). The one-to-one line is presented in green and the average with the standard deviation (magenta) for each morphology in \citetalias{DominguezSanchez2022}. {\it Lower panel} show our relative offset of the average with respect to the one-to-one line (red), and scatter with respect to the average (standard deviation; blue) for a given morphological bin in \citetalias{DominguezSanchez2022}. Dotted-lines (blue) are the median estimated from all values of scatter for types $\leq$Scd, and the zero point reference for the offset (red). Later types than Scd (dashed and open circles) are omitted due to the low number of galaxies in these bins.}
 \label{fig:Histclass_compDS22}
\end{figure}

Figure~\ref{fig:Histclass_compDS22} compares our final visual classification  with the results of an automatic classification for the entire MaNGA sample, based on machine learning (ML) methods by \citet[][hereafter DS22]{DominguezSanchez2022}. 
We compare our average T-type with the corresponding T-type in \citetalias{DominguezSanchez2022} by measuring an offset, defined as the difference relative to the one-to-one line, while the scatter indicates the 1$\sigma$ deviation from that comparison. We find that, on average, our visual classification and the ML-based classification in \citetalias{DominguezSanchez2022} are within the expected range of dispersion observed when individual observers compare their classifications (usually between 1.3 and 2.3 types, with an average of around 1.8 types), as discussed in \citet{Naim1995}.    
Several points are worth noting from this comparison. At any given morphological type, the observed dispersion reflects not only the inherent variability among classifiers, but also an additional contribution from a broader range of morphological details identified in our visual classification after post-processing the DESI images. This differs from the SDSS $gri$ composite images used as a training set in \citetalias{DominguezSanchez2022}, which capture fewer details. The difficulty in distinguishing inner and outer morphological features in early types in \citetalias{DominguezSanchez2022} could explain, for instance, the different fractions of early types reported in each case. 
For intermediate and late types, the greater dispersion observed in our classifications also suggests an increased contribution from a wider variety of features seen in the discs of galaxies. Our visual classification reflects the contribution of a broader spectrum of morphological features across the different types.

These differences most likely arise from (i) the varying depth and quality of the images used in each case, and from (ii) the combination of our digital post-processing and the light distribution models subtracted from the original DESI images, suited to generate the residual images of the legacy pipelines. These residual images, along with our contrast-enhanced images, proved to be very useful for visually identifying structural components against dominant central light distributions, dusty, clumpy background discs, as well as faint structures in the outer regions of galaxies, which are more difficult to detect in the RGB non-processed SDSS images.

\subsubsection{Bars}

\begin{figure}
 \includegraphics[width=0.9\columnwidth]{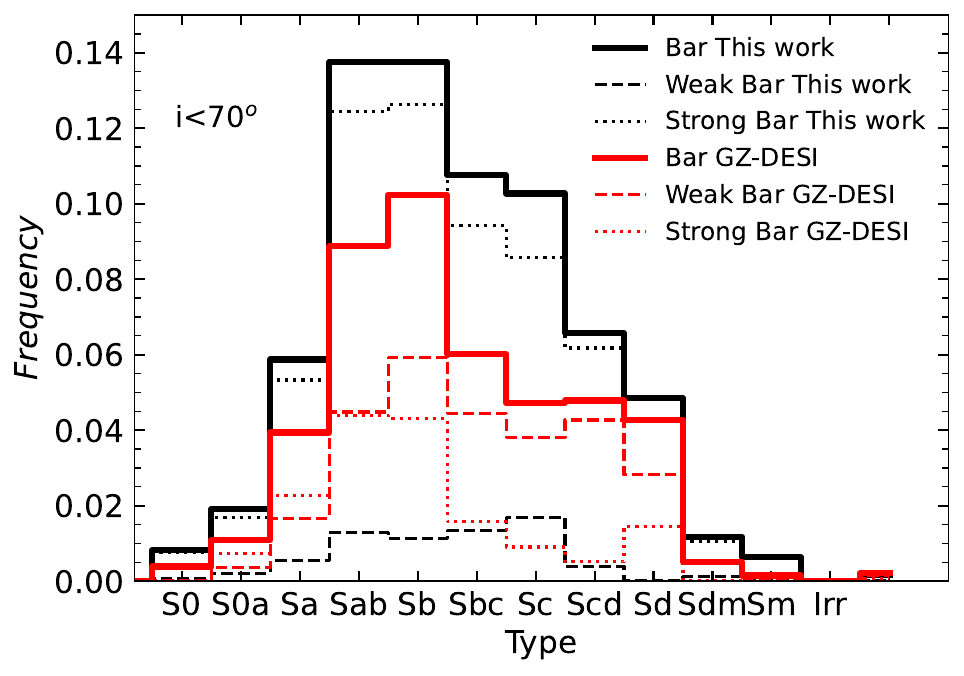}
 \caption{Barred galaxies distribution as a function of morphology, for bars detected in this work (solid-black) and those identified in the GZ-DESI catalogue (solid-red), in the common disc sample. We also separate the weak (dashed-line) and strong (dotted-line) bars according to the GZ-DESI (red) and this work (black) bar identification. These distribution are normalized to the 3,298 discs in common with the GZ-DESI catalogue. }
 \label{fig:bar_comp}
\end{figure}

\begin{table*}
  \begin{center}
    \caption{Number of bars identified in this work and GZ-DESI, for disc galaxies in common and bar information in GZ-DESI, only considering galaxies with $i<$ 70$^{\circ}$.}
    \label{tab:bar-comp}
    \begin{tabular}{c c c c c c c}
    \hline
    \hline    
      \multicolumn{7}{c|}{disc sample in common = 3,298} \\
      \multicolumn{7}{c|}{Barred galaxies in this work = 2,323} \\
      \multicolumn{7}{c|}{Barred galaxies in GZ-DESI = 1,486} \\
    \hline      
      & {\bfseries GZ-DESI} &{\bfseries This work Weak} & \multicolumn{3}{c}{\bfseries This work Strong} & {\bfseries This work}\\
      &  & \textbf{$\underline{A}$B} & \textbf{AB} & \textbf{A$\underline{B}$} & \textbf{B} & \textbf{No bar} \\
      
      \hline
\textbf{This work}    & 1419 & 226 & 966 & 294 & 837 & 975\\
\textbf{Weak GZ-DESI} & 937 & 35 & 402 & 110 & 325 & 64\\
\textbf{Strong GZ-DESI} & 549 & 1 & 105 & 85 & 356 & 2\\
\textbf{No bar GZ-DESI} & 1812 & 190 & 459 & 99 & 154 & 909\\

    \hline
    \end{tabular}
  \end{center}
\end{table*}

We compare our results on the identification of bars and bar families with the more recent ML inferences in the Galaxy-Zoo DESI catalogue \citep[GZ-DESI,][]{Walmsley2023}, trained using DESI images, yielding 9865 galaxies in common. These results are based on a new GalaxyZoo-DECALS decision tree \citep{Walmsley2022} that assists volunteers in identifying bars and bar families. As a reference, we consider the \texttt{gz\_desi\_deep\_learning\_catalog\_friendly} table, which includes vote fractions for answers to questions that the majority of volunteers would have been asked, as described in \citet{Walmsley2023}.

According to the definitions and fraction thresholds proposed by \citet{Walmsley2022} and \citet{Geron2024}, a bar can be identified in oblique disc galaxies, defined as $P_\emph{features/disc}$ $\geq$ 0.27, and in those that are not edge-on $P_\emph{notedge-on}$ $\geq$ 0.68. Each galaxy is also classified based on its bar type (strong, weak, or no bar). Therefore, if the combined $P_\emph{strong bar + weak bar} <$ 0.5, then the galaxy has no bar, but if this sum is $\geq$ 0.5, then it is a barred galaxy. If $P_\emph{strong bar} \geq P_\emph{weak bar}$, then the galaxy has a strong bar; otherwise, it has a weak bar.
The reader should note that in GZ-DESI, the definition of bar families is based on both their length and brightness contribution, 
such that strong bars extend across a large part of the galaxy, contributing significantly to its brightness, whereas weak bars are proportionally smaller in size and brightness relative to the galaxy \citep[][]{Walmsley2023, Geron2021, Geron2024}. 

In the present work, however, our definition of bar families differs from that in GZ-DESI. While a weak bar, denoted as $\underline{A}$B, is an oval structure with its corresponding brightness contribution, strong bar families, denoted as AB, A$\underline{B}$, and B, are well-defined bars, regardless of their size, also considering their brightness contribution, as described in Sec~\ref{Sec:BarClasificaion}.
Following \citet{Walmsley2022} and \citet{Geron2021,Geron2024}, as well as the morphological information in our catalogue, we find 3,298 disc galaxies of all types (\type $\geq$S0) and inclination below 70$^\circ$ in common with GZ-DESI with bar (or no-bar) information. Among those, GZ-DESI reports 1,486 barred galaxies, while we identify 2,323 bars, recovering 95\% (1,419/1,486) of the barred galaxies reported in GZ-DESI (see Table~\ref{tab:bar-comp}). This suggest that, despite using a similar set of images as ours and an improved decision tree for bar identification in GZ-DESI, we are still identifying a larger number of bars, most likely due to the use of a combined set of processed DESI images.

Figure~\ref{fig:bar_comp} shows the frequency distribution of barred galaxies based on their morphology. Note that for galaxies in  GZ-DESI that overlap with our sample, we used our classification since GZ-DESI does not include detailed Hubble morphologies. Barred galaxies in this work (solid-black) and GZ-DESI bars (solid-red) are normalized to the total number of disc galaxies in common. A trend of increasing bar frequency from early types to a peak near Sab-Sb (for the MaNGA sample) and Sb (for GZ-DESI sample) types is evident. From Sb to later types, both samples show a decreasing trend; however, while our bar frequency decreases steadily, GZ-DESI exhibits a less steep decrease up to Sd-Sdm types, where both fractions drop sharply. Figure~\ref{fig:bar_comp} also highlights the contribution of different bar families to the overall bar frequency. Weak and strong bars identified in GZ-DESI are shown with red dashed and red dotted lines, respectively. Similarly, our identification of strong and weak bars in the MaNGA sample is represented with black dotted and black dashed lines, respectively.

Our results, shown in Figures~\ref{fig:Bar_fam} and \ref{fig:bar_comp}, reveal a small contribution of weak bars ($<$12\%) to the MaNGA bar frequency and a clear dominance of strong bars across the entire morphological domain. In contrast, for GZ-DESI, the contribution of weak bars is more comparable to that of strong bars. In Table~\ref{tab:bar-comp}, out of 937 weak bars identified in GZ-DESI that overlap with our sample, we classify 837 as strong bars and only 35 as weak bars. These differences naturally arise from the different definitions used in each sample. Table~\ref{tab:bar-comp} also shows the results of the bar family classifications in GZ-DESI compared to our findings.
Columns 3-6, lower line, list the number of mismatched cases (190+459+99+154 = 902) that GZ-DESI reported as non-barred, but which we classify as barred. These discrepancies are most likely due to our image processing, which, combined with the residual images from the DESI legacy pipeline, allows us to detect "hidden" structures within strong or dominant central light concentrations, as well as providing a clearer identification of structures amid a clumpy and dusty background (see Sec~\ref{sec:methods} and Paper1).

The last column in Table~\ref{tab:bar-comp} reports 64 galaxies that, according to GZ-DESI, are weakly barred but were not identified as barred after our visual analysis. A careful examination of the corresponding images shows that ML inference model in GZ-DESI often misidentifies bars, resulting in false positive in most cases. This is due to the presence of clumpy features creating apparent linear structures in the central regions, strong central dust lines generating confusing structures, or interacting galaxies showing some degree of central distortion. Additionally, for the 2 strong bars reported in GZ-DESI that we did not detected (shown in the last column of Table~\ref{tab:bar-comp}), we found misidentifications on our part.

Also note the apparent absence of bars in S0-S0a types when we limit the samples to a common number. Considering 2923 S0-S0a galaxies in MaNGA that overlap with GZ-DESI, we identified 1011 (35\%) as barred galaxies. Conversely, GZ-DESI reports bar/no-bar information for 378 (13\%) galaxies, indicating bars in only 183 of them (6.3\%). Part of this discrepancy may be related to the decision tree scheme used in GZ-DESI \citep[see][Figure 4]{Walmsley2022}, where galaxies lacking apparent signs of discs are unlikely to be classified as barred. 

Finally, looking at the state-of-the-art simulations, \citet{Roshan2021} have tested the bar fraction measured in TNG50, TNG100 and EAGLE100, and compared it to a bar fraction inferred from observations \citep[][]{Oh2012,Erwin2018}. In general, the bar fraction distributions in terms of stellar mass resulting from these simulations are quite different from those observed in observational studies, especially in this work. Among the reasons for these differences, the resolution of the simulations could be one factor, but a more detailed analysis to understand these results is beyond the scope of the present paper.

\subsection{Implications of the bars and tides distributions}

\subsubsection{Bars: Morphology, Mass, and Colour Dependencies}
\label{sec:bar_dependencies}

\begin{figure}
\centering
 \includegraphics[width=0.8\columnwidth]{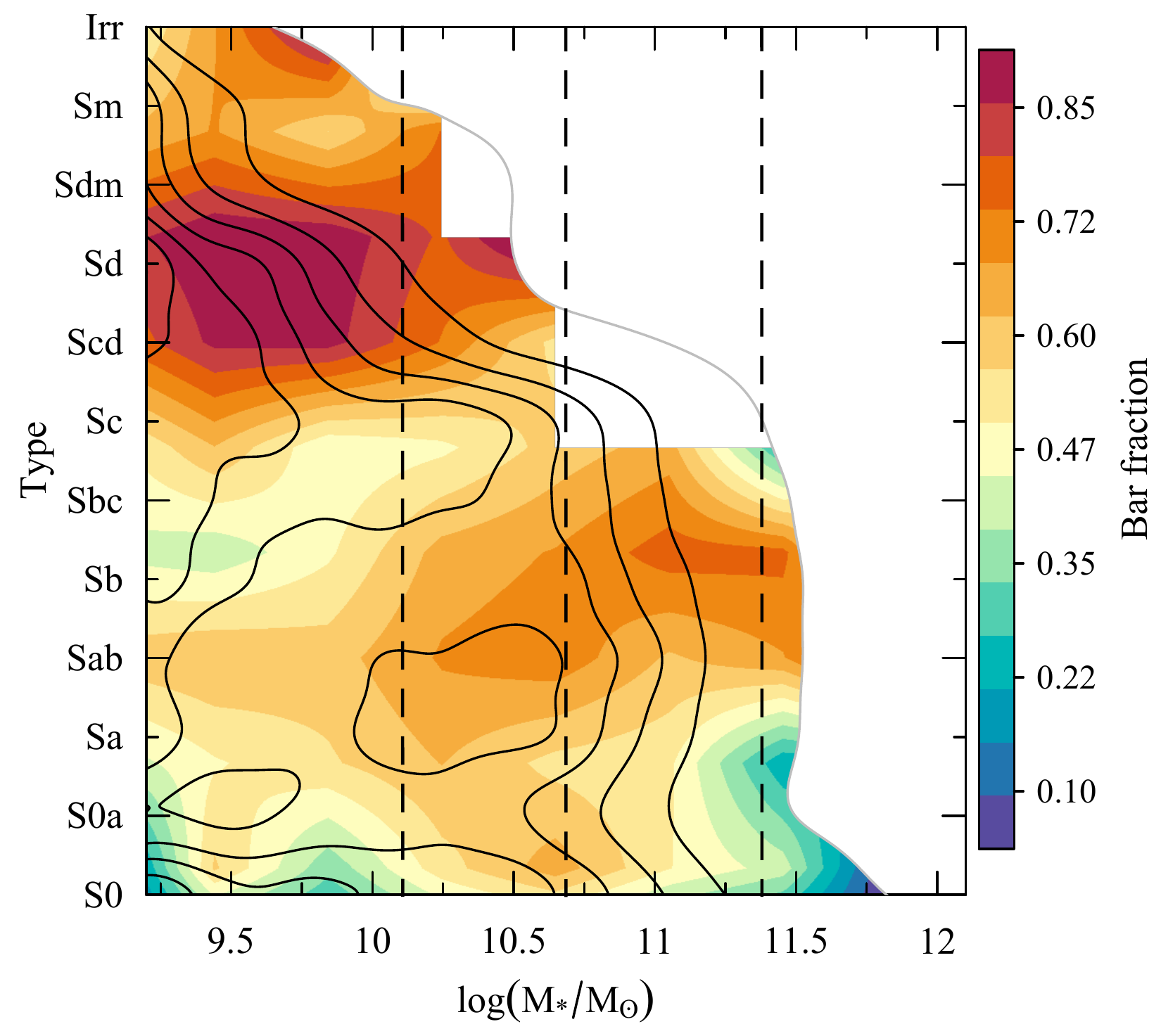}
 \caption{Bivariate distributions in the \type--\ms\ diagram (isodensity contours and dashed lines are the same as right-hand panel in Fig.~\ref{fig:mass_morpho}) coloured by the bar fractions.}
 \label{fig:Bar_mass_morp}
\end{figure}

Figure~\ref{fig:Bar_mass_morp} shows the bivariate \ms-\type\ distribution for the MaNGA sample, colour-coded by the bar fraction. We highlight the following from this figure:  

\begin{itemize}
    \item  Earlier discs (S0-S0a) have the lowest bar fractions at higher stellar masses; however, at intermediate and low masses, the bar fraction slightly increases for these types. 

\item Sa-Sb types display bar fractions above 50\% across the entire stellar mass range, with an increasing fraction towards higher stellar masses. 

\item Interestingly, the highest bar fractions are seen in Scd-Sd types, low-mass galaxies, with fractions exceeding 80\%. Later types (Sdm-Irr) show a slightly lower bar fraction around 60\%. 

\item There is also a transitional region, or valley, displaying a low bar fraction in Sbc-Sc types but within the low-mass regime. This valley has been noted by other authors \citep[e.g.][]{DiazGarcia2016}, as mentioned in Sec~\ref{sec:bars}.  

\end{itemize}

The distribution of MaNGA barred galaxies in this 2D diagram, combined with other physical information such as the characteristic stellar masses inferred in section (GMSF) and results from galaxy structural analysis, could help clarify possible scenarios for bar formation and maintenance.
Table~\ref{tab:bars} in Appendix~\ref{sec:TableBars}, provides a detailed summary of our visual inspection results for bars and bar families based on their morphological types. 

\subsubsection{Tidal features: Morphology and Mass Dependencies}

\begin{figure}
 \centering
 \includegraphics[width=0.8\columnwidth]{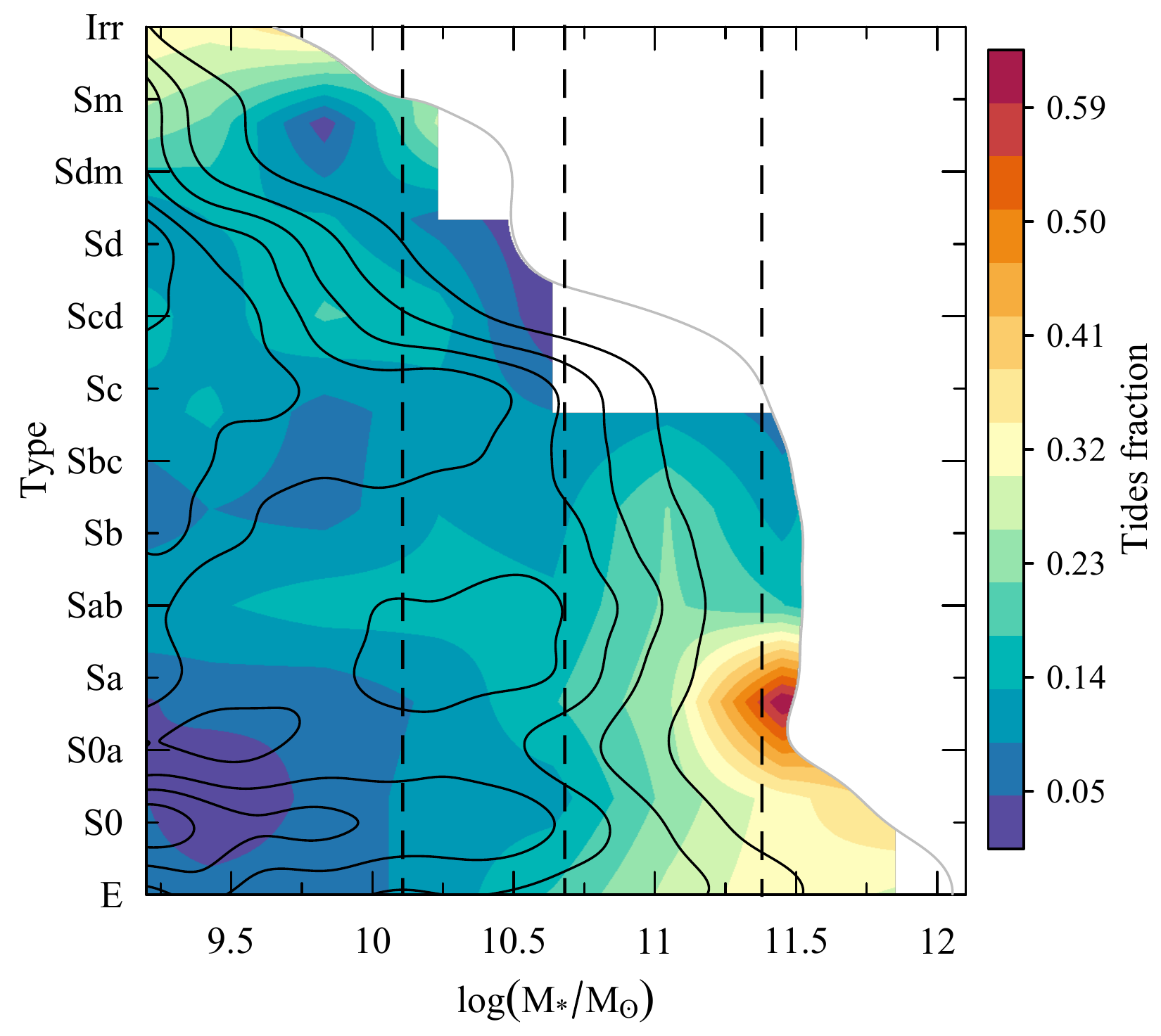}
 \caption{Bivariate distributions in the \type--\ms\ diagram (similar to Fig.~\ref{fig:Bar_mass_morp}) coloured by the tides fractions. }
 \label{fig:tides_frac}
\end{figure}

Figure~\ref{fig:tides_frac} displays the bivariate \type--\ms\ distribution, colour-coded according to the tidal fraction (right palette). The variety of tidal features and their distribution within this 2D diagram can offer insight into potential tidal formation scenarios in the MaNGA sample.
Evidence of merging events, observed through tidal features \citep[e.g.][]{Alladin1965,Toomre1972, Dubinski1996, Cooper2010}, helps infer the history of recent galaxy accretion events, which are likely connected to their local environment and also to their more large-scale environment.  We highlight the following from Fig.~\ref{fig:tides_frac}:

\begin{itemize}
    
\item  For ETGs (E--S0), the higher the mass, the larger the fraction of tidal features, increasing from less than 1\% at log($M_{*}/\msun) \lesssim 10.5$ to over 30\% at log($M_{*}/\msun) > 11.5$.  Thus, the presence of tidal features in ETGs strongly depends on the stellar mass.

\item For intermediate LTGs (Sa--Sc), the fractions of tidal features also rise with mass, but at each mass bin, these fractions are higher than those in ETGs. Massive Sa galaxies have the highest fractions among all galaxy types, exceeding 40\%.

\item For very LTGs galaxies (Scd--Irr), the fractions are generally higher than in earlier-type galaxies. Interestingly, within these galaxies, the less massive ones tend to have higher fractions than the more massive ones. Sdm--Irr galaxies with lower masses have fractions ranging from 30\% to 40\%.

\item Given a stellar mass bin, particularly in the lowest mass bin (log($M_{*}$) < 9.5), there is 
a significant increase in the fraction of tidal features when going from E to Sm-Irr types.

\end{itemize}

In general, we observe that, on one end, the most massive ETGs, including Sa types, show the highest fractions of tidal features in the MaNGA sample. These are most likely gas-poor galaxies with low sSFR. On the other end, the less massive and latest-type galaxies exhibit the second-highest tidal fractions. These galaxies are probably gas-rich with high sSFR values.

A possible explanation for the presence of tidal features, in these two extreme groups, could be related to the nature of the tidal features themselves and their environments, leading to different formation mechanisms in each case. 
As discussed in subsect. \ref{sec:implications}, the GSMF of the MaNGA sample can be described by a triple Schechter function with characteristic stellar masses $M^*_{1} \sim 1.2 \times 10^{10} M_{\odot}$, $M^*_{2} \sim 5.3 \times 10^{10} M_{\odot}$ and $M^*_{3} \sim 2.4 \times 10^{11} M_{\odot}$. The first characteristic mass $M^*_{1}$ marks the transition at the low-mass end of the GSMF, where galaxies are suggested to grow significantly through internal processes such as in situ star formation and secular evolution. The nature of the tidal features in this case could be, either of resonant nature related to close interactions of similar mass galaxies, flyby encounters with gas-rich systems or gas-rich minor mergers.  
According to \citet{Paudel2018} and \citet[]{Subramanian2024}, and references therein, the low-mass galaxies are the most abundant objects in the universe and the interactions, in this mass regime, are found to be carried out in low-density environments but in groups or poor groups.  
Late type galaxies at lower masses are preferably central galaxies (see Fig.~\ref{fig:frac-sat_all}), which can indicate the relatively low density environment where they inhabit. However, a more detailed study is necessary to confirm these results.

The third characteristic mass $M^*_{3}$, in turn, dominates the high-mass end, and is attributed to E and S0 galaxies, which are suggested to be more influenced by ex-situ events such as strong environmental interactions and major mergers. Therefore, the nature of the tidal features in this case could be more likely related to major mergers in denser environments.

\subsection{Implications of the breakdown of the galaxy mass function into morphological types}
\label{sec:implications}

\begin{figure*}
    \includegraphics[width=1\textwidth]{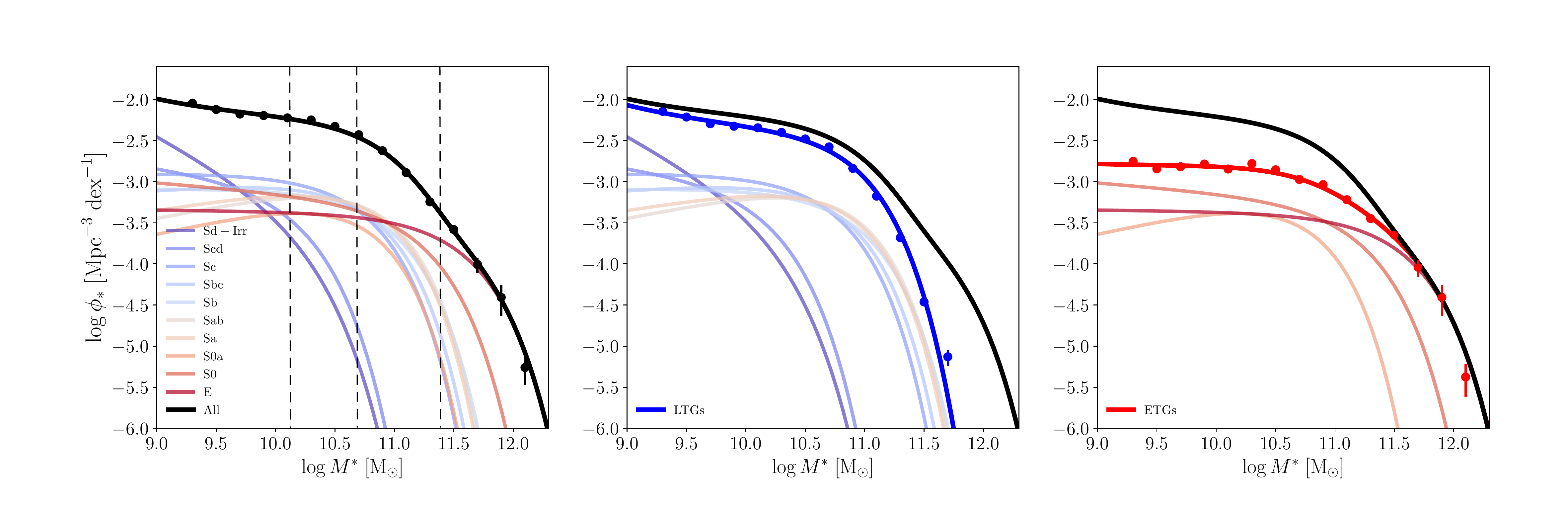}
    \caption{Contribution of each morphological type to the total GSMF (left panel), of types Sa--Irr to the LTG GSMF (medium panel), and of types E--S0a to the ETG GSMF (right panel). The light coloured lines represent the best single Schechter fits for each morphological type as in Fig. \ref{fig:SMF_all}, black line is the sum of all of them, while blue and red lines are the sum of the fits to only the Sa--Irr and E--S0 GSMFs, respectively. The black, blue and red filled circles show the observed total, LTG and ETG GSMFs. The former is best described by a triple Schechter function (see Sect.~\ref{sec:SMF} and Table \ref{tab:fits-triple}) with characteristic masses of $M_1^\ast \approx 1.2 \times 10^{10} \msun$, $M_2^\ast \approx 5.3 \times 10^{10} \msun$, and $M_3^\ast \approx 2.4 \times 10^{11} \msun$ (dashed-lines). The first mass is dominated by galaxies with morphologies later than Scd, the intermediate mass is dominated by S0a-Sc types, and the last mass is primarily dominated by E galaxies. The GSMF for LTGs follows a double Schechter function, while ETGs exhibit a more complex shape but can be roughly approximated by a Schechter function with a large characteristic mass, $\approx 2.4\times 10^{11}$ \msun. At lower masses, the total and LTG GSMFs are dominated by the steep slope of the Sd-Irr morphological types.}
    \label{fig:GSMF_by_morphology}
\end{figure*}

As discussed in Sect.~\ref{sec:SMF} and as seen in Fig. \ref{fig:SMF}, the shape of the total GSMF is complex and requires intricate functions to describe it. With the contributions of each morphological type shown in Fig. \ref{fig:SMF_all} to the total GSMF, we are now better positioned to understand this complexity. The left panel of Figure \ref{fig:GSMF_by_morphology} illustrates how our ten morphological types combine to form the total GSMF. We highlight several important features, some of which, to our knowledge, have not been previously discussed in the literature.

As shown  in Sect.~\ref{sec:SMF}, the total GSMF can be described by a triple Schechter function with characteristic masses of $M^*_1\sim1.2\times10^{10} \msun$,  $M^*_2\sim5.3\times10^{10} \msun$, and $M^*_3\sim2.4\times10^{11} \msun$. 
The first characteristic mass, $M^*_1$, marks the transition at the low-mass end of the GSMF. The second, $M^*_2$, defines the knee of the GSMF, close to the point where ETGs become more abundant than LTGs. The third characteristic mass, previously unidentified, dominates the high-mass end, attributed to elliptical galaxies; it is approximately 6 times larger than $M^*_2$, most likely driven by dry mergers. 

The third Schechter function, with its larger characteristic mass, explains why recent fits to the GSMF suggest a shallower decline at the high-mass end, or even a power-law, instead of an exponential drop-off \citepalias[see e.g.,][]{Rodriguez-Puebla+2020}. When analyzing LTGs and ETGs, the middle panel of the figure shows that LTGs are represented by a double Schechter function, with characteristic masses $M^*_1$ and $M^*_2$, with the low-mass end slope mainly influenced by Sd-Irr types. ETGs are primarily described by a single Schechter function, with lower masses dominated by S0 galaxies and the high-mass end by E galaxies.

\begin{figure}
 \includegraphics[width=\columnwidth]{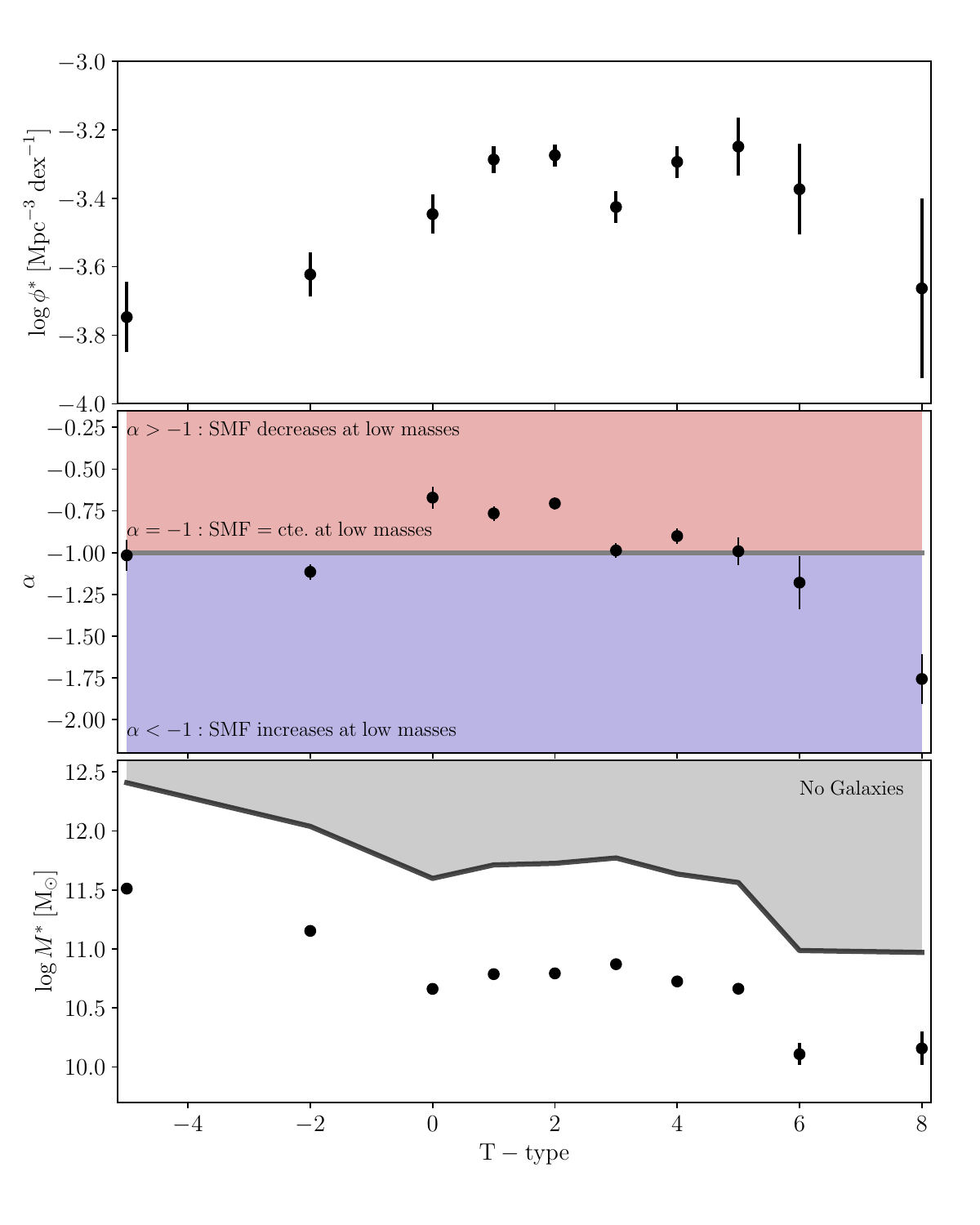}
 \caption{Upper panel: The GSMF fitted normalization from a single Schechter function fit, $\phi^\ast$. Middle panel: The GSMF fitted slope ($\alpha$). The gray solid line represents the case when $\alpha=-1$, for values $\alpha>-1$, the number density of galaxies decreases as mass decreases, whereas for $\alpha<-1$, the number density of galaxies increases. Note that Scd-Irr morphological types, which dominate the low-mass end of the GSMF, become more prevalent at lower masses. Lower panel: The GSMF fitted characteristic stellar mass ($M^*$). The solid dark gray line represents the stellar mass at which the GSMF decreases by three orders of magnitude compared to the number density of $M^*$ galaxies. The probability to observe such galaxies is less than $\sim2\times 10^5$ when accounting galaxies with stellar masses 
 $M_* \geq 10^{9} M_\odot$.
 Equivalently, to observed at least one of such galaxies volumes of ($\sim0.4$ Gpc)$^3$ are needed. These values are reported in Table~\ref{tab:SMF_all}.}
 \label{fig:param}
\end{figure}

Finally, in Figure~\ref{fig:param}, we display the normalization ($\phi^\ast$; upper panel), the slope ($\alpha$; middle panel), and the characteristic mass ($M^*$; lower panel) from the Schechter fits for each morphological type. Generally, the normalization of LTGs is higher than that of ETGs. Specifically, galaxies with types between Sab and Scd have a normalization of $\phi^\ast$ approximately 5$\times 10^{-4}$ Mpc$^{-3}$. In contrast, ETGs have a normalization of $\phi^\ast$ around 2$\times 10^{-4}$ Mpc$^{-3}$, similar to later types like Sd-Irr. Regarding the slopes, the ones for the latest types, Sd-Irr and Scd, are negative. It is encouraging to note that the slope of Sd-Irr galaxies is $\alpha\sim-1.8$, which aligns with the low-mass slope of the halo mass function \citep[see e.g.,][]{Mo+2010,RodriguezPuebla2016}. Since the GSMF is dominated by the very late-type galaxies at low masses, this suggests a flattening in the stellar-to-halo mass relation at these masses. 

Morphological types ranging from S0a to Sbc have slopes greater than $\alpha = -1$, indicating that their GSMFs decrease toward lower masses, meaning their abundance diminishes as mass decreases. In contrast, E and S0 galaxies have values with $\alpha\lesssim-1$, which means their abundance (slightly) increases as mass decreases. In fact, most E and S0 galaxies at these low masses are satellites (see subsect. \ref{sec:environment}).  
Figure~\ref{fig:param} also shows that the characteristic mass, $M^*$, decreases as the galaxy types become later, although it remains roughly constant from S0a to Sc types, with a value of log($M^*$/\msun)$\approx 10.7$. As discussed earlier, this figure reveals at least three characteristic masses that shape the overall GSMF. At lower masses, the characteristic mass for Scd and Sd-Irr galaxies, about 1.2$\times10^{10} \msun$, is smaller than the characteristic masses identified in previous studies. For example, \citet{Stephenson+2024} found a mass of roughly 2.4$\times10^{10} \msun$ where there is a bend in the star-forming main sequence. These authors also noted significant changes in the structural parameters of galaxies at this mass. Another characteristic mass, identified by \citet{Kauffmann2003}, is $3\times 10^{10} \msun$, based on the analysis of galaxy star formation and structural properties, while \citet{Rodriguez-Puebla+2017} found a mass of approximately 3.2$\times 10^{10} \msun$, corresponding to the maximum of the stellar-to-halo mass relation at $z\sim0.1$. 

All of these characteristic masses are about 2–3 times larger than those of Sd-Irr galaxies. In contrast, these masses are smaller than the plateau observed around $M_\ast\sim5\times 10^{10}\msun$ for morphological types between S0a and Sc galaxies, which are more consistent with the characteristic masses reported for the fraction where quiescent galaxies exceed 50\% of the population, ranging from $4$ to $8\times10^{10}\msun$ \citep[see][for further discussion]{Stephenson+2024}. Notice that the plateau is predominantly influenced by later-type galaxies and aligns with the characteristic masses of quiescent galaxies. A possible interpretation suggests a characteristic maximum mass beyond which galaxies can no longer grow significantly through internal processes, such as in-situ SF and secular evolution. 

Surprisingly, our findings indicate that S0 and E galaxies have characteristic masses approximately three and six times larger, respectively, than the plateau observed for Scd--S0a galaxies. This indicates that S0 and E galaxies have experienced additional processes beyond internal evolution, and a possible scenario involves ex-situ events such as environmental interactions and mergers. This aligns with the two-phase scenario \citep[see e.g.,][]{Naab+2009,Oser+2010,Peng+2010,RodriguezGomez2016}, where in-situ stars form during an early rapid phase of SF, followed by a prolonged phase of ex-situ stellar accretion through mergers. Under this assumption, we can use the characteristic masses from Table \ref{tab:SMF_all} to estimate the approximate fraction of mass gained through mergers as follows:

\begin{equation}
    f_{\text{ex-situ},i} = \frac{M_{i}^\ast - \bar{M_{2}^\ast}}{M_{i}^\ast}.
\end{equation}
Here, $\bar{M_{2}^\ast}$ represents the geometric mean of the characteristic masses for morphological types between S0a and Sc, resulting in $\bar{M_{2}^\ast} = 5.63\times10^{10}\msun$, with the subscript $i$ denoting the characteristic masses for S0 and E galaxies. Our results indicate the following ex-situ fraction for these morphological types.
\begin{equation}
	   f_{\text{ex-situ},i} = \left\{ 
			\begin{array}{c l}
				0.827, & i=\mbox{E}\\
				0.604, & i=\mbox{S0}
			\end{array}.\right.
\end{equation}

Indeed, this is consistent with what it is expected from those galaxies \citep{Oser+2010,RodriguezGomez2016}. 

What are the most massive galaxies expected for each morphological type? In Figure \ref{fig:mass_morpho}, we note that when plotting type as a function of stellar mass, there is a 'forbidden region' where galaxies of a specific morphology cannot exceed a certain mass. This is clearly related to the shape of the GSMF, and we describe this relationship quantitatively. To do so, we estimate the stellar mass at which the GSMF drops by three orders of magnitude compared to the number density of $M^*$-galaxies. This is shown as the solid dark line in the bottom panel of Figure \ref{fig:param}. Based on this definition, those galaxies will have a probability of less than approximately 2$\times10^{-5}$ for galaxies with $M_\ast> 10^{9}\msun$ and a number density below approximately 10$^{-7}$ Mpc$^{-3}$. To observe even one such galaxy, a volume of approximately (0.4 Gpc$)^{3}$ would be required. Therefore, the region above the dark solid line in Figure \ref{fig:param} remarks where we do not expect to find massive galaxies. These massive galaxies are approximately 0.9 dex larger than the characteristic mass, consistent with the conclusion of the existence of a forbidden region in Figure \ref{fig:mass_morpho}.

\section{Summary and conclusions}
\label{sec:conclusions}

The MaNGA DR17 dataset presents an unprecedented opportunity to investigate the relationships between morphology, structure, and both global and local stellar population, as well as the properties of ionized gas in galaxies.
In this work, we provide a detailed morphological classification of 10,059 galaxies in the final MaNGA DR17 sample, following the standard Hubble sequence (E, S0, S0a, Sa, Sab, Sb, Sbc, Sc, Scd, Sd, Sm, Irr). Our classification relies on careful visual inspection of mosaics created for each galaxy, which include the combined $gri$ SDSS image, the combined $grz$ DESI image, the filtered-enhanced $r-$band DESI image, and the residual DESI image after subtracting the best-fitting 2D surface brightness model. This strategy is useful for identifying structures that are difficult to recognize in SDSS images or similar images that are not properly processed, making our classification more robust, reliable and abundant in morphological details. 

Although we aimed for a definitive morphological assignment, a flag was given in cases of uncertainty, such as interacting or diffuse objects. We were also able to identify and classify bars, bars families (following the scheme of \citealp{Buta2015}) and bright tidal structures. Additionally, we provide a new estimate of the CAS parameters based on the DESI images.

Our results are compiled in the MaNGA Visual Morphology (MVM) catalogue, included as an SDSS Value-Added Catalogue (see Sec~\ref{sec:VAC}). The MVM catalogue is expected to serve as a reference for studies involving morphological analysis and form the basis for future automatic classifications.

We carried out a series of studies examining morphology and its relationship with bar and tidal fractions, the galaxy stellar mass function (GSMF), and stellar population properties provided by MaNGA. To address sample incompleteness, we used a volume correction. Our main results are summarized as follows.

\begin{itemize}

\item The overall (volume-corrected) distribution of galaxy types follows a bimodal trend with peaks in S0 (12\%) and Scd (35\%) types, and a valley around the S0a and Sa types (Fig. \ref{fig:hist_morpho}). In the bivariate \ms--\type\ plane (Fig. \ref{fig:mass_morpho}), an approximate bimodal clustering is observed, with a transition zone around S0a-Sa types.

\item After volume correction, 54\% of disc galaxies with inclinations $<70^\circ$ exhibit bars, mostly strong, emphasizing their importance in structural evolution and SF regulation. The bar fraction follows a bimodal distribution as a function of \type, with peaks at the Sab-Sb and Scd-Sd types; however, as a function of \ms, the distribution is approximately flat, with an incipient bimodality observed when separating early and late type discs, and a strong decrease for $\ms\gtrsim 10^{11}$ \msun\ (Fig. \ref{fig:Bar_fam}).  

\item Bright tidal features were identified in approximately 13\% of galaxies (after volume-correction) across all morphological types and stellar masses, with higher incidences in massive E--Sa galaxies and low-mass Sdm--Irr galaxies (Fig. \ref{fig:tides_hist}). 

\item We present our estimate of the $CAS$ parameters using DESI images, which offer a more robust estimate due to the improved image quality. The $CAS$ parameters correlate with our morphological classification, with the early types exhibiting higher concentration and lower asymmetry than the late types (Fig. \ref{fig:CAS}), supporting their utility as structural descriptors and predictors of morphology.

\item The GSMFs obtained for all galaxies, and separated into LTGs and ETGs, exhibit complex shapes that were fit using a composition of subexponential Schechter and double power-law functions (Fig. \ref{fig:SMF}). LTGs increasingly dominate at lower masses. 
In contrast, ETGs dominate the GSMF starting from $\msun\sim 10^{11}$ \msun. We find that the GSMF of ETGs does not decrease at lower masses and that most of these low-mass ETGs are actually satellites.

\item We present the GSMFs for each morphological type (Fig. \ref{fig:SMF_all}), a result rarely achieved in the literature. Single Schechter functions provide a good fit to these GSMFs. The low-mass slope increases steeply with decreasing \ms\ for Sd-Irr types ($\alpha_S\sim -1.8$), becomes flatter for earlier types, and from Sbc to S0a types, the slope actually decreases with decreasing \ms. For S0 and E galaxies, the slope  (slightly) increases again (see also Fig. \ref{fig:param}). 

\item The final GSMF, after summing the Schechter GSMFs of each type, requires a triple Schechter function to be adequately described. The characteristic masses of each Schechter are respectively associated with Irr-Sd types, $M_1^\ast\approx1.3\times10^{10}M_\odot$, Sc-S0a types, $M_2^\ast\approx5.3\times10^{10}M_\odot$ (defining the knee of the GSMF), and E types, $M_3^\ast\approx2.4\times10^{11}M_\odot$, which dominate the high-mass end. The `anomalous' high characteristic mass for ETGs suggests significant ex-situ growth through mergers. We estimate that 83\% of the mass in E galaxies and 60\% in S0 galaxies result from this process of mergers.

\end{itemize}

The MaNGA spectroscopic survey allows us to analyse the stellar populations both globally and locally. These data, combined with multiband photometric studies (SDSS and DESI) and the detailed visual morphological analysis presented in this paper, offers valuable insights for understanding galaxy evolution. As an example of the potential of these studies, we discussed some relationships between global stellar population properties and morphology, mass, and environment of galaxies. Our main conclusions from this study are as follows.

\begin{itemize}
    \item Less massive later-type galaxies have higher sSFRs, are bluer, and younger than more massive and earlier-type galaxies (Fig. \ref{fig:AGE}). For a given \type: the more massive the galaxy, the lower the sSFR, the redder the colour, and the older the age, these trends being more pronounced for sSFR and colour than for age, at least for types earlier than Sc. Age (both light and mass weighted), and its ratio, $Age_{\rm mw}$/\age, depend substantially more on \type\ than on mass, suggesting that the mean SF history of galaxies is mainly related to their morphology. The late SF history is influenced by both morphology and stellar mass, and the processes of cessation of SF are more dominated by the mass effect than by morphology. 
    
    \item On average, E--Sa galaxies show evidence of early and coeval SF, Sab--Sc galaxies appear to have more sustained SF histories and possible contribution from recent SF bursts, especially for the less massive galaxies, while Scd--Irr galaxies show evidence of delayed SF with very recent SF episodes.

    \item In general, less massive and later-type galaxies have lower $Z_{\rm lw}$ values than the more massive and earlier-type galaxies. For most of galaxies more massive than $\sim 10^{11}$ \msun, $Z_{\rm mw}>Z_{\rm lw}$, independently of \type, implying that chemical enrichment was very efficient in their early active phases of SF, while younger populations formed from accreted pristine gas and/or these populations, especially for the most massive galaxies, come from dry mergers with lower-mass, younger, and less metallic galaxies. For less massive and late-type galaxies, $Z_{\rm mw}>Z_{\rm lw}$, suggesting, especially for types later than Sbc, that young stellar populations formed from gas that underwent chemical enrichment, while the old populations formed from more pristine gas. 

    \item  Satellite galaxies less massive than $\sim 4 \times 10^{10} M_{\odot}$ and types earlier than Sbc have lower sSFRs, are older, and slightly more metallic than their central counterparts (Fig. \ref{fig:Age-ZH-env}). The fraction of satellites increases significantly for the earlier types at lower masses, while Sc and later-type galaxies are mostly central at all masses $\sim 70-80\%$ (Fig. \ref{fig:frac-sat_all}). These results suggest that the halo environment enhances the cessation of SF and chemical enrichment in low-mass galaxies that become satellites and probably induces their structural/morphological transformation.

\end{itemize}


The present study aims to enhance our understanding of galaxy morphology and its connection to other physical properties at different spatial scales, while also laying the foundation for future research into the role of environment, feedback, and secular evolution in shaping galaxies. The publicly available MaNGA Visual Morphology Value-Added Catalogue (MVM-VAC) will also facilitate studies in machine learning-based galaxy classification and detailed studies of galaxy dynamics.

\section*{Acknowledgements}

JAVM and MHE acknowledge support from the CONAHCYT Postdoctoral program \textit{Estancias Postdoctorales por Mexico}.
JAVM, HMHT, VAR and ARP acknowledge financial support from projects DGAPA-PAPIIT IN106823 and CONAHCyT ``Ciencia de Frontera'' G-543. 
JAVM and HMHT acknowledge financial support from DGAPA-PAPIIT project AG101725.
HMHT acknowledges CONAHCYT project CF-2023-G-1052 and
ARP acknowledges financial support from DGAPA-PAPIIT grants IN106924.
DFM was supported by the Universidad Nacional Autónoma de México Postdoctoral Program (POSDOC).


This research made use of Montage, funded by the National Aeronautics and Space Administration's Earth Science Technology Office, Computational Technnologies Project, under Cooperative Agreement Number NCC5-626 between NASA and the California Institute of Technology. The code is maintained by the NASA/IPAC Infrared Science Archive.
This service also uses software developed by IPAC for the US National Virtual Observatory, which is sponsored by the National Science Foundation.

We acknowledge the use of the Legacy Surveys data; full acknowledgments can be found here: \url{http://legacysurvey.org/acknowledgment/}

Funding for the Sloan Digital Sky Survey IV has been provided by the Alfred P. Sloan Foundation, the U.S. Department of Energy Office of Science, and the Participating Institutions. SDSS-IV acknowledges support and resources from the Center for High-Performance Computing at the University of Utah. The SDSS web site is www.sdss.org.

SDSS-IV is managed by the Astrophysical Research Consortium for the 
Participating Institutions of the SDSS Collaboration including the 
Brazilian Participation Group, the Carnegie Institution for Science, 
Carnegie Mellon University, the Chilean Participation Group, the French Participation Group, Harvard-Smithsonian Center for Astrophysics, 
Instituto de Astrof\'isica de Canarias, The Johns Hopkins University, 
Kavli Institute for the Physics and Mathematics of the Universe (IPMU) / 
University of Tokyo, Lawrence Berkeley National Laboratory, 
Leibniz Institut f\"ur Astrophysik Potsdam (AIP),  
Max-Planck-Institut f\"ur Astronomie (MPIA Heidelberg), 
Max-Planck-Institut f\"ur Astrophysik (MPA Garching), 
Max-Planck-Institut f\"ur Extraterrestrische Physik (MPE), 
National Astronomical Observatories of China, New Mexico State University, 
New York University, University of Notre Dame, 
Observat\'ario Nacional / MCTI, The Ohio State University, 
Pennsylvania State University, Shanghai Astronomical Observatory, 
United Kingdom Participation Group,
Universidad Nacional Aut\'onoma de M\'exico, University of Arizona, 
University of Colorado Boulder, University of Oxford, University of Portsmouth, 
University of Utah, University of Virginia, University of Washington, University of Wisconsin, 
Vanderbilt University, and Yale University.

This project makes use of the MaNGA-Pipe3D dataproducts. We thank the IA-UNAM MaNGA team for creating this catalogue, and the CONACyT-180125 project for supporting them.

\section*{Data Availability}

The catalogue presented in this paper is part of the final data
release of the MaNGA survey, and has been released as part of
the SDSS DR17. This catalogue is available at \url{https://www.sdss.org/dr18/data access/value-added-catalogs/?vac_id=80}. Additional data underlying this article can be shared on reasonable request to the corresponding author.




\bibliographystyle{mnras}
\bibliography{MaNGApaper} 




\appendix

\section{Morphology and Stellar Mass}
\label{App:morph-mass}

Table~\ref{tab:mass} summarizes the relation between stellar mass and morphological types \type\ for the full MaNGA DR17 sample. Two broad panels are shown; the upper one with the original quantities and the lower one with the corresponding quantities after volume corrections. Column 1 shows the stellar mass in intervals of 0.5 dex bins from $\log (\ms/M_{\sun})$ = 9 to $\log (\ms/M_{\sun})$ = 12. Columns 2 to 14 show the numbers and fractions (in parentheses) corresponding to each morphological type \type\ from E to Irr. A lower sub-panel provides for each column, the number and fraction of galaxies per type \type, and the corresponding average and sigma values of $\log (\ms/M_{\sun})$.

\begin{landscape}
\begin{table}
  \begin{center}
    \caption{Top panel: Number of galaxies and fractions to the total sample (in parentheses) in bins of morphology and stellar mass. The lower sub-panel present the total number of galaxies and fraction, the mean stellar mass and the standard deviation of the stellar mean, for every morphological type and sub-class. Bottom panel: the volume-corrected fractions in morphology and stellar mass bins. Lower sub-panel present the total fraction, mean stellar mass and its standard deviation per morphological bin after applying volume corrections. }
    \label{tab:mass}
    \setlength\tabcolsep{1.5pt}
    \begin{tabular}{c c c c c c c c c c c c c c}
      \hline
      \multicolumn{14}{c|}{Original sample} \\
      \hline
      $\log (M_{\ast}/M_{\sun})$ & E & S0 & S0a & Sa & Sab & Sb & Sbc & Sc & Scd & Sd & Sdm & Sm & Irr \\
      \hline
      
$\leq$ 9 & 7 ( 0.001 ) & 1 ( 0.0 ) & 3 ( 0.0 ) & 9 ( 0.001 ) & 5 ( 0.0 ) & 5 ( 0.0 ) & 9 ( 0.001 ) & 12 ( 0.001 ) & 10 ( 0.001 ) & 17 ( 0.002 ) & 9 ( 0.001 ) & 16 ( 0.002 ) & 8 ( 0.001 ) \\ 
9 - 9.5 & 47 ( 0.005 ) & 77 ( 0.008 ) & 29 ( 0.003 ) & 56 ( 0.006 ) & 51 ( 0.005 ) & 85 ( 0.008 ) & 81 ( 0.008 ) & 133 ( 0.013 ) & 135 ( 0.013 ) & 133 ( 0.013 ) & 33 ( 0.003 ) & 42 ( 0.004 ) & 6 ( 0.001 ) \\ 
9.5 - 10 & 115 ( 0.011 ) & 200 ( 0.02 ) & 126 ( 0.012 ) & 167 ( 0.017 ) & 147 ( 0.015 ) & 215 ( 0.021 ) & 231 ( 0.023 ) & 315 ( 0.031 ) & 222 ( 0.022 ) & 167 ( 0.017 ) & 26 ( 0.003 ) & 18 ( 0.002 ) & 2 ( 0.0 ) \\ 
10 - 10.5 & 148 ( 0.015 ) & 264 ( 0.026 ) & 155 ( 0.015 ) & 277 ( 0.027 ) & 237 ( 0.023 ) & 253 ( 0.025 ) & 328 ( 0.032 ) & 295 ( 0.029 ) & 80 ( 0.008 ) & 30 ( 0.003 ) & 10 ( 0.001 ) & 5 ( 0.0 ) & 0 ( 0.0 ) \\ 
10.5 - 11 & 255 ( 0.025 ) & 318 ( 0.031 ) & 194 ( 0.019 ) & 338 ( 0.033 ) & 346 ( 0.034 ) & 334 ( 0.033 ) & 248 ( 0.025 ) & 212 ( 0.021 ) & 12 ( 0.001 ) & 3 ( 0.0 ) & 1 ( 0.0 ) & 2 ( 0.0 ) & 0 ( 0.0 ) \\ 
11 - 11.5 & 845 ( 0.084 ) & 328 ( 0.032 ) & 92 ( 0.009 ) & 220 ( 0.022 ) & 239 ( 0.024 ) & 209 ( 0.021 ) & 124 ( 0.012 ) & 49 ( 0.005 ) & 2 ( 0.0 ) & 1 ( 0.0 ) & 0 ( 0.0 ) & 0 ( 0.0 ) & 0 ( 0.0 ) \\ 
11.5 - 12 & 379 ( 0.038 ) & 119 ( 0.012 ) & 14 ( 0.001 ) & 16 ( 0.002 ) & 24 ( 0.002 ) & 19 ( 0.002 ) & 12 ( 0.001 ) & 5 ( 0.0 ) & 0 ( 0.0 ) & 1 ( 0.0 ) & 0 ( 0.0 ) & 0 ( 0.0 ) & 0 ( 0.0 ) \\ 
$\geq$ 12 & 7 ( 0.001 ) & 5 ( 0.0 ) & 0 ( 0.0 ) & 0 ( 0.0 ) & 1 ( 0.0 ) & 3 ( 0.0 ) & 0 ( 0.0 ) & 2 ( 0.0 ) & 0 ( 0.0 ) & 0 ( 0.0 ) & 0 ( 0.0 ) & 0 ( 0.0 ) & 0 ( 0.0 ) \\ 
\hline
N gal & 1803 ( 0.179 ) & 1312 ( 0.13 ) & 613 ( 0.061 ) & 1083 ( 0.107 ) & 1050 ( 0.104 ) & 1123 ( 0.111 ) & 1033 ( 0.102 ) & 1023 ( 0.101 ) & 461 ( 0.046 ) & 352 ( 0.035 ) & 79 ( 0.008 ) & 83 ( 0.008 ) & 16 ( 0.002 ) \\ 
$\langle \log M_{\ast} \rangle$ & 11.05 & 10.63 & 10.44 & 10.49 & 10.55 & 10.43 & 10.31 & 10.12 & 9.69 & 9.56 & 9.49 & 9.33 & 8.97 \\ 
$\sigma \langle \log M_{\ast} \rangle$ & 0.62 & 0.68 & 0.58 & 0.6 & 0.61 & 0.63 & 0.58 & 0.56 & 0.4 & 0.39 & 0.43 & 0.46 & 0.56 \\ 

    \hline
      \multicolumn{14}{c|}{Volume-corrected fractions} \\
    \hline
      $\log (M_{\ast}/M_{\sun})$ & E & S0 & S0a & Sa & Sab & Sb & Sbc & Sc & Scd & Sd & Sdm & Sm & Irr \\
      \hline

9.2 - 9.5 & 0.016 & 0.022 & 0.008 & 0.014 & 0.012 & 0.022 & 0.024 & 0.036 & 0.031 & 0.029 & 0.007 & 0.007 & 0.0 \\
9.5 - 10 & 0.019 & 0.034 & 0.017 & 0.024 & 0.024 & 0.036 & 0.036 & 0.051 & 0.035 & 0.029 & 0.003 & 0.003 & 0.0 \\
10 - 10.5 & 0.016 & 0.033 & 0.019 & 0.034 & 0.029 & 0.029 & 0.04 & 0.04 & 0.011 & 0.004 & 0.002 & 0.0 & 0.0 \\
10.5 - 11 & 0.016 & 0.023 & 0.012 & 0.023 & 0.024 & 0.02 & 0.017 & 0.019 & 0.0 & 0.0 & 0.0 & 0.0 & 0.0 \\
11 - 11.5 & 0.012 & 0.006 & 0.002 & 0.004 & 0.005 & 0.004 & 0.003 & 0.002 & 0.0 & 0.0 & 0.0 & 0.0 & 0.0 \\
11.5 - 12 & 0.003 & 0.001 & 0.0 & 0.0 & 0.0 & 0.0 & 0.0 & 0.0 & 0.0 & 0.0 & 0.0 & 0.0 & 0.0 \\
$\geq$ 12 & 0.0 & 0.001 & 0.0 & 0.0 & 0.0 & 0.0 & 0.0 & 0.0 & 0.0 & 0.0 & 0.0 & 0.0 & 0.0 \\
\hline
Total frac & 0.084 & 0.119 & 0.058 & 0.099 & 0.094 & 0.111 & 0.12 & 0.148 & 0.077 & 0.063 & 0.011 & 0.011 & 0.0 \\
$\langle \log(M_{\ast}/M_{\sun}) \rangle$ & 10.28 & 10.09 & 10.1 & 10.14 & 10.16 & 10.02 & 10.0 & 9.93 & 9.64 & 9.56 & 9.54 & 9.44 & 9.42 \\
$\sigma (\log(M_{\ast}/M_{\sun})$ & 0.71 & 0.58 & 0.49 & 0.51 & 0.54 & 0.54 & 0.49 & 0.51 & 0.31 & 0.27 & 0.33 & 0.25 & 0.12 \\

      \hline
     \multicolumn{14}{l|}{NOTE: When the values of the volume-corrected fractions are below 1$\times10^{-4}$, the values appear as 0.0 as we are rounding to three decimal places.} \\ 
     \multicolumn{14}{l|}{However, given a mass and morphology bin, if the original number of galaxies in the upper panel is 0, then the volume-corrected fraction is definitely 0.0.} \\
    \end{tabular}
  \end{center}
\end{table}
\end{landscape}


\section{Colouring the \ms\--\type\ diagrams.}
\label{App:colour}

The \ms\--\type\ diagrams presented in this paper were coloured according to the number density (Figure~\ref{fig:mass_morpho}), by the fraction of bars (Figure~\ref{fig:Bar_mass_morp}), the tides fraction (Figure~\ref{fig:tides_frac}) and by the stellar population properties of galaxies (Figure~\ref{fig:AGE}). 

The raster figures are constructed from a regular grid of 10 x 10 bin in the \ms\--\type\ diagram, spanning masses between 8.5 < $\log(\ms/\msun)$ < 12.3 and 
T-types
from -2 to 10. At the centre of each pixel, we obtained the weighted median of a certain property of the galaxy population, considering the entire sample, by associating weights to each galaxy according to a bivariate normal distribution. Weights related to the Vmax corrections are additionally considered. The kernel widths of the bivariate distribution in the x and y directions are obtained automatically following Silverman's rule \citep[][]{Silverman1986}, divided by a factor of 4. This approach has an advantage over traditional methods that consider statistics limited to the geometry of a pixel. The raster matrix of the medians is then interpolated to a resolution of 1000 x 1000 pixels. Subsequently, those pixels corresponding to densities corrected by Vmax in the plane, lower than a certain limit, are removed from the image, avoiding to display extrapolated values of properties in empty regions of the plane. In our case, we define this limit arbitrarily at 10$^{-3}$.

Depending on the property to be shown in the map, the colour palette can be centred on convenient values, for example, in the transition region of the sSFR distribution, the Green Valley in the \textit{g-i} colour distribution, etc. To define an intermediate region, where applicable, we applied the entropic thresholding method \citep[][]{Pandey24}. Briefly, this method is based on two fundamental steps where in the first step the \citet{Otsu75} technique is applied to obtain an optimal threshold for a bimodal distribution, which minimizes the 'intra-class' variance and maximizes the 'inter-class' variance. The last step only considers the region between the means of the modes obtained previously, and iterates between all possible boundaries of the intermediate region until finally keep those that maximize the total entropy, e.g. the sum of the three regions delimited by the boundaries (see more details in \citealt{Pandey24}). These plots were coded in \texttt{R} \citep[][]{RCoreTeam2023}.


\section{Bars families}
\label{sec:TableBars}

Table~\ref{tab:bars} summarizes the results of our visual inspection for bars and bar families in terms of morphological types. Column (1) presents the morphological T-types. Column (2) is the total number of galaxies in the barred ($\underline{A}$B+AB+A$\underline{B}$+B) families. Column (3) presents, for each morphological bin, the number of galaxies in the barred families normalized to the total number of bars in the MaNGA sample, before and after volume corrections (right-after in parenthesis). Columns (4-11) report similar fractions for the individual bar families. At the bottom panel, galaxies were separated into five morphological groups emphasizing the fractions of bars and their corresponding families at early, intermediate, late, very late type and irregular galaxies. All the corresponding volume-corrected fractions are reported in parentheses. 

\begin{table*}
  \begin{center}
    \caption{Distribution of galaxies with bars according to the bar family and morphology. \textbf{N$_{i}$} refers to the number of barred galaxies, 
    followed by the fraction of
    barred galaxies in the MaNGA sample (and volume-corrected fraction in parentheses) for each morphological type and bar family. Bottom panel shows the fractions in terms of 
    5 morphological groups.}
    \label{tab:bars}
    \begin{tabular}{c|c|c|c|c|c|c|c|c|c|c}
      \hline
      \textbf{Type} & \textbf{N$_{\underline{A}B+AB+A\underline{B}+B}$} & \textbf{\underline{A}B+AB+A\underline{B}+B} & \textbf{N$_{\underline{A}B}$} & \textbf{$\underline{A}B$} & \textbf{N$_{AB}$} & \textbf{AB} & \textbf{N$_{A\underline{B}}$} & \textbf{$A\underline{B}$} & \textbf{N$_B$} & \textbf{B}  \\
      \hline
S0 & 274 & 0.07  ( 0.07 ) & 42 & 0.01  ( 0.01 ) & 67 & 0.02  ( 0.02 ) & 48 & 0.01  ( 0.01 ) & 117 & 0.03  ( 0.03 ) \\
S0a & 263 & 0.07  ( 0.06 ) & 55 & 0.01  ( 0.01 ) & 64 & 0.02  ( 0.02 ) & 44 & 0.01  ( 0.01 ) & 100 & 0.03  ( 0.02 ) \\
Sa & 493 & 0.13  ( 0.11 ) & 81 & 0.02  ( 0.01 ) & 163 & 0.04  ( 0.05 ) & 84 & 0.02  ( 0.02 ) & 165 & 0.04  ( 0.04 ) \\
Sab & 657 & 0.17  ( 0.13 ) & 70 & 0.02  ( 0.01 ) & 239 & 0.06  ( 0.05 ) & 120 & 0.03  ( 0.02 ) & 228 & 0.06  ( 0.04 ) \\
Sb & 604 & 0.16  ( 0.12 ) & 48 & 0.01  ( 0.01 ) & 272 & 0.07  ( 0.05 ) & 82 & 0.02  ( 0.01 ) & 202 & 0.05  ( 0.04 ) \\
Sbc & 468 & 0.12  ( 0.12 ) & 50 & 0.01  ( 0.02 ) & 245 & 0.06  ( 0.06 ) & 33 & 0.01  ( 0.01 ) & 140 & 0.04  ( 0.03 ) \\
Sc & 407 & 0.11  ( 0.13 ) & 62 & 0.02  ( 0.02 ) & 237 & 0.06  ( 0.08 ) & 22 & 0.01  ( 0.01 ) & 86 & 0.02  ( 0.03 ) \\
Scd & 272 & 0.07  ( 0.11 ) & 17 & 0.0  ( 0.01 ) & 139 & 0.04  ( 0.06 ) & 15 & 0.0  ( 0.01 ) & 101 & 0.03  ( 0.04 ) \\
Sd & 234 & 0.06  ( 0.1 ) & 3 & 0.0  ( 0.0 ) & 78 & 0.02  ( 0.03 ) & 4 & 0.0  ( 0.0 ) & 149 & 0.04  ( 0.06 ) \\
Sdm & 53 & 0.01  ( 0.02 ) & 5 & 0.0  ( 0.0 ) & 20 & 0.01  ( 0.01 ) & 1 & 0.0  ( 0.0 ) & 27 & 0.01  ( 0.01 ) \\
Sm & 43 & 0.01  ( 0.01 ) & 1 & 0.0  ( 0.0 ) & 11 & 0.0  ( 0.01 ) & 3 & 0.0  ( 0.0 ) & 28 & 0.01  ( 0.01 ) \\
Irr & 3 & 0.0  ( 0.0 ) & 0 & 0.0  ( 0.0 ) & 1 & 0.0  ( 0.0 ) & 0 & 0.0  ( 0.0 ) & 2 & 0.0  ( 0.0 ) \\
\hline
Total & 3779 & 1.0  ( 1.0 ) & 436 & 0.12  ( 0.1 ) & 1542 & 0.41  ( 0.44 ) & 456 & 0.12  ( 0.09 ) & 1345 & 0.36  ( 0.37 ) \\
\hline
S0-S0a & 537 & 0.14  ( 0.13 ) & 97 & 0.03  ( 0.02 ) & 131 & 0.03  ( 0.04 ) & 92 & 0.02  ( 0.02 ) & 217 & 0.06  ( 0.06 ) \\
Sa-Sab & 1150 & 0.3  ( 0.25 ) & 151 & 0.04  ( 0.03 ) & 402 & 0.11  ( 0.1 ) & 204 & 0.05  ( 0.04 ) & 393 & 0.1  ( 0.08 ) \\
Sb-Sbc & 1072 & 0.28  ( 0.24 ) & 98 & 0.03  ( 0.03 ) & 517 & 0.14  ( 0.11 ) & 115 & 0.03  ( 0.02 ) & 342 & 0.09  ( 0.08 ) \\
Sc-Sd & 913 & 0.24  ( 0.35 ) & 82 & 0.02  ( 0.03 ) & 454 & 0.12  ( 0.17 ) & 41 & 0.01  ( 0.02 ) & 336 & 0.09  ( 0.13 ) \\
Sdm-Irr & 99 & 0.03  ( 0.03 ) & 6 & 0.0  ( 0.0 ) & 32 & 0.01  ( 0.01 ) & 4 & 0.0  ( 0.0 ) & 57 & 0.02  ( 0.02 ) \\
      \hline
    \end{tabular}
  \end{center}
\end{table*}

Bar families are present along all the Hubble sequence, similar to that reported in, e.g., \cite{Buta2015}.
From Table~\ref{tab:bars}, we observe that the underline families ($\underline{A}$B, A$\underline{B}$) are under-represented compared to the main categories. As mentioned previously, this fact was also observed by \citet{Buta2015}, who explained it as a natural consequence of the visual classification of bars.

The tendency observed in Table~\ref{tab:bars} is to increase the number and fraction 
of bars towards the B (strong bar) category in the very early (S0-Sa) and very late types 
(Sd to Sm and Irr) while for types in between (Sab to Scd) the more frequent family is the AB bar family.
The total contribution of each bar family to the morphological bins in the original and (volume corrected) barred sample amounts to 12\% (10\%) for the $\underline{A}$B family, 41\% (44\%) for the $AB$ family, 12\% (9\%) for the A$\underline{B}$ family and 36\% (37\%) for the $B$ bar family, indicating that almost half of the bars in our identification belong to the intermediate AB family, followed by the strong bar family.

The contribution of the bar families to different morphological groups in the bottom panel of Table~\ref{tab:bars} is higher for the intermediate and late type spirals from Sa to Sd, with 30\% (25\%) for the Sa-Sab group, 28\% (24\%) for the Sb-Sbc group and with 24\% (35\%) for the Sc-Scd group, and diminishing for the early (S0-S0a) group with 14\% (13\%) and very late types (Sdm-Irr) with 3\% (3\%). 
Note also that AB and B families are more predominant in the Sa-Sd families, with a corrected percentage of 39\% and 29\% respectively.
According to the definition of AB given in Section~\ref{Sec:BarClasificaion}, these bars can be formed by the inner parts of spiral arms.

Also and equally important is to study the change in bar fraction with galaxy types and the properties of those bars in galaxies of different Hubble types. Some previous works point that the bars in early- and late-type galaxies, appear different in nature, i.e., bars in late type galaxies are rather  knotty and they do not show vertical structure, while bars in early type galaxies appear smooth and show vertical structure apart from barlenses or ansae \citep{Elmegreen1985,Buta2013,Laurikainen2014,Buta2015,HerreraEndoqui2017}. Numerical simulations of the bar instability successfully explain the type of bars in early-type galaxies \citep[e.g.,][]{Athanassoula2015}, but not the type of bars present in late-type galaxies. This fact, together with the observed bimodality in bar fraction for early- and late-type galaxies seem to point to something more fundamental than just a change in bar fraction with respect to galaxy types. Trying to disentangling this require more detailed studies on the properties of bars that will be presented in forthcoming papers.


\section{The MaNGA galaxies that are difficult to classify}
\label{Sec:diffuse}

In Section \ref{sec:morpho_results} we mentioned that $\approx 10\%$ of the galaxies in the MaNGA DR17 sample were difficult to classify. Although we attempted a very crude classification, it is  important to understand the nature of these galaxies flagged as ``unsure''. They refer to intrinsically faint or diffuse objects, some of them barely showing signs of an apparent disc (S-like) at different inclinations, and a few of concentrated appearance with very few resolution elements. Similar to that reported in Paper 
I, cases like pre-mergers (binary almost fusing nuclei), mergers (two or more galaxies already collided plus tidal signatures) and bright dwarf galaxies (when recognized) were omitted from this category as much as possible. Our results indicate that 1021 MaNGA galaxies in the DR17 sample were identified as ``unsure'' objects, with tentative morphologies along the whole Hubble sequence, with an apparent peak at Sbc types. Notice however that when these galaxies are volume-corrected, their morphological distribution moves to later types.

\begin{figure}
 \includegraphics[width=\columnwidth]{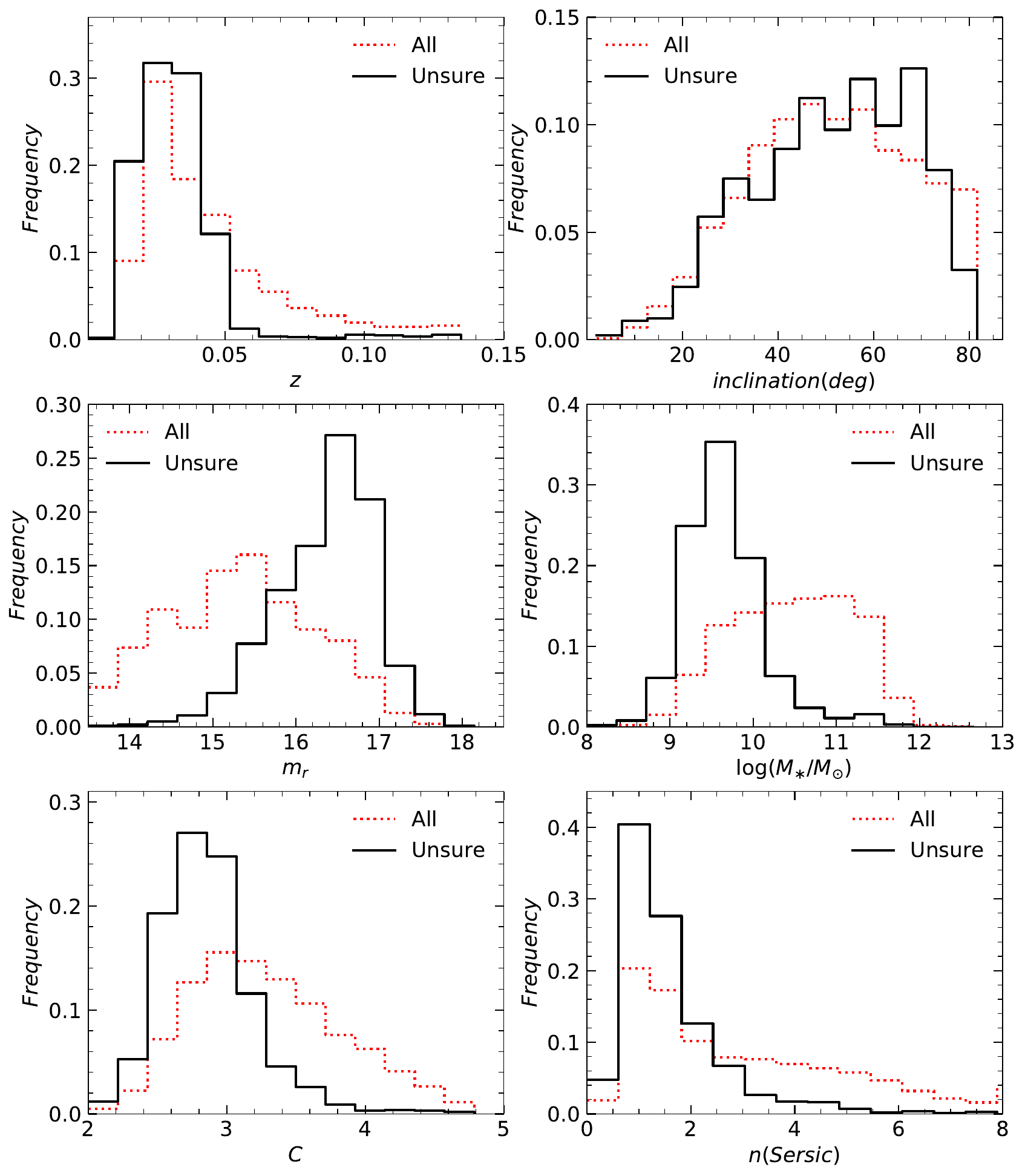}
 \caption{Frequency distribution of galaxies difficult to classify ("unsure"; solid-line) as a function of redshift, inclination, $r-$band apparent magnitude, stellar mass, concentration (C), and S\'ersic index. As reference, red dotted-lines show the distributions for the whole MaNGA DR17 sample. All these distributions are normalized to the total number in every sample. 
 }
 \label{fig:hist_dif}
\end{figure}

Figure~\ref{fig:hist_dif} shows the frequency distribution of the 1021 galaxies identified as ``unsure'' in the full MaNGA sample (black solid line), in terms of redshift (upper left panel), inclination (under the assumption of a measurable external disc; upper right panel), apparent $r-$band magnitude (considering the {\it modelmag} magnitudes from the SDSS database corrected by extinction and K-corr=0.0, middle left panel), stellar mass (middle right panel), concentration ($C$, this comes from our estimate in Section~\ref{sec:cas}, lower left panel) and S\'ersic index $n$ (lower right panel). S\'ersic indices were taken from \citet{DominguezSanchez2022} (S\'ersic fit) after a 2D surface brightness modelling. The red dotted-line histograms correspond to the whole MaNGA sample. 

Figure~\ref{fig:hist_dif} shows that most (95\%) of the MaNGA galaxies identified as ``unsure'' are in the low redshift range, with a median value $z \approx 0.028$, compared to the median $z = 0.038$ for the whole MaNGA DR17 sample.  These are visually faint (90\% having $m_r$ $>$ 15.5 mag) with a median value $m_r \approx 16.52$ mag, compared to a median $m_r$ = 15.32 mag for the whole sample. Although not particularly biased by inclination, most of them are intrinsically of low masses, with a median value of $\log(\ms/\msun)=9.61$, lower than the median for the whole sample, $\log(\ms/\msun)=10.49$.
They are also low concentration galaxies with a median $C$ = 2.82 compared to  $C$ = 3.23 for the whole sample. Finally, most of them are disc-like with a median S\'ersic index $n$ = 1.3 compared to a median $n$ = 2.45 for the whole MaNGA sample.

\bsp	
\label{lastpage}
\end{document}